\input{psfig.sty}
\documentclass[12pt,preprint]{aastex} 
\usepackage{maple2e}
\newcommand{\be}{\begin{equation}}
\newcommand{\ee}{\end{equation}}
\newcommand{\bea}{\begin{eqnarray}}
\newcommand{\eea}{\end{eqnarray}}
\shorttitle{Cusp Core} \shortauthors{A. Del Popolo et al.
}
\begin{document}
\def\kms {\ km s$^{-1}$}
\def\msol{\ifmmode {\>M_\odot}\else {$M_\odot$}\fi}
\def\cmsq{\ifmmode {\>{\rm\ cm}^2}\else {cm$^2$}\fi}
\def\psqcm{\ifmmode {\>{\rm cm}^{-2}}\else {cm$^{-2}$}\fi}
\def\psqpc{\ifmmode {\>{\rm pc}^{-2}}\else {pc$^{-2}$}\fi}
\def\pcsq{\ifmmode {\>{\rm\ pc}^2}\else {pc$^2$}\fi}
\def\Tkev{\ifmmode{T_{\rm kev}}\else {$T_{\rm keV}$}\fi}
\def\hubunits{\ifmmode {\>{\rm km\ s^{-1}\ Mpc^{-1}}}\else {km
s$^{-1}$ Mpc$^{-1}$}\fi}
\def\gta{\;\lower 0.5ex\hbox{$\buildrel > \over \sim\ $}}
\def\lta{\;\lower 0.5ex\hbox{$\buildrel < \over \sim\ $}}
\def\phiIV{\ifmmode{\varphi_4}\else {$\varphi_4$}\fi}
\def\phiI{\ifmmode{\varphi_i}\else {$\varphi_i$}\fi}
\def\be{\begin{equation}}
\def\ee{\end{equation}}
\def\bea{\begin{eqnarray}}
\def\eea{\end{eqnarray}}
\def\beas{\begin{eqnarray*}}
\def\eeas{\end{eqnarray*}}
\def\gtrapprox{\;\lower 0.5ex\hbox{$\buildrel >\over \sim\ $}}
\def\lessapprox{\;\lower 0.5ex\hbox{$\buildrel < \over \sim\ $}}
\def\deg   {$^\circ$}
\def\Ftwo  {$F_{-21}$}
\def\Pcos  {$\Phi^0$}
\def\Jtwo  {$J_{-21}$}
\def\Fcos  {$F_{-21}^0$}
\def\Jcos  {$J_{-21}^0$}
\def\Em    {${\cal E}_m$}
\def\ALL   {A_{\scriptscriptstyle LL}}
\def\JLL   {J_{\scriptscriptstyle LL}}
\def\nuLL  {\nu_{\scriptscriptstyle LL}}
\def\sigLL {\sigma_{\scriptscriptstyle LL}}
\def\tauLL {\ifmmode{\tau_{\scriptscriptstyle LL}}\else 
           {$\tau_{\scriptscriptstyle LL}$}\fi}
\def\nuOB  {\nu_{\scriptscriptstyle {\rm OB}}}
\def\aB    {\alpha_{\scriptscriptstyle B}}
\def\nH    {n_{\scriptscriptstyle H}}
\def\Em{\ifmmode{{\rm E}_m}\else {{\rm E}$_m$}\fi}
\def\NH{\ifmmode{{\rm N}_{\scriptscriptstyle\rm H}}\else {{\rm N}$_{\scriptscriptstyle\rm H}$}\fi}
\def\Ha    {H$\alpha$}
\def\Hb    {H$\beta$}
\def\HI    {H${\scriptstyle\rm I}$}
\def\HII    {H${\scriptstyle\rm II}$}
\def\eg    {{\it e.g.}}
\def\ie    {{\it i.e.}}
\def\cf    {{\it cf. }}
\def\qv    {{\it q.v. }}
\def\etal  {\ et al.}
\def\kms{\ifmmode {\>{\rm\ km\ s}^{-1}}\else {\ km s$^{-1}$}\fi}
\def\Em{\ifmmode{{\cal E}_m}\else {{\cal E}$_m$}\fi}
\def\Dm{\ifmmode{{\cal D}_m}\else {{\cal D}$_m$}\fi}
\def\fesc{\ifmmode{\hat{f}_{\rm esc}}\else {$\hat{f}_{\rm esc}$}\fi}
\def\fescs{\ifmmode{f_{\rm esc}}\else {$f_{\rm esc}$}\fi}
\def\rsolar{\ifmmode{r_\odot}\else {$r_\odot$}\fi}
\def\emunit{\ifmmode{{\rm cm}^{-6}{\rm\ pc}}\else {
cm$^{-6}$ pc}\fi}
\def\intensity{\ifmmode{{\rm erg\ cm}^{-2}{\rm\ s}^{-1}
      {\rm\ Hz}^{-1}{\rm\ sr}^{-1}}
      \else {erg cm$^{-2}$ s$^{-1}$ Hz$^{-1}$ sr$^{-1}$}\fi}
\def\flux{\ifmmode{{\rm erg\ cm}^{-2}{\rm\ s}^{-1}}\else {erg
cm$^{-2}$ s$^{-1}$}\fi}
\def\fluxdensity{\ifmmode{{\rm erg\ cm^{-2}\ s^{-1}\ Hz^{-1}}}\else {erg
cm$^{-2}$ s$^{-1}$ Hz$^{-1}$}\fi}
\def\phoflux{\ifmmode{{\rm phot\ cm}^{-2}{\rm\ s}^{-1}}\else {phot
cm$^{-2}$ s$^{-1}$}\fi}
\def\phorate{\ifmmode{{\rm phot\ s}^{-1}}\else {phot s$^{-1}$}\fi}
\def\apj{{\it Ap.J.~}}
\def\apjs{{\it Ap.J.Suppl.~}}
\def\apss{{\it Astrophys.Sp.Science~}}
\def\aj{{\it Astron.J.~}}
\def\aph{{\it astro-ph}}
\def\mn{{\it MNRAS~}}
\def\araa{{\it Annu.Rev.Astron.Astrophys.~}}
\def\pasp{{\it PASP.~}}
\def\aaa{{\it Astron.Astrophys.~}}
\def\aaas{{\it Astron.Astrophys.Suppl.~}}
\def\astroph{{\it astro-ph}}
\def\rmp{{\it Rev.Mod.Phys.~}}
%\newcommand{\simgreat}{\mbox{$\,\mathrel{\mathpalette\oversim>}\,$}} % >~ sign
%\newcommand{\simless} {\mbox{$\,\mathrel{\mathpalette\oversim<}\,$}} % <~ sign
%\newcommand{\simgreat}{\mbox{$\,\mathrel{\mathpalette\oversim>}\,$}} % >~ sign
%\newcommand{\simless} {\mbox{$\,\mathrel{\mathpalette\oversim<}\,$}} % <~ sign
%%% SOME SPECIAL DEFINITIONS %%%%%%%%%%%%%%%%%%%%%%%%%%%%%%%%%%%%%
\renewcommand{\baselinestretch}{1.0}
\def\pd#1#2{{\upartial #1 \over \upartial #2}}
\def\spd#1#2#3{{\upartial ^2 #1 \over \upartial #2 \upartial #3}}
\def\sspd#1#2{{\upartial ^2 #1 \over \upartial #2^2}}
\def\tfrac#1#2{{\textstyle\frac{#1}{#2}}}
\def\vect#1{{\mathbfit{#1}}}
%%%%%%%%%%%%%%%%%%%%%%%%%%%%%%%%%%%%%%%%%%%%%%%%%%%%%%%%%%%%%%%%%%
\def\eq{equation}
\def\fig{Fig.}
\def\figs{Figs.}
\def\etal{{\it et al.~\/}}
\def\cf{{\it cf.}}
\def\ie{{\it i.e.}}
\def\eg{{\it e.g.}}
\def\ltsima{$\; \buildrel < \over \sim \;$}
\def\simlt{\lower.5ex\hbox{\ltsima}}
\def\gtsima{$\; \buildrel > \over \sim \;$}
\def\simgt{\lower.5ex\hbox{\gtsima}}
\def\fesc{{$\langle f_{\rm esc}\rangle$}\xspace}
\def\h2{H$_2$\xspace}
\def\ion#1#2{\text{#1\,\sc #2}}
\def\HI{{\ion{H}{i} }}
\def\HII{{\ion{H}{ii} }}
\def\GI{{\ion{He}{i} }}
\def\GII{{\ion{He}{ii} }}
\def\GIII{{\ion{He}{iii} }}
\def\Mpc{h$^{-1}$ Mpc\xspace}
\def\kpc{h$^{-1}$ kpc\xspace}
\title{The Cusp/Core problem and the Secondary Infall Model}
\author{A. Del Popolo\altaffilmark{1,}\altaffilmark{2,}\altaffilmark{3}
%, N.Hiotelis\altaffilmark{2} \& J. Pe\~{n}arrubia\altaffilmark{3}
}

\altaffiltext{1}{Argelander-Institut f\"ur Astronomie, Auf dem H\"ugel 71, D-53121 Bonn
}
\altaffiltext{2}{
Dipartimento di Fisica e Astronomia, Universit\'a di Catania, Viale Andrea Doria 6, 95125 Catania, Italy
}
\altaffiltext{3}{Istanbul Technical University, Ayazaga Campus, Faculty of Science and Letters, 34469 Maslak/ISTANBUL, Turkey
}

%\pagerange{\pageref{firstpage}--\pageref{lastpage}} \pubyear{2000}
\begin{abstract}
We study the cusp/core problem using a secondary infall model (SIM) that takes into account the effect 
of ordered and random angular momentum, dynamical friction and baryons adiabatic contraction. The model is applied to 
structures on galactic scales (normal and dwarfs spiral galaxies) and on clusters of galaxies scales. 
Our analysis suggest that angular momentum and dynamical friction are able, on galactic scales, to overcome the competing effect of adiabatic contraction eliminating the cusp. 
The slope of density profile of inner haloes flattens with decreasing halo mass and 
the profile is well approximated by a Burkert's profile. In order to obtain the NFW profile, 
starting from the profiles obtained from our model, 
the magnitude of  angular momentum and dynamical friction must be reduced 
%of a factor $2-3$ 
with respect to the values predicted by the model itself.
%switched off. 
The rotation curves of four LSB galaxies from Gentile et al. (2004) are compared to
the rotation curves obtained by the model in the present paper obtaining a good fit to the observational data. 
The time evolution of the density profile of a galaxy of $10^8-10^9 M_{\odot}$ shows that 
after a transient steepening, due to the adiabatic contraction, the density profile flattens to $\alpha \simeq 0$.
On cluster scales we observe a similar evolution of the dark matter density profile but in this case the density profile slope 
flattens to $\alpha \simeq 0.6$ for a cluster of $\simeq 10^{14} M_{\odot}$. The total mass profile, differently from that of dark matter,
shows a central cusp well fitted by a NFW model.
%............(CAMBIARE I in we)
\end{abstract}
%\begin{keywords}
%cosmology: theory - large scale structure of Universe - galaxies: formation
%\end{keywords}
%\newpage
\section{Introduction}
The formation and structure of dark matter (DM) haloes around galaxies
and clusters of galaxies poses one of the great challenges to theories
of structure formation in an expanding universe. The basic problem of the collisionless 
collapse of a spherical perturbation in an expanding background was first addressed in the 
two seminal papers by Gunn \& Gott (1972) and Gunn (1977), where the cosmological expansion
and the role of adiabatic invariance were first introduced in the
context of the formation of individual objects.  
%The next step was
%accomplished by Fillmore \& Goldreich (1984) and Bertschinger (1985), who
%found analytical predictions for the density of collapsed objects
%seeded by scale-free primordial perturbations in a flat
%universe. Hoffmann \& Shaham (1985) generalised these solutions to
%realistic initial conditions in flat as well as open Friedmann
%models. 
%
%%{ \it To overcome the problem of excessively steep density profiles,
%%$\rho\propto r^{-4}$, obtained
%%in numerical experiments of simple gravitational collapse, Gunn \& Gott (1972),
%%Gott (1975), Gunn (1977), were able to produce shallower profiles, 
%%$\rho\propto r^{-2}$ through the ${\it secondary}$ ${\it infall}$ process.}
%
Self-similar solutions were found by Fillmore \& Goldreich (1984) (hereafter FG84) and
Bertschinger (1985) who found a profile of $\rho\propto r^{-2.25}$.
Hoffman \& Shaham (1985) (hereafter HS), showed that
%, for scale-free initial perturbation spectra,
if structures form from density extrema 
%(assuming a scale-free initial perturbation 
%spectra, $P(k) \propto k^n$, that
%local density extrema are the progenitors of cosmic structures and that the
%density contrast profile around maxima is proportional to
%the two-point correlation function) 
%showed that 
the density profile is $\rho\propto {r^{-\alpha}}$ with
$\alpha=\frac{3(3+n)}{(4+n)}$, recovering Bertschinger's (1985) profile for $n=0$ and
$\Omega=1$. Hoffman (1988) refined the
calculations of HS and made a detailed comparison
of analytical predictions of the 
%secondary infall model 
SIM with the
simulations of Quinn, Salmon \& Zurek (1986) and Quinn \& Zurek (1988). 
%FG84 (NUSSER)

In the attempt to relax the restrictive assumptions underlying the analytic
investigations, to incorporate the full range of non-linear
gravitational effects, and to
study the role of initial conditions in shaping the final structure of the dark
matter halos, Quinn, Salmon \& Zurek (1986) pioneered the use
of N-body simulations to study halo formation.
While Quinn et al. (1986) and
Efstathiou et al. (1988)  found a connection
between the density profiles of collapsed objects and the initial fluctuation spectrum
for Einstein-de Sitter universes (in particular Efstathiou et al. (1988)
found density profiles steepening with increasing spectral index $n$),
West et al. (1987)
arrived at the opposite conclusion. In any case, the previous studies showed that the mass
density profiles steepen with decreasing $\Omega$, 
%This result is 
in agreement with the result of HS.
More recent studies (Voglis et al. 1995; Zaroubi, Naim \& Hoffman 1996) showed a
correlation between the profiles and the final structures. Finally
Dubinski \& Carlberg (1991), Lemson (1995), Cole \& Lacey (1996), 
Navarro et al. (1996, 1997) (NFW), Moore et al. (1998), Jing \& Suto (2000), Klypin et al. (2001), 
Bullock et al. (2001), Power et al. (2003) and Navarro et al. (2004) 
found that although the spherically-averaged density profiles of the
N-body dark matter halos are similar, regardless of the mass of the
halo or the cosmological model, their profiles are significantly different from
the single power laws predicted by the theoretical studies.
The N-body profiles are characterized by an $r^{-3}$ decline at large
radii and a cuspy profile of the form $\rho(r)\propto r^{-\alpha}$,
where $\alpha < 2$ near the center. The actual value of the inner density 
slope $\alpha$ is a matter of some controversy, with NFW 
%Power et al. (2003) 
suggesting $\alpha=1$, but with Moore et al.
(1998), Ghigna et al. (2000) and Fukushige \& Makino (2001) arguing for
$\alpha=1.5$, while 
%Klypin et al. (2001) argue that NFW and Moore et al. 
%profiles are compatible with each other, within their range of applicability:
%the profile can only be trusted beyond radii which are 2-4 times larger than 
%the formal resolution of the simulation. 
Jing \& Suto (2000) and Klypin et al. (2001) claimed that the actual value of $\alpha$
may depend on halo mass, merger history, and substructure.
Power et al. (2003) pointed out that the
logarithmic slope becomes increasingly shallow inwards, with little
sign of approaching an asymptotic value at the resolved radii. In that
case, the precise value of $\alpha$, at a given cut-off scale,
would not be particularly meaningful.
This result has been later confirmed by Hayashi et al. (2003) and Fukushige et al. (2004), 
and it is predicted by several analytical
models (e.g., Taylor \& Navarro (2001), Hoeft et al. 2003).
Finally, Navarro et al. (2004) proposed a new fitting formula having a logarithmic 
slope that decreases inward more gradually than the NFW profile. 
%Sumarizing, multi-mass simulation techniques developed in last years,
%which allow one to investigate the formation of individual halos with
%high numerical resolution, revealed systematic differences 
%with the NFW profile.
%As previously reported, in particular, Fukushige \& Makino (2001) reported that NFW
%fits tend to underestimate the dark matter density within the scale
%radius.  Moore et al. (1999) (M99) reached a similar conclusion and interpreted this result
%as indicating a density cusp steeper than that of the NFW profile. 

The previous different results show
%One should note, however, 
that there is no consensus amongst N-body practitioners on the shape of the density profile.
%for such a modified profile (see, e.g., Klypin et al
%2001 and Power et al 2003).
This unsettled state of
affairs illustrates the difficulties associated with simulating the
innermost structure of CDM halos in a reliable and reproducible
manner. The high density of dark matter in such regions demands large
numbers of particles and fine time resolution, pushing to the limit
even the largest supercomputers available at present. 
%As a result,
%many of the studies mentioned above are either of inadequate
%resolution to be conclusive or are based on results from a handful of
%simulations where computational cost precludes a systematic assessment
%of numerical convergence.

While numerical simulations universally produce a cuspy density profile,
%***
observed rotation curves of dwarf spiral and LSB galaxies
 %\citep[e.g.][]{FloresPrimack94,Moore94,Burkert95,KKBP98,BorrielloSalucci01,Blok01,BlokBosma02,Marchesini02, deBlok:03}
seem to indicate that the shape of the density profile at small scales
is significantly shallower than what is found in numerical simulations
(Flores \& Primak 1994; Moore 1994; Burkert 1995; Kravtsov et al. 1998; Salucci \& Burkert 2000; Borriello \& Salucci 2001; de Blok et al. 2001; de Blok \& Bosma 2002; Marchesini et al. 2002; de Blok 2003; de Blok, Bosma \& McGaugh 2003). 
%What is also apparent is
It seems that the data generally favor logarithmic density slopes close to
$0.2$ (de Blok 2003; de Blok, Bosma \& McGaugh 2003).
%$\citep[\eg,][]{deBlok:03}. 
%The recent identification of a kinematically cold stellar substructure in the Ursa Minor dSph
%\citep{Kleyna:03} strongly suggests that the dark halo of that galaxy
%has a central core. 
Using $N$-body simulations Kleyna et al. (2003)
%\cite{Kleyna:03} 
showed
that Ursa Minor dSph would survive for less than $1$ Gyr if the DM
core were cusped. Additionally, Magorrian (2003) found $\alpha = 0.55^{+0.37}_{-0.33}$ for the
Draco dSph. Gentile et al. (2004) (and similarly Gentile et al. 2006)  
decomposed the rotational curves of five spiral galaxies into their stellar, gaseous and dark matter components and fitted the inferred density distribution with various models and found that models with a 
constant density core are preferred.
%(AGGIUNGERE ALTRI).
The largest part of studies of galaxies arrive at the conclusion that cores are preferred to cuspy profiles.
Two exceptions are the papers of van den Bosch \& Swaters (2001) and Swaters et al (2003). van den Bosch \& Swaters (2001)
studied the rotation curves of 19 dwarfs, and claimed that CDM halos are consistent with data. However, as pointed out by Moore (2001), to justify this claim ``they had to throw away half of the Galaxies and adopt unphysicsl (zero) L/M ratios".   
Spekkens et al. (2005) derived inner dark matter halo density profiles for a sample of 165 low-mass galaxies using rotation
curves obtained from high-quality, long-slit optical spectra assuming minimal disks and spherical
symmetry. They measure median inner slopes ranging from $\alpha= 0.22 \pm 0.08$ to $0.28 \pm 0.06$ for various subsamples of the data.

%This discrepancy 
The discrepancy between smulations and observations has been often signaled as a genuine crisis of the CDM scenario and has become known as the ``cusp/core" problem. 
Since LSB galaxies are thought to be ideal for the comparison with theory, as their dynamics are dominated 
by dark matter with little contribution from baryons (Bothun \etal 1997), 
the discrepancy with simulations is particularly troublesome.
The significance of this disagreement, though,
remains controversial and different solutions have been proposed.
A number of authors attribute the problem to
a real failure of the CDM model, or to that of simulations (de Blok et
al. 2001a; de Blok, McGaugh, \& Rubin 2001b; Borriello \& Salucci
2001; de Blok, Bosma, \& McGaugh 2003). This has led to suggestions that dark matter properties may deviate from standard
CDM and  several alternatives have been
suggested, such as warm (Colin et al. 2000; Sommer-Larsen \& Dolgov 2001),
%\citep{Colin00,SommerLarsenDolgov01}
repulsive (Goodman 2000)
%\citep{Goodman00},
fluid (Peebles 2000),
%\citep{Peebles00},
fuzzy (Hu et al. 2000),
%\citep{Hu00},
decaying (Cen 2001),
%\citep{Cen01},
annihilating (Kaplinghat et al. 2000),
%\citep{Kaplinghat00},
or self-interacting (Spergel \& Steinhardt 2000; Yoshida et al. 2000; Dave et al. 2001)
%\citep{SpergelSteinhardt00,Yoshida00,Dave01}
dark matter. 
Others argue that
the inconsistency may reflect the finite resolution of
the observations that has not been properly accounted for in the analysis of
the HI rotation curves  (van den Bosch \etal\ 2000; van den Bosch \& Swaters
2001). Alternatively, it has been suggested that stellar feedback from the first generation of stars formed 
in galaxies was so efficient that the remaining gas was expelled on a timescale comparable to, or less than, the local dynamical timescale. The dark matter subsequently adjusted to form an approximately constant density core (e.g., Gelato \& Sommer-Larsen 1999). This is however unlikely to affect cluster
cusps.

On cluster scales, X-ray analyses have led to wide ranging results,
from $\alpha=0.6$ (Ettori et al. 2002)
%\citep{EttoriFAJ02} 
to $\alpha=1.2$ (Lewis et al. 2003)
%\citep{Lewis03}
or even $\alpha=1.9$ (Arabadjis et al. 2002).
%\citep{Arabadjis02}.
Measurements based on gravitational lensing yield conflicting
estimates as well, either in rough agreement with the results of
numerical simulations (e.g. Dahle et al. 2003; Gavazzi 2003),
%\citep[e.g.][]{Dahle03,Gavazzi03}, 
or finding
much shallower slopes, $\alpha=0.5$ (e.g. Sand et al. 2002; Sand et al. 2004).
%\citep[e.g.][]{Sand02,Sand_03}.
%Gravitational lensing observations show that the inner slope of the DM
%profile in clusters is $1 \simlt \alpha \simlt 1.5$.
%\cite{Dahle:03} studied six massive clusters with masses $M \sim
%10^{15}$ M$_\odot$ at $z=0.3$.
%They find $\alpha=1.4^{+0.2}_{-0.1}$ and $c=1.7^{+0.9}_{-1.3}$ at the
%68 \% confidence level.  \cite{Smith:01} finds $\alpha = 1.3$ for the
%lensing cluster A383 that has a mass $M \sim 10^{14}$ M$_\odot$
%($z=0.188$).  The lensing cluster MS 2137-2353 has been studied
%independently by two groups. Their conclusions are quite different.
%\cite{Gavazzi:03} find $0.7 \le \alpha \le 1.2$, but they note that if the
%identification of a fifth lensed image is confirmed by HST
%observations, an isothermal profile with a flat core fits the
%data better.
%\cite{Sand:02} find a best-fit $\alpha = 0.35$ and $\alpha < 0.9$ at
%99 \% confidence level for the same cluster. These results are
%extended and strengthened by the analysis of six clusters presented in
%\cite{Sand:04}. X-ray studies based on Chandra observations find that
%the inner slope of the DM profile is consistent with $\alpha=1$
%\citep{Arabadjis:02}.  \cite{Lewis:03} use a generalised profile to
%fit their observations and find $\alpha \approx 1.2 \pm 0.04$ for the
%central slope of A2029.
Ricotti's (2003) N-body simulations suggest that density profile of DM haloes 
is not universal (in agreement with Jing \& Suto 2000; Subramanian et al. 2000; Simon et al. 2003b; Cen et al. 2004; Ricotti \& Wilkinson 2004; Ricotti et al. 2007), presenting shallower cores 
in dwarf galaxies and steeper cores in clusters. 
Thus, it seems that there is some evidence coming from observations and simulations for a dependence of the inner slope of dark matter haloes
on halo mass, and hence for the
non-universality of the dark matter profile.
%, in agreement with the findings of R03.
%This uncertainty is the consequence of the fact that the origin of such
%a universal profile is poorly understood.  Two extreme points of view
%have been envisaged. In one of these, it would be caused by repeated
%significant mergers \citep{sw98,rgs98,SCO00,DDH03}, while in the other
%it would be essentially the result of smooth accretion or secondary
%infall \citep{arfh,ns99,pgrs,kull99,M03,wbd04,As04}.
A conclusive theoretical prediction of the central mass distribution
of CDM haloes is therefore an important check for any model of structure
formation. \\

The controversy regarding the ``universal" density profile
and its logarithmic slope at the centre has stimulated a great deal of analytical work.
Apart the theoretical work previously described, starting fron Gunn \& Gott (1972) model,
modifications of  the self-similar collapse model to
include more realistic dynamics of the growth process have been
proposed (e.g.  Avila-Reese et al. 1998; Nusser \& Sheth 1999; Henriksen \& Widrow 1999; Subramanian et al. 2000; Del Popolo et al. 2000 (hereafter DP2000)). 
%proposed \citep[e.g.][]{AFH98, Nusser \& Sheth 1999, HenriksenWidrow99,Lokas00,Kull99,SCO00, DP2000}. 
Several authors (e.g. Syer \& White 1998; Salvador-Sol\'e et al. 1998; Manrique et al. 2003)
%\citep[e.g.][]{syerWhite98, Salvador98, NusserSheth99, Manrique03} 
argue that the central density profile is linked to the
merging history of dark matter substructure, and baryons have been
invoked both to shallow (El-Zant et al. 2001 (hereafter EZ01), 2004; Romano-Diaz et al. 2008)
%\citep[e.g.][]{ElZant01,ElZant_03}
and to steepen (Blumenthal et al. 1986)
%\citep{Blumenthal86} 
the dark matter profile.
Numerous authors have emphasized the effect of
an isotropic velocity dispersion (thus of non-radial
motion) in the core of collisionless haloes. 
Ryden \& Gunn (1987) (hereafter RG87)
%were the first to adopt peaks in the density field as the
%density profile of the proto-halos. RG87 
were the first  to relax the assumption of purely radial self-similar collapse by including
non-radial motions arising from secondary perturbations in the halo.
Gurevich \& Zybin (1988a,b) developed a formalism based on the theory of adiabatic capture (e.g., Lifshitz \&
Pitaevskii 1981), to estimate the mass profile in a spherical collapse
including non-radial motions. White \& Zaritsky (1992) introduced a heuristic source term that switches off at turnaround,
or in another context simply by assigning an angular momentum
distribution at turn-around time (Sikivie et al 1997).
In Avila-Reese \etal (1998) 
%differs from most other works because
%their halos are assembled through a series of discrete merger episodes,
%described by mass aggregation histories (MAH), which are generated from 
%the Gaussian fluctuation field. 
%Their 
dark matter particles were endowed with 
``thermal motions'' resulting in non-radial velocities.
Huss et al. (1999b) showed an NFW-type density cusp flattening relative to
the isothermal profile just where the velocity dispersion changes from predominantly
radial to isotropic. This flattening did not appear, in Huss et al (1999a),
%(\cite{Huss99a}),
for the case of pure radial force. A similar result was obtained
by Tormen et al. (1997), and Teyssier et al. (1997).
%(\cite{Tormen97}), and (\cite{Teyssier97}).
%who also found that singular isothermal
%profiles arise during radial infall while isotropic velocity dispersions
%are associated with flatter profiles. Similarly, Moutarde
%\etal (\cite{Moutarde}) correlate flat density profiles with higher dimensionality
%of the available phase space during infall and they confirm the
%natural development of self-similarity after turn-around.  Thus we
%are motivated to consider a simple extension of the SSIM that
%includes the effects of angular momentum (i.e. of each orbiting
%particle producing in general a smooth velocity `anisotropy' --- the
%limits being purely radial or spherically symmetric in velocity space --- at 
%each point on a spherical `shell').
Nusser (2001) tried two different analytical schemes for non-radial velocities in his halos,
and concluded that angular momentum is most effective when added to a
particle at the time of maximum expansion.
%Lokas \& Hoffman (\cite{LH00}) have studied the consequences of
%abandoning the self-similar aspect of the SSIM in favour of an
%adiabatic invariant calculation, and of introducing a more detailed
%form of the initial power spectrum. However the self-similarity has been
%found to arise naturally in most shell code simulations (i.e. it is not
%an assumption) and in any case the results of Lokas \& Hoffman are very close to those
%cited above for the SSIM. The authors conclude by suggesting that it
%is prominently the presence of angular momentum that flattens the central cusp.
Hiotelis (2002) found that larger amount of angular momentum leads to shallower final density profiles
in the inner region.
EZ01 showed that if the gas is distributed in clumps, dynamical friction 
acting on these clumps moving in the background of dark matter particles, dissipate the clumps orbital energy
and deposit it in the dark matter with the final effect of erasing the cusp. 
Le Delliou \& Enriksen (2003) studied the effects of angular momentum on density profile 
in the SIM
%Self-Similar Infall Model (SSIM) 
showing that 
%for higher and higher angular momentum, there is a point at which 
angular momentum induces an inner
turn around radius at the size of the self-similar core. Particles with
smaller angular momentum will be able to enter the core but with
a reduced radial velocity compared with the purely radial
SIM. 
%This gradual extinction of the mass flux due to increasing particle angular
%momentum (velocity anisotropy) gradually shifts the system from its intermediate quasi-static phase to its
%final virialized phase.
Ascasibar et al. (2004) included non-radial motions obtaining the result that 
dark matter profile is entirely determined by the initial conditions.
Williams et al. (2004), using a semi-analytic scheme based 
on RG87, explored the 
relationship between the specific angular momentum distribution in a halo 
and its density profile. 
Compared to those formed in N-body simulations, their ``semi-analytic'' halos 
are more extended, have flatter rotation curves and have higher specific angular momentum.
%%%%%, even 
%%%%%though they have not yet taken into account the effects of tidal torques.     

Simulations and semi-analytic models agree on outer parts of the haloes' structure and also on properties of haloes substructure; 
the disagreement in the inner regions could be connected to limits in numerical simulations (de Blok 2003; Taylor et al. 2004; see also Section 3 of the present paper for a discussion) or to the fact that dissipationless N-body simulations does not take into account the effects of baryons on dark matter evolution (see Section 3).
%%%in the inner regions, simulations may still suffer from overmerging (Taylor et al. 2004), the artificial disruption of 
%%%substructure due to numerical effects, or other limits of numerical simulations 
%%%(de Blok 2003). 
%\footnote{The processes of relaxation is difficult to quantify, but in the large $N$ limit the discreteness effects inherent to %the N-body technique should tend to vanish, so one tries to use as large a number of particles as computationally possible. 
%Diemand et al. (2004a) explored the effect of resolution on the degree of relaxation
%finding that increasing $N$ slowly reduces the degree of relaxation $\propto N^{-0.25}$
%rather than proportional to $N$ as expected from the collisionless Boltzmann equation.
%Diemand et al. (2004b), performed convergence analysis demonstrating that for their code and integration scheme,
%the radius beyond which they could trust the density profiles scale according to
%the mean interparticle separation. In the best case they reached
%a resolution of about 0.3 $\% r_v$. More recently Stadel et al. (2008) by means of simulations using several billions 
%particle measured the density profile to a distance of 120 pc ($0.05 \%$ of $r_v$).
%Convergence in the density profile and the halo shape scales as
%$N^{-1/3}$, but the shape converges at a radius three times larger at which point the halo
%becomes more spherical due to numerical resolution.
%The effects of particle discreteness in N-body simulations of $\Lambda$CDM models are still an intensively debated issue (Romeo %et al. 2008).
%}. 
Interestingly, the amount of central substructure seen in the semi-analytic haloes 
%considered here 
is consistent with the amount of substructure inferred from strong lensing experiments (Taylor et al. 2004).
%(0312086)). 
Thus the semi-analytic haloes might provide a more accurate picture of the
spatial distribution of substructure around galaxies even if 
an analytical method, no matter how sophisticated, will never be 
able to capture the full extent of complexity of a non-linear process.

Quite apart from the apparent discrepancies between numerical results
and observations, there is yet another problem, on which very little light 
has been shed in spite of the tremendous progress achieved by numerical 
work: namely the problem of the physics underlying the universal density profiles 
predicted by simulations.
%There have been several attempts in the literature to explain the
%shapes of virialized halos.  
Intuitively, one could argue that the
major difference between earlier analytical work and numerical
simulations is that halo formation in N-body simulations proceeds
through repeated mergers. 
Syer \& White (1998) and Nusser \& Sheth (1999)
claimed that the universal profile is a result of hierarchical clustering by mergers of smaller halos into bigger ones.
However, Moore et al. (1999) performed N-body simulations with a cut-off in the power spectrum at small scales
and also obtained halos with cuspy density profiles. This
proves that merging and substructure does not play a critical role in the formation of density cusps.
Huss et al. (1999a,b) found that
simulations of isolated halos collapsing more or less spherically also
result in universal profiles, thus suggesting that hierarchical
merging is not crucial to the outcome.  Instead, they suggested that the
profile is a consequence of a near universal angular momentum
distribution of the halos. Unfortunately, it is unclear what
circumstances lead the halos to this universal angular momentum
distribution. Thus the issue remains unsolved.
%%%Simulations of isolated halos collapsing more or less spherically
%%%also result in universal profiles, thus suggesting that hierarchical merging is not crucial to
%%%the outcome.
%
%There have been a number of attempts to identify a
%mechanism responsible for the formation of the cusp.
%Syer \& White (1998) and Nusser \& Sheth (1999)
%claimed that the universal profile is a result of hierarchical clustering by mergers of smaller halos into bigger ones.
%However, Moore et al. (1999) performed N-body simulations with a cut-off in the power spectrum at small scales
%and also obtained halos with cuspy density profiles. This
%proves that merging and substructure does not play a critical role in the formation of density cusps.
%Huss, Jain \& Steinmetz 1999a,b, 
%results show that the universality of dark halo density profiles
%does not depend crucially on hierarchical merging as has been suggested recently in the literature.
%Rather, it arises because apparently different collapse histories produces a near universal angular
%momentum distribution among the halo particles. 
%Simulations of isolated halos collapsing more or less spherically
%also result in universal profiles, thus suggesting that hierarchical merging is not crucial to
%the outcome.
Nevertheless the very large amount of work carried out by many researchers to date 
using N-body simulations has met with limited success in elucidating the
physics of halo formation. 
The reason is due to the fact that the  point of force of numerical simulations 
(namely to capture the full extent of complexity of a non-linear process) 
is also their weakness: 
%And therein lies the power of the analytical approach: 
numerical simulations yield little physical insight beyond empirical findings precisely because 
they are so rich in dynamical processes, which are hard to disentangle and 
interpret in terms of underlying physics.
%{\it Because of the ``black-box" character of numerical simulations it is not easy to explore intermediate 
%physical processes important for structure formation. }
Analytical and semi-analytical models are much more flexible than N-body simulations (see Williams et al. 2004).
So even if analytical models like SIM treat collapse and virialization of halos that are spherically simmetric, 
that have suffered no major mergers, and that have suffered quiescent accretion, 
they are worth investigating (see next Sections for a discussion). 

In this paper, I shall present an analytical model for haloes formation based on SIM. 
The paper is an extension of DP2000 in which we showed that a simple spherical infall
model gives rise to profiles which are not power laws. So, again, in the present paper, 
we will follow the HS spirit assuming that objects form around maxima of the smoothed density field 
and the model of Zaroubi and Hoffman (1993) (hereafter ZH93).
Differently from DP200 and ZH93, we shall study the collapse in presence of non-radial motions (ordered and random angular momentum), dynamical friction and baryons adiabatic contraction.
%dissipative infall. 

The plan of the paper is the following: in Section 2, we introduce the model that shall be used to calculate the density profiles. Section 3 deals with results and discussion. Section 4 is devoted to conclusions. 
Appendices describe the initial set-up, namely the way initial conditions, angular momentum, dynamical friction and adiabatic contraction are introduced in the model and, how the adiabatic contraction is calculated.

\section{Model}

The simplest version of SIM considers an initial point mass, which acts
as a nonlinear seed, surrounded by a homogeneous uniformly expanding
universe. Matter around the seed slows down due to its gravitational
attraction, and eventually falls back in concentric spherical shells with
pure radial motions. The assumptions of SIM that are most often questioned
are the spherical symmetry and the absence of peculiar
velocities (non-radial motions): in the ``real" collapse, accretion does not
happen in spherical shells but by aggregation of subclumps of matter which
have already collapsed; a large fraction of observed clusters of galaxies
exhibit significant substructure (Kriessler et al. 1995). Motions are not
purely radial, especially when the perturbation detaches from the general
expansion. Nevertheless the SIM gives good results in describing the formation
of dark matter haloes, because in energy space the collapse is ordered
and gentle, differently from the chaotic collapse that is seen in N-body simulations
(Zaroubi, Naim \& Hoffman 1996). 
This is confirmed in other studies (T\'{o}th \& Ostriker 1992; Huss, Jain \& Steinmetz 1999a,b; Moore et al. 1999).
Moreover, judging by the commonness of extended thin spiral disks in the Universe, 
Dark halo formation of disk galaxies may reasonably be described by the SIM because
dynamical disk fragilty implies that major mergers could not have played a significant role in
these cases (e.g., T\'{o}th \& Ostriker 1992). 
We should also add that analytical and semi-analytical models have some advantages on N-body simulations: a) they
are flexible (one can study the effects of physical processes one at a time); b) one   
can incorporate many physical effects at least in a schematic manner; c)  
they are computationally efficient (it takes about 10 s to compute the density profile of a given object at a given epoch on a desktop PC (Ascasibar et al. 2007).

As I showed in DP2000,
the discrepancies between the SIM and some high resolution N-body
simulations are not due to the spherical symmetry assumption
%the previous quoted intrinsic limits
of the SIM but arises because of some non-accurate assumptions
used in its implementation. In the quoted paper, I improved HS model showing that SIM predicts non power-law profiles. 
HS considered a scale-free
initial perturbation spectra, $P(k) \propto k^{n}$ (where $n$ is the spectral index) and assumed that local density extrema are the
progenitors of cosmic structures and that the density contrast profile around maxima is
proportional to the two-point correlation function. They thus showed that $\rho \propto r^{-\alpha}$ with
$\alpha= 3(3+n)/(4+n)$, being $n$ the spectral index.
% , thus recovering Bertschinger's (1985) profile for $n = 0$ and $\Omega= 1$. 
%
%Hoffman (1988) refined the calculations of HS and made a detailed comparison of the analytical
%predictions of the secondary infall model (hereafter SIM) with the simulations by Quinn
%et al. (1986) and Quinn \& Zurek (1988).

%As I showed in DP2000, the predictive power of the SIM is greatly improved when some problems of the previous implementations
%are removed. \\
The modification of HS model in DP2000 are connected to the following observations: \\
%To begin with, 
a) the conclusion that the final density of the profile is $\rho \propto r^{-2}$ for $n <-1$, claimed
by HS, is not a direct consequence of the SIM
model, but it is an assumption made by the quoted authors,
following the study of self-similar gravitational collapse by FG84.
%Fillmore \& Goldreich (1984).  
%In fact,
%as reported by the same authors, in deriving the relation between the density at maximum expansion and
%the final one, HS assumed that each new shell that collapses
%can be considered as a small perturbation to the gravitational
%field of the collapsed halo. This assumption breaks down for $n<-1$. \\
%Secondly, 
b) The assumption made by HS that the initial mean fractional density excess inside the shell, $\delta(r)=\frac{\rho(r)-\rho_b}{\rho_b}$  
\footnote{$\rho_b$ is the mean background (critical) density.} 
%density profile of dark matter haloes 
is proportional to the correlation function $\xi(r)$ (namely  
$\delta(r) \propto \xi(r) \propto r^{-(3+n)}$) is not good for regions internal
to the virial radius, $r_{\rm v}$ 
\footnote{The virial radius of the halo is defined by the radius of a sphere enclosing a given over-density, $\Delta_v$. This last value is obtained in our case from Bryan \& Norman (1998). The relation between the virial radius and the virial mass is given by $M_v = 4 \pi/3 \rho_b \Delta_v r_v^3$}
(see Peebles 1974; Peebles \& Groth 1976;
Davis \& Peebles 1977; Bonometto \& Lucchin 1978; Peebles 1980; Fry 1984; DP2000).
%In the inner regions of the halo, scaling arguments plus the stability
%assumption tell us that $\xi(r) \propto r^{-\frac{3(3+n)}{(5+n)}}$, and
%we expect a slope different from that of HS.
In other words, HS's solution applies only to
the outer regions of collapsed haloes, and consequently the conclusion,
obtained from that model,
that dark matter haloes density profiles
can be approximated by power-laws on their overall radius range
is not correct. 
%It is then necessary to
%introduce a model that can make predictions also on the inner parts of
%haloes. 
c) According to Bardeen et al. (1986),
(hereafter BBKS), the mean
peak profile depends on a sum involving the initial correlation function,
$\xi(r) \propto r^{-(3+n)}$,
and its Laplacian, ${\bf \bigtriangledown}^2 \xi(r) \propto r^{-(5+n)}$
(BBKS; RG87) (see Appendix B), not only the correlation function $\xi(r)$ as in HS. 

As shown in DP2000 (their Eq. 20, and Fig. 6 of the present paper), 
%can be seen for example in the case of a scale-free density perturbation
%spectrum (DP2000, equation (20)), 
the initial mean density obtained 
using the model of that paper is extremely different from that obtained and used in HS, and  
differently from HF's  model the density profiles
are not power-laws but have a logarithmic slope that increase from the
inner halo to its outer parts.
%In the outer parts of the halo the
%density profiles are steeper than that found by HF
%and are consistent with $\rho \propto r^{-3}$ while in the inner part of the
%halo we find $\rho \propto r^{-0.95}$ for $n=-1$ and
%$\rho \propto r^{-1}$ or $n=0$.
% 
%The analytic model proposed by Navarro et al. (1995) is a good fit to the halo profiles. 
%
%The radius, $a$, at which the slope equals $-2$ is a function
%of the mass of the halo and of the spectral index $n$. 
%Lower mass halos are more centrally concentrated than higher ones.
%For a given mass $M$ halos having larger values of $n$ have denser cores. 
%The good agreement of our model (spherical simmetric)
%with several N-body simulations led us to think, in agreement to
%Huss et al. (1998) paper, that the role
%of merging in the formation of halos is not crucial as generally
%believed. 

In summary the model assumes that the initial probability distribution of the density field is Gaussian.
The dynamical evolution of matter at the distance $x_i$ from the peak is determined by the mean cumulative density
perturbation within $x_i$ (see Appendix A Eq. (\ref{eq:overd})) and the maximum radius of expansion can be obtained knowing $x_i$ and 
the mean cumulative density of the perturbation (Eq. (\ref{eq:overd})). 
After reaching maximum radius, a shell collapses and will start oscillating and it will contribute to the inner 
shells with the result that energy will not be an integral of motion any longer. 
The dynamics of the infalling shells is obtained by assuming that the potential well near the center varies adiabatically (Gunn 1977, FG84).
The details of the model, which is an extension of DP2000 to take account of the effects of angular momentum, dynamical friction and dissipative infall on the density profile, are exposed in Appendix A.

The quoted model needs that the initial density profile is given together with angular momentum and dynamical friction. The way these quantities are calculated and introduced in the model are described in Appendix B, C and D, respectively. Appendix E shows how we take into account adiabatic contraction.

%%%%%In  the next section, we will show how the initial conditions, angular momentum and dynamical friction are introduced into the model, and in %%%%%Section 4, we describe the results of the model concerning the density profiles at galactic and cluster scales. 

\section{Results and discussion}

After fixing the initial conditions and describing how to calculate angular momentum, dynamical friction and adiabatic contraction, we can use 
the model in Section 2 to obtain the density profile of haloes. 
{\bf As reported in Appendix E,  
%In order to use the adiabatic invariant, one has to know the initial mass distribution $M_i (r_i)$ and the final distribution of dissipational %baryons $M_b (r)$. 
we employ the usual assumption that initially baryons had the same density profile as the dark matter (Mo  et al. 1998; Cardone \& Sereno 2005; Treu \& Koopmans 2002; Keeton 2001; Klypin 2002; Tonini et al. 2006).
%TOLTO
%%and that initial dark matter density profile is described by a NFW 
%%profile. This is the reason why in Fig. 3, and Fig. 5a the initial density profile is in both cases a NFW profile.}
%
}
In panels (a)-(d) of Fig. 1, the solid line represents the NFW profile for haloes having masses equal to $10^8 M_{\odot}$ (panel a), $10^{10} M_{\odot}$ (panel b), $10^{11} M_{\odot}$ (panel c), $10^{12} M_{\odot}$ (panel d). 
The NFW profile for the given mass was calculated by means of the relationships connecting the concentration parameter, $c$, and the virial mass, $M_v$, to the shape of NFW profile. 
%%%%%%%%%%%%%%%%%The solid line in Fig. 7 represents NFW profile for a value of the concentration parameter, $c=10$. 
%%%This last quantity is connected to 
%%%the virial mass, $M_v$, 
%through 
We used the following equation for $c$:
\begin{equation}
c \simeq 13.6 \left( \frac{M_v}{10^{11} M_{\odot}} \right)^{-0.13}
\label{eq:cvirr}
\end{equation} 
(Gentile et al. 2007),
%%%%and to the scaling radius, $r_s$, of NFW profile through $c=r_v/r_s$, where:
%%%%\begin{equation}
%%%%r_s \simeq 8.8 \left( \frac{M_v}{10^{11} M_{\odot}} \right)^{0.46}  {\rm kpc }
%%%%\end{equation} 
%%%%(Gentile et al. 2007)
%%%%and the NFW profile is given by:
%PROFILO IN APPENDIX D
%\begin{eqnarray}
%\label{eq:NFWa}
%	\rho_{\rm halo}(r) &=& \frac{\rho_s}{x(1+x)^2}, \quad x = r/r_s \\
%	M_{\rm halo}(r) &=& 4\pi\rho_sr_s^3f(x) \\ &=& M_{\rm vir}f(x)/f(C), \\
%	f(x) &=& \ln(1+x) -\frac{x}{1+x},\\  C &=& r_{\rm vir}/r_s,\\
%	M_{\rm vir} &=&\frac{4\pi}{3}\rho_{cr}\Omega_0\delta_{\rm th}r_{\rm vir}^3
%\label{eq:NFWz}
%\end{eqnarray}
%{\it
%\noindent where $C$ and $M_{\rm vir}$ are the halo concentration and
%virial mass, and $r_{\rm vir}$ is the virial radius. In the above
%equations, the parameter $\rho_{cr}$ is the critical density of the
%Universe and $\delta_{\rm th}$ is the overdensity of a collapsed
%object in the ``top-hat'' collapse model ($\delta_{\rm th}\approx 340$
%for our cosmological model). Two independent parameters --- $C$ and
%$M_{\rm vir}$ --- completely define all relevant halo properties. 
%}
and the usual one for the NFW profile
\begin{equation}
\rho(r)=\frac{\rho_s}{r/r_s(1+r/r_s)^2}=\frac{\rho_b \delta_v}{r/r_s(1+r/r_s)^2}
\label{eq:navar}
\end{equation}
where
\begin{equation}
\delta_v=\frac{\Delta_v}{3}
\frac{c^3}{\log(1+c)-c/(1+c)}
\label{eq:navarr}
\end{equation}
and $\Delta_v$ is the virial overdensity (see Bryan \& Norman 1998).
The scaling radius, $r_s$, of NFW profile is connected to the virial radius, concentration parameter and virial mass through $c=r_v/r_s$, where:
\begin{equation}
r_s \simeq 8.8 \left( \frac{M_v}{10^{11} M_{\odot}} \right)^{0.46}  {\rm kpc }
\end{equation} 
(Gentile et al. 2007).

{\bf Notice that we took the redshift dependence in the model using the technique described in Del Popolo (2001) (the reader is referred to the quoted paper to have more insights).}

NFW profiles change slope rapidly from $\alpha=1$ to $\alpha=3$ at the characteristic radius $r_s$. 
%Fig. 5 shows haloes obtained using the model in Section 2.
For example, in the case of the $10^{12} M_{\odot}$, the NFW profile have a characteristic scale length, equal to $0.1 r_v$, 
%in this case, 
beyond which the density profile steepens, so that much of the mass is piled up within 10\% of the virial radius.
The haloes obtained using the model in Section 2 are different in character from the profiles predicted by numerical simulations, like those of NFW. 
Within the virial radius the log-log density slope changes gradually and the slopes of the inner part of haloes flattens with decreasing mass.
%from $\alpha \simeq 2$ to $\alpha \simeq 0$.
%Always in Fig. 5, 
%
%%%%The short-dashed line, long-dashed line, dot-short-dashed line, dot-long-dashed line, represents the density profile of haloes of decreasing mass: $10^{11} M_{\odot}$, $5 \times 10^{10} M_{\odot}$, $10^{10} M_{\odot}$, $10^{9} M_{\odot}$, respectively. 
%%%%Our results shows a steepening of the density profile with increasing mass with slopes $\alpha \simeq 0$ for $M \simeq 10^{9} M_{\odot}$,
%%%%$\alpha \simeq 0.2$ for $M \simeq 10^{10} M_{\odot}$, $\alpha \simeq 0.4$ for $M \simeq 5 \times 10^{10} M_{\odot}$, 
%%%%$\alpha \simeq 0.6$ for $M \simeq 10^{11} M_{\odot}$, $\alpha \simeq 0.8$ for $M \simeq 10^{12} M_{\odot}$. 
%%%%The short-dashed-long-dashed line in Fig. 7 (almost indistinguishable from the density profile of the $10^{9} M_{\odot}$ halo), represents a fit %%%%to the density profile by means of a Burkert's profile considered a 
%%%%good fit to the dark matter rotation curves inferred from observations (e.g., Salucci \& Burkert 2000).
%
The dotted line represents the density profile of haloes calculated according to the model of Section 2 and have 
%of increasing mass: 
masses: $10^{8} M_{\odot}$ (panel a), $10^{10} M_{\odot}$ (panel b), $10^{11} M_{\odot}$ (panel c), $10^{12} M_{\odot}$ (panel d).
Our results shows a steepening of the density profile with increasing mass with slopes $\alpha \simeq 0$ for $M \simeq 10^{8}- 10^9 M_{\odot}$,
$\alpha \simeq 0.2$ for $M \simeq 10^{10} M_{\odot}$, $\alpha \simeq 0.6$ for $M \simeq 10^{11} M_{\odot}$, 
$\alpha \simeq 0.8$ for $M \simeq 10^{12} M_{\odot}$. 
The dashed line in Fig. 1 (almost indistinguishable from the density profile of the $10^{8} M_{\odot}$ halo), represents a fit 
to the density profile by means of a Burkert's profile considered a good fit to the dark matter rotation curves inferred from observations (e.g., Salucci \& Burkert 2000).
The functional form of this profile is characterized by:
\begin{equation}
\rho(r)= \frac{\rho_o}{(1+r/r_o)[1+(r/r_o)^2]}
\end{equation}
%\begin{equation}
%\rho(r)=\frac{\rho_o}{(1+r/r_o)(1+(r/r_o)^2)}
%\end{equation}
where $\rho_o \simeq \rho_s$ and $r_o \simeq r_s$ (EZ01).
The dark matter within the core is given by $M_o=1.6 \rho_o r_o^3$.
Although the dark matter parameters $r_o$, $\rho_o$ and $M_o$ are in principle independent, the observations reveal a clear connection (Burkert 1995):
\begin{equation}
M_o= 4.3 \times 10^7 \left ( r_o/kpc\right)^{7/3} M_{\odot}
\end{equation}
which indicates that dark haloes represent a one parameter family that is completely specified, e.g., by the core mass (Salucci \& Burkert 2000).

%{\bf We note that the change in the density profile from a NFW profile to a burkert profile is the result of 
%a decrease in central density accompanied by "heating" of the central region and that the region where this mechanism is effective 
%lies inside $r_s$  (quest'ultimo non va bene perche implica che h non ha nessun ruolo)
%}

Two important things must be noticed: a) less massive haloes are less concentrated; b) the halo's inner slope is smaller for smaller mass. \\

The first point can be explained as follows: higher peaks (larger $\nu$), which are progenitors of more massive haloes\footnote{ \bf This affirmation is discussed in Peacock \& Heavens 1990; Del Popolo \& Gambera 1996 or Gao \& White (2007) (Fig.1). Modelling the peaks as triaxial spheroids one obtains
a peak mass $M=\frac{2^{3/2} (4 \pi/3) \rho_b R_{\ast}^3}{\gamma^3+(0.9/\nu)^{3/2}}$ for $0.5 \leq \gamma \leq 0.8 $, where $R_{\ast}$ and $\gamma$ are given in Eq. (\ref{eq:gammm}) and Eq. (\ref{eq:rrr}), showing an increase of mass with $\nu$. It is reasonable that lower $\nu$ peaks should have lower mass; peaks with $\nu \simeq 0$ will tend to sit in regions of larger scale underdensity (cancelling the small-scale overdensity), and hence the $\sim \rho_b R_{\ast}^3$ of material, which initially surrounds the peak, may not be accreted following central collapse.}, have greater density contrast at their center, and so shells do not expand far before beginning to collapse. This reduces $j$ and allows haloes to become more concentrated. An alternative explanation is connected to the quoted angular momentum-density anti-correlation showed
by Hoffman (1986): $j \propto \nu ^{-3/2}$. So, density peaks having low (high) value of $\nu $ acquire a larger (smaller) angular momentum than high 
$\nu $ peaks and consequently the halo will be less (more) concentrated. It is important to notice that the quoted trend of increased central concentration as a function of mass applies only to halos that started out as peaks in the density field smoothed with a fixed $R_f$ scale. Our conclusions do not mean that, for example, clusters of galaxies will be very much more centrally concentrated 
than galaxies, since different smoothing scales would apply in the two cases.

Point (b) can be explained in a similar way to (a), as described previously. 
Less massive objects are generated by peaks with smaller $\nu$, which acquire 
more angular momentum ($h$ and $j$). Angular momentum sets the shape of the density profile at the inner regions. For pure radial orbits, the core is dominated by particles from the outer shells. As the angular momentum increases, these particles remains closer to the maximum radius, 
resulting in a shallower density profile. 
Particles with smaller angular momentum will be able to enter the core but with a reduced radial velocity compared with the purely radial
radial SIM. For some particles the angular momentum is so large that they will never fall into the core (their rotational kinetic energy makes them unbound). Summarizing, particles with larger angular momenta are prevented from coming close to the halo's center and so contributing to the central density. This has the effect of flattening the density profile.
%%%%At the same time, this gradual extinction of the mass flux due to increasing particle angular momentum (velocity
%%%%anisotropy) gradually shifts the system from its intermediate
%%%%quasi-static phase to its final virialized phase. 
This result is in agreement with  
%For example, according to 
the previrialization conjecture (Peebles \& Groth 1976; Davis
\& Peebles 1977; Peebles 1990), according to which initial asphericities and tidal
interactions between neighboring density fluctuations induce
significant non-radial motions that oppose the collapse. 
%This means
%that virialized clumps form later, with respect to the predictions
%of the linear perturbation theory or the spherical collapse model,
%and that the initial density contrast needed to obtain a given final
%density contrast must be larger than that for an isolated spherical
%fluctuation.
% 
%%%%This kind of conclusion was supported by Barrow \&
%%%%Silk (1981), Szalay \& Silk (1983), Villumsen \& Davis (1986),
%%%%Bond \& Myers (1993a, 1993b), and Lokas et al. (1996). Arguments
%%%%based on a numerical least-action method led Peebles
%%%%(1990) to the conclusion that irregularities in the mass distribution,
%%%%together with external tides, induce nonradial motions that
%%%%slow down the collapse. In a more recent paper, Audit et al.
%%%%(1997) conclude that spherical collapse is the fastest. This result
%%%%is in agreement with Peebles (1990) and more recent papers,
%%%%namely, 
%%%%Del Popolo (2002).
%(AGGIUNGERE BIBLIGRAFIA PER QUESTA PARTE)
%
In order to reproduce the NFW profile, we performed an experiment similar to that performed by Williams et al. (2004), namely we 
reduced the magnitude of the $h$ and $j$ angular momentum and dynamical friction, $\mu$. 
The experiment was performed on the halo of mass $10^{12} M_{\odot}$, and in order to reproduce the NFW profile having $c=10$ and mass 
$\simeq 10^{12} M_{\odot}$,
%In the case of a $10^{12} M_{\odot}$ halo, 
we had to reduce the magnitude of $h$ of a factor of 2, $j$ and $\mu$ of a factor 2.5. 
The result of the quoted experiment is the dashed line in Fig. 1d, which closely reproduces the NFW profile.
%%%%%(AGGIUNGERE l'alone di 10 alla 12 nella figura) 
%almost to zero the three quoted quantities so that the collapse was a pure radial one as in DP2000, Lokas (2000). 
Similarly, Williams et al. (2004) had to 
%considerably 
reduce random velocities, which amount to reducing the angular momentum, in order to obtain a NFW profile. 
%In the case of pure radial motions they obtained a limiting slope of $\alpha=2$, larger than what we obtained.
%%%The result of the quoted experiment is the dotted line in Fig. 6, which closely reproduces NFW profile. 
%and switch off angular momentum, dynamical friction 
With each reduction of the random velocities, the profiles get steeper at the center. This effect can be understood, as already reported, as follows: the central density 
is built up by shells whose pericenters are very close to the center of the halo. Particles with larger angular momenta are prevented from 
coming close to the halo's center and so contributing to the central density. 
%However, no matter how much angular momentum is reduced, it is impossible to obtain profiles steeper than log-log slope of $\alpha=2$ (nel nostro %caso meno) in the central part of the haloes. 
%%%%%{ \bf 
The correlation between increasing angular momentum and the reduction of inner slopes in halos has been also noticed by several other authors 
(Avila-Reese et al. 1998, 2001; Subramanian et al. 1999; Nusser 2001; Hiotelis 2002; Le Delliou \& Henriksen 2003; Ascasibar et al. 2003).  
%(ADD FIGURE angular momentum distribution).

{\bf Before going on, we want to add that haloes of a given mass corresponding to higher peaks (larger $\nu$) in the initial density profile, 
collapse at earlier epochs than those having smaller $\nu$ and they give rise to more massive and concentrated haloes. Higher peaks, being more concentrated, feel less the tidal torque and as a consequence the central part of the profile is steeper than in peaks having smaller $\nu$ which are more torqued and collapse in flatter profiles. }

{\bf As Shown in Fig. 2, the solid histogram representing the total specific angular momentum distribution of the density profile reproducing the NFW halo (described in the previously quoted experiment) is more centrally concentrated than the total specific angular momentum distribution of our reference haloes (dashed histogram), and is closer to those of typical halos 
emerging from numerical simulations. In Fig. 2, the dotted-dashed and dashed line represents the quoted 
distribution for the halo n. 170 and n. 081, respectively, of van den Bosch et al. (2002). The halo n. 170 resembles most of the specific angular momentum distributions, while the halo n. 081 has the shallowest distribution in their simulations. 
This may suggests, in agreement with Williams et al. (2004),   
that haloes in N-body simulations lose a considerable amount of 
angular momentum between 0.1 and 1 $r_{v}$. 
Since virialization proceeds from inside out, this means that the angular momentum loss takes 
place during the later stages of the halos' evolution, rather then very early on. 
This is somehow confirmed by the so called angular momentum catastrophe, 
namely the fact that dark matter halos generated through
gas-dynamical simulations
%N-body 
are too small and have too little angular momentum compared to the halos of real disk galaxies, possibly because it was lost during repeated collisions through dynamical friction or other mechanisms (van den Bosch et al. 2002; Navarro \& Steinmetz 2000). 
The problem can be solved invoking stellar feedback processes (Weil et al. 1998), but part of the angular momentum problem seems due to numerical effects, most likely related to the shock capturing, artificial viscosity used in smoothed particle hydrodynamics (SPH) simulations (Sommer-Larsen \&Dolgov 2001).
%%%%%}
}
We discussed the effect of changing the magnitude of angular momentum but we did not speak of the effect of changing the magnitude 
of $\mu$ (dynamical friction). The effect of changing this last quantity is very similar to changing the magnitude of angular momentum: 
an increase in the term $\mu$ produces shallower profiles as larger values of angular momentum does. This is expected from Fig. 11, showing that 
%As shown in Fig. 3?, 
dynamical friction influence the dynamics of collapse in a similar way to that
of angular momentum slowing down the collapse of outer shells and so compelling the particles to remain closer to the maximum radius.

Fig. 3 shows the evolution of a $10^9 M_{\odot}$ halo. {\bf The evolution was obtained by calculating the profile at different redshifts (see Del Popolo 2001). We started the evolution at $z=50$. The solid line represents 
%%%%%%%%%%%%%%%%%%%%%%%%%%%%%%%%%%%%%%%%%%%the initial NFW 
the profile at $z=10$. This is the epoch at which the profile virializes\footnote{The virialization is calculated as usual, by means of the virial theorem, as the epoch at which the ratio between the kinetic and potential enrgy is $\simeq 1/2$.}
and the shape of the profile at virialization is obviously given by
the line corresponding to $z=10$.
} 
At subsequent times, the profile at $z=3$, $z=2$, $z=1$ and $z=0$ is represented by the long-dashed line, short-dashed line, dot-dashed line and 
dotted-line, respectively.
%At subsequent times ($z=3$ to $z=0$) 
{\bf The evolution after virialization is produced by secondary infall, two-body relaxation, dynamical friction and angular momentum.}
The cusp is slowly eliminated and within $ \simeq 1$ kpc a core forms. 
At an early redshift, $z \simeq 5$, the dark matter density experiences the adiabatic contraction by baryons producing a slightly more cuspy profile 
than that represented by the solid line (not represented in the plot).
%(dot-short-dashed line). 

The previous result is similar to what found by Romano-Diaz et al. (2008) who studied the dark matter cusp evolution using N-body simulations with and without baryons. The ``erasing" of the cusp is associated by them to the heating up of the cusp region via dynamical friction (EZ01) or influx of subhaloes into the innermost region of the dark matter halo. 
%
%%%%The erasing is not associated to the energy feedback from stellar evolution (Mashchenko et al. 2006) or angular momentum transfer from the %%%%stellar bar- DM interaction (Weinberg \& Katz 2002). Major mergers cannot produce the erasing of the cusp since the epoch in which they are %%%%important ends %at around $z \simeq 1.5$ and most of the cusp levelling happens later (after $z \simeq 1$).
%
%dot-long-dashed line
%long-dashed line, dotted line, and short-dashed line, respectively.
In our model the erasing of the cusp is connected to the joint effect of dynamical friction and angular momentum.
{\bf As previously discussed, larger amount of angular momentum leads to shallower final density profiles
in the inner region because particle with larger angular momentum have lower probabilities to enter the center. 
The effects of dynamical friction 
%The effect of changing this last quantity 
can be interpreted in two different fashions: (a) an increase in the term $\mu$ is very similar to changing the magnitude of angular momentum (see Fig. 11) with the final result of producing shallower profiles; (b) dynamical friction can act on gas moving in the background of dark matter particles, dissipate the clumps orbital energy and deposit it in the dark matter with the final effect of erasing the cusp (similarly to EZ01; El-Zant et al. 2004; TLS; Romano-Diaz et al. 2008;). 
Baryons have another effect, at an early redshift, the dark matter density experiences the adiabatic contraction by baryons producing a slightly more cuspy profile. This last is overcome from the previous two effects. As shown by Fig. 11, the magnitude of dynamical friction effect is a bit larger than that due to angular momentum and that these two effects add to improve the flattening of the profile.}

In Fig. 4, we plot the rotation curves obtained by our model and we compare them to four LSB galaxies studied by Gentile et al. (2004), namely ESO 116-G12, ESO 79-G14, ESO 287-G13 and NGC 1090. The physical parameters of the quoted galaxies are given in Gentile et al. (2004) (Table 1). In the case of ESO 116-G12 the total HI mass is $M_{HI} \simeq 1.5 \times 10^9 M_{\odot}$ and the dynamical mass\footnote{The dynamical mass $M_{dyn}$ is determined at the farthermost radius with data.} $M_{dyn} \simeq 3.3 \times 10^{10} M_{\odot}$. For the other three galaxies ESO 79-G14, ESO 287-G13 and NGC 1090, the masses are: $M_{HI} \simeq 3.5 \times 10^9 M_{\odot}$, $M_{dyn} \simeq 1.3 \times 10^{11} M_{\odot}$; 
$M_{HI} \simeq 1.1 \times 10^{10} M_{\odot}$, $M_{dyn} \simeq 1.9 \times 10^{11} M_{\odot}$; 
$M_{HI} \simeq 8.5 \times 10^{9} M_{\odot}$, $M_{dyn} \simeq 1.8 \times 10^{11} M_{\odot}$, respectively. In all the four cases, the data are compared with 
the rotation curve obtained using our model (solid line) and with rotation curves obtained from NFW profile (dotted lines), given by:  
\begin{equation}
V(r)=V_v
\left \{
\frac{\ln (1+cx)-cx/(1+cx)}
{x [\ln(1+c)-c/(1+c)]}
\right \}^{1/2}
\end{equation}
where $x=r/r_v$ and $V_v$ is the virial velocity \footnote{The value of the characteristic velocity $V_v$ of the halo is defined in the same way as the virial radius $r_v$. 
%$\Delta_v$ fig. 7
%$r_v$ and $V_v$ depend is obtained by means
%of Bryan \& Norman (1998) formulas.????????
}. 
Fig. 4 shows that NFW haloes are higher than the rotation curves obtained using our model, in which more massive haloes tend to be more centrally concentrated and have flatter rotation curves. Less massive haloes are less concentrated, ad have slowly rising rotation curves. In contrast
NFW rotation curves rise very steeply and as a consequence NFW fits to dwarf galaxy rotation curves have too low concentration parameters (van den Bosch \& Swaters 2001) compared to N-body predictions. 
NFW fails to reproduce velocities and shape of the observed rotation curves, especially in the case of ESO 116-G12, ESO 79-G14. NFW haloes predict
too high velocities in the central part of haloes, and  
%Our rotation curves are more flat
even leaving $c$ as free parameter, instead of using Eq. (\ref{eq:cvirr}), there is no appreciable improvement in the fit. Using Eq. (\ref{eq:cvirr}), one obtains very low values of $c$. 
The result is similar to that described by Gentile et al. (2004): data are much better described by core-like profiles, like the Burkert profile generating flatter rotation curves:
\begin{equation}
V(r)= \left 
(\frac{2 \pi G \rho_0 r_o^3}{r})
\right)^{1/2}
\left \{
\ln
\left[
(1+r/r_o) \sqrt(1+(r/r_o)^2)
\right]-
\arctan (r/r_o)
\right \}^{1/2}
\end{equation}
Our rotation curves are very similar to those generated by the Burkert profile and the residuals and discrepant points for our rotation curves are close to that given in Gentile et al. (2004) for the Burkert's fit to their data.
%The previous result suggest the existence of a scale-length connected to the baryonic component which is not expected in the CDM theory: %non-cosmological effects destroy the cusp. 

As noticed in the introduction, while on galactic scales a large number of studies predicts central cores and it seems that the cusp/core problem
is a real problem not attributable to systematic errors in the data (de Blok, Bosma \& McGaugh 2003), on cluster scales the situation is less clear with slopes ranging from $\alpha=0.5$ (e.g. Sand et al. 2002; Sand et al. 2004) to $\alpha=1.9$ (Arabadjis et al. 2002).
%CONTINUARE

%The baryonic-to-total mass ratio is 0.16, reported by the WMAP collaboration (Spergel et al. 2003)

In order to study the problem on cluster scales, we have calculated the density profile evolution of dark matter and that of the total matter 
distribution for halo of $\simeq 10^{14} h^{-1} M_{\odot}$.

%The density profile evolution of a $10^{14} M_{\odot}$ halo.

Fig. 5 plots the evolution of a density profile of $10^{14} h^{-1} M_{\odot}$. 
%The profile at $z=3$, $z=1.5$, $z=1$ and $z=0$ is represented by the solid line, dotted-line, short-dashed-line, long-dashed-line,
%respectively.
{\bf 
%Togliere sistemare \\
The solid line represents the density profile at $z=3$ which slightly steepens due to baryon settling in virialized dark matter haloes (AC) %(dot-short-dashed line) 
at $z=2$ (not shown in the plot). 
Once the baryons condense to form stars and galaxies, they experience a dynamical friction force from the less massive dark matter particles as they move through the halo. Energy and angular momentum is transferred to dark matter, increasing its random motion. Moreover, angular momentum acquired 
in the expansion phase gives rise to non-radial motions in the collapse phase. The effect of angular momenta and dynamical friction overcomes that of the AC and the profile starts to flatten ($z=1.5$ dotted line; $z=1$ short-dashed line; $z=0$ long-dashed line). The final dark matter 
profile (long-dashed line) is characterized by a log-log slope of $\alpha \simeq 0.6$ at $0.01 r_s$. So the situation is similar to that of 
haloes on galactic scales but the slope is larger than for dwarf galaxies. In this case the profile virializes at $z \simeq 0$ and the corresponding line respresents the profile at virialization.
Comparing the two profiles at virialization in Fig. 1 and Fig. 3, (line corresponding to $z=10$ for the profile having $10^9 M_{\odot}$; line corresponding to $z \simeq 0$ for the profile having $10^{14} M_{\odot}$), it is clear that the profile at virialization is flatter in the case of smaller masses. The difference is due to the fact that angular momentum starts to change the profile shape starting already from turn-around epoch.
}
%%%%%%(PERCHE?). 
The dot-dashed line in Fig. 5 (uppermost line), represents the final total density profile of a $10^{14} M_{\odot}$ halo.
%TOLTO
%%%In Fig. 5b, we plot the evolution of the total density profile at initial redshift, $z=3$, (solid line) and at the final redshift $z=0$ (dotted %%%line). 
%%%The plot shows that the total density profile slightly changes 
%%%with time and the cusp is not ``erased" as in the case of the dark matter profile. 
%
The plot shows that the cusp in the total density profile 
%slightly changes 
%with time and the cusp 
is not ``erased" as in the case of the dark matter profile. 
This result also implies that the baryonic component becomes steeper than the original NFW profile. The behavior of total mass is in agreement with X-ray observations by Chandra and XMM (Buote 2003, 2004; Lewis et al. 2003) weak lensing (Dahle et al. 2003) and strong lensing (Bartelmann 2002), which are consistent with a cusp having $\alpha=1$ or larger. The behavior of the dark matter halo is in agreement with analysis of Sand et al. (2002, 2004) 
%showing that if one separates the dark matter distribution from the total matter, one obtains a nearly flat core $\alpha=0.35$. Sand et al. (2002) 
who fitted the baryonic and dark matter profiles only in the very inner part of the cluster MS 2137-23 within $\simeq 50 h^{-1} $ kpc by means of a generalized NFW profile:
\begin{equation}
\rho(r)= \frac{\rho_b \delta_v}{(r/r_s)^{\alpha}(1+r/rs)^{3-\alpha}}
\end{equation} 
obtaining a nearly flat core $\alpha=0.35$. The steepening of the baryonic component, is consistent with what found by Brunzendorf \& Meusinger (1999), who found that the projected galaxy distribution in Perseus cluster diverge as $r^{-1}$.

%So, the apparent disagreement between dissipationless N-body simulations and observations of the central density profiles of %galaxies and clusters can be resolved within the CDM paradigm (cosmogony). 

The results previously reported have several implications on the effort to test predictions of the CDM model observationally. 
The test that received much attention in the last decade, as several times stressed, is the density distribution in the inner regions of galaxies and clusters. 

{\bf 
As previously stressed, in our model the cusp/core problem is solved by the effects of dynamical friction and angular momentum overcoming 
that of the AC, however, at the same time, the effect of baryons is non-negligible (e.g. when compared to collisionless N-body simulations), in the sense that we will describe.
The effect of baryons is two-fold:
a) in the inner parts of the haloes baryons are more present and their adiabatic collapse, from one side, steepens the density profile because the amount of dark matter in the central region will always be increased. b) From the other side, one has to take account of the exchange of angular momentum between the baryons and the dark matter. Dynamical friction can result in a transfer of angular momentum from the baryons to the dark matter. Because the dark matter gains angular momentum, it moves further from the galactic center reducing the steepeness of the inner profile (see also Klypin et al. 2001). 
So baryons have orbital energy that is transferred through dynamical friction and deposited in the dark matter, giving a partial solution 
to the cusp/core problem (EZ01; El-Zant et al. 2004; Romano-Diaz et al. 2008). 
%play the role of  
In collisionless N-body simulations, this complicated interplay between different effects is not taken into account and 
makes it necessary to run N-body simulations that repeat the mass modeling including a self-consistent treatment of the baryons and dark matter component.    
So, on galactic scales, where dark matter dynamics and baryons dynamics are entangled, the cusp/core problem seems to be a ``genuine" one, in the sense that the disagreement between observations and N-body simulations is not due to numerical artifacts or problems with simulations.  
At the same time it is an apparent problem, since the disagreement between observations and dissipationless simulations is related to the 
the fact that this last are not 
%can be explained 
taking account of baryons physics. This means that we are comparing two different systems, one dissipationless (i.e., DM) and the other dissipational (i.e., inner part of structures), and we cannot expect them to have the same behavior. 
}

As our results show, baryons at early times steepens the cusp due to the AC and after the cusp is erased through dynamical friction and non-radial motions effects. This results are in agreement with other analyses which studies separately the effects of dynamical friction (e.g., EZ01) and those of angular momentum (non-radial motions) (e.g., Nusser 2001; Hiotelis 2002; Ascasibar et al. 2003; Williams et al. 2004) and with the recent Sph simulations of Romano-Diaz et al. (2008).
%As noticed in the introduction, while on galactic scales a large number of studies predicts central cores and it seems that the cusp/core problem
%is a real problem not attributable to systematic errors in the data (de Blok, Bosma \& McGaugh 2003), on cluster scales the situation is less clear %with slopes ranging from $\alpha=-0.5$ (e.g. Sand et al. 2002; Sand et al. 2004) to $\alpha=-1.9$ (Arabadjis et al. 2002).
Going to larger scales the situation changes. The analysis of density distribution for bright galaxies is complicated by the uncertain contribution of stars to the total mass profile (Treu \& Koopmans 2002; Mamon \& Lokas 2004). Some analyses tend to favor inner slopes shallower than predicted by CDM (e.g., Gentile et al. 2004) but other deduce slopes of the inner profiles that are at least marginally consistent with predictions (Treu \& Koopmans 2002, 204; Koopmans \& Treu 2003; Jimenez et al. 2003). 
As previously reported, our results shows a steepening of the density profile with increasing mass with a density profile of haloes of mass $ > 10^{12} M_{\odot}$ having slopes $>0.8$. 
This is in agreement with recent N-body simulations having a 
logarithmic slope that decreases inward more gradually than the NFW profile (Hayashi et al. 2003; Navarro et al. 2004; Stadel et al. 2008).  
In the case of Stadel et al. (2008) the logarithmic slope is $0.8$ at $0.05 \%$ of $r_v$.

The density distribution in clusters of galaxies can, in principle, provide a cleaner test of the models because the effects 
of the baryons and gas on the dark matter distribution are expected to be smaller and simpler. However also in this case observations predict slopes ranging from $\alpha \leq 0.5$ ($\alpha=0.35$,  Sand et al. 2002, 2004) to values larger than one (Arabadjis et al. 2002) and in some cases different results even for the same object. It is the case of the cluster MS2137-23 studied by Sand et al. (2004), who found a shallow density slope ($\leq 0.5$)  
while Dalal \& Keeton (2003), Bartelman \& Meneghetti (2003) and Gavazzi (2003) contested Sand's results, which according to them is neglecting lens ellipticity, and found consistency of the inner slope with a NFW profile. 
Our result concerning cluster scales makes a difference between dark matter and total mass distribution: the first tend to be less cuspy than observations in agreement with some observations (e.g., Sand et al. 2002, 2004), while the second is cuspy and well described by a NFW profile (in agreement with Brunzendorf \& Meusinger 1999) .

One question that may arise at this point is: why the results of analytical or semi-analytical models are different from
those of N-body simulations?
%and why one should trust models like SIM?

The discrepancy between N-body simulations and observations at small scales has led some authors to attribute the problem to a real failure of the CDM model, or to that of simulations (de Blok et al. 2001a; de Blok, McGaugh, \& Rubin 2001b; Borriello \& Salucci 2001; de Blok, Bosma, \& McGaugh 2003; de Blok 203; Taylor 2004). One of the reasons why N-body simulations could give unreliable results are connected to two-body relaxation. The processes of relaxation is difficult to quantify, but in the large $N$ limit one expects that the discreteness effects inherent to the
N-body technique vanish, so one tries to use as large a number of particles as computationally possible.
Unfortunately in most cosmological simulations the importance of two body interactions does not vanish if one
increases $N$, since structure formation in the cold dark matter (CDM) model occurs hierarchically since there is power
on all scales, so the first objects that form in a simulation always contain only a few particles (Moore et al. 2001),
(Binney \& Knebe 2002).  With higher resolution the first structures form earlier and have higher physical densities
because they condense out of a denser environment. Two body relaxation increases with density, so it is not clear if
increasing the resolution can diminish the overall amount of two body relaxation in a CDM simulation, i.e. if testing for
convergence by increasing the mass resolution is appropriate.
Diemand et al. (2004a,b) explored the effect of resolution on the degree of relaxation finding that increasing $N$ slowly reduces the degree of relaxation $\propto N^{-0.25}$ rather than proportional to $N$ as expected from the collisionless Boltzmann equation. This means that if to resolve 
$10 \%$ of the virial radius we need 1000 particles, we need $10^6$ particles to resolve $1 \%$ of the virial radius.
To have such a high number of particles per halo the use of the multi-mass technique is required. This technique consists of simulating with high mass resolution only the particles that will end up in the halo of interest at $z = 0$, while having less mass resolution for the particles that end up far away from the halo of interest. The task of achieving reliable profiles with such a high spatial resolution is sensitive to numerical
integration errors and requires careful resolution studies (Power et al. 2003); perhaps this is a reason for the disagreement
between groups, using different codes, on the slope of the inner profile (Ricotti 2003).  
Moreover, this has the drawback that each simulation can resolve only one halo at a time, therefore selection effect biases,
(difficult to control, determined by the criteria for picking the halos to
re-simulate) and a poor statistical sample could affect the reliability of the final result even if the simulation is very accurate and has high resolution (Ricotti 2003).
%astro-ph/0212146).
More recently Stadel et al. (2008) by means of simulations using several billions particle measured the density profile to a distance of 120 pc ($0.05 \%$ of $r_v$).
Convergence in the density profile and the halo shape scales as $N^{-1/3}$, but the shape converges at a radius three times larger at which point the halo becomes more spherical due to numerical resolution. 
This last simulation has surely enough resolution ($\leq 1 $ kpc) to distinguish between the core and cusp model, which was one of the problems de Blok (2003) enumerated in the reasons why simulations produce just cuspy profiles.
Other problems due to numerical artifacts are the over-merging problem, the artificial disruption of substructure due to numerical effects, that according to Taylor et al. (2004) has not been solved 
as claimed by some authors (e.g. Ghigna et al. 2000).
%and the claim that the problem was solved (e.g., Ghigna et al. 2000) 
%has not been fully tested (Taylor et al. 2003).
Even if the effects of particle discreteness in N-body simulations of $\Lambda$CDM are still an intensively debated issue (Romeo et al. 2008), 
it is highly probable that N-body simulations are correctly predicting the density profiles of the CDM haloes but on
small scales other effects not taken into account by dissipationless simulations, for example the presence of baryons and the 
effect that they have on dark matter, change the halo shape. In other words, at this state of the art of N-body simulations and observations, the discrepancy between simulations and observations is not the fault of numerical artifacts of simulations or problems of the observations (see de Blok (2003) to have a list of the problems imputed to observations and its confutation) but it is just connected to the fact that we are trying to force dissipationless simulations to predict the same behavior for the density profile of a system whose physics is not just the dissipationless physics typic of dark matter. 
It is interesting to note that the problems of CDM only become clear on length scales where
the baryons start playing a role and that this applies not only to the cusp/core problem but also to the missing dwarfs problem. 
Moreover, it is noteworthy that if, indeed, most star-forming galaxies in the early universe lost their DM cusps because of stellar feedback, the missing dwarfs (satellites) problem could also be solved. Dwarf galaxies without a central cusp have a lower average core density than cuspy ones, and are
hence much easier to disrupt tidally during the hierarchical assembly of larger galaxies (Mashchenko \& Sills 2005). As a consequence, the removal of galactic cusps by stellar feedback in the early universe would result in fewer satellites today.
This again indicates that baryon physics is one of the missing pieces of the puzzle, and will very likely make a major contribution toward a solution. If this is true, it would be unwise to ignore the conclusions to which data are leading us, namely that small scales tells us more about galaxy formation than it does about CDM\footnote{
%Another possibility is 
Other possibilities are that the observed dark matter cores are telling us that dark matter has pressure at small scales or something unexpected.}. 
In other terms, the centers of galaxies are, special places, the only places where we can study dark matter under peculiar conditions. 

The predictions of large amounts of small-scale structure and substructure in CDM cosmologies is perhaps startling, but it is not in and of itself a reason to reject CDM.

Another important point on which our results can tell something is the debate on the universality of the density profiles of dark matter haloes.
The result that the density profiles of haloes in CDM and other hierarchical clustering cosmologies have a universal form which is well represented
by the simple fitting formula given by NFW96, 97, has been confirmed in almost all the subsequent papers dealing with the subject except some of them (Jing \& Suto 2000; Subramanian et al. 2000; Ricotti 2003, Ricotti \& Wilkinson 2004; Cen et al. 2004; Ricotti et al. 2007). 
%In the last decade the debate focused on wheter the slope approaches values close to 1 or 1.5 at small radii. 
Ricotti's papers, using the same type of simulations in NFW97, found that at virialization the central logarithmic slopes $\alpha$ at $5 \%$-$10 \%$ of the virial radius are correlated with the halo mass, with $\alpha=0.2$ for $ \simeq 10^{8} M_{\odot}$, $\alpha=1$ for $\simeq 10^{13} M_{\odot}$, $\alpha \simeq 1.3 $ for $ \simeq 10^{15} M_{\odot}$, and that there is no reason to believe that the value of $\alpha$ converges to any asymptotic value as also suggested by some high-resolution N-body simulations (e.g., Navarro et at al. 2004; Graham et al. 2005; Stadel et al. 2008). This leads to the conclusion that density profiles do not have a universal shape. Moreover in agreement with Subramanian (2000), the halo shape at a given mass or spatial scale depends on the slope of the power spectrum at that scale. Similar correlations have been found by other authors (e.g., Jing \& Suto 2000; Taylor \& Navarro 2001; Cen et al. 2004). Cen et al. (2004) confirmed Ricotti's result; in addition they identify a redshift dependence of the typical halo profile. 
%More recent work by 
Graham et al. (2005) and Merrit et al. (2005) also find a correlation between halo mass and the shape of  the density profile, parameterizing it in terms of the S\'ersic profile index.  
There are also observational evidences of a mass dependence of the dark matter density profile, as reported in Ricotti \& Wilkinson (2004) and in the introduction of the present paper. Other examples of a possible non-universality of density profiles
comes from galaxies observations. Galaxies like NGC2976, NGC6689, NGC5949, NGC4605, NGC5963 have very different values of the slope: 
$\alpha \simeq 0.01$, 0.80, 0.88, 0.88, 1.28, respectively
%: NGC2976 $\alpha \simeq 0.01$,
%NGC6689 $\alpha \simeq 0.80$, NGC2976 $\alpha \simeq 0.01$, NGC5949 $\alpha \simeq 0.88$, NGC4605 $\alpha \simeq 0.88$, NGC5963 %$\alpha \simeq 1.28$
(see Simon 2003a,b). Moreover observed slopes on galactic scales have large scatter compared to simulations and mean slope shallower than simulations.
%and several others reported in Ricotti \& Wilkinson (2004) and in the introduction of the present paper.
Thus there is also some observational evidence for a dependence on halo mass of the inner slope of dark matter haloes, and hence for the non-universality of the dark matter profile.

Our results in Fig. 1 show clearly different slopes flattening for decreasing values of the halo mass in agreement with Ricotti (2003), 
Ricotti \& Wilkinson (2004) and Williams et al. (2004) (eg., their Fig. 1), showing a flattening of the profile with decreasing mass. 
{\bf As shown in Fig. 3, the density profile of a halo of $10^9 M_{\odot}$ at virialization is different from a NFW profile, the inner slope 
is less steep than the NFW model. Dynamical friction and angular momentum contribute to flatten the density profile which is well fitted by a 
Burkert's profile at $z=0$. In the case of the $10^{14} M_{\odot}$ halo, as shown in Fig. 5, the situation is similar except that the final density profile is more steep than in the case of a $10^9 M_{\odot}$ halo.
}

%
%%%Our previous result suggest the existence of a scale-length connected to the baryonic component which is not expected in the CDM theory, and that %%%non-cosmological effects destroy the cusp. ?????
%%%This is also confirmed by Romano-Diaz et al. (2008) by means of numerical simulations. 
%

{\bf 
The idea coming out from the previous arguments is that one should not expect universality of  
density profiles\footnote{As previously stressed, 
the situation is different for the total mass profile for which the NFW seems to give a good fit.}. 
%Our results shows that the final inner slopes of haloes of different masses are different. As previously described, in the inner parts of the %haloes baryons are more present and their adiabatic collapse, from one side, steepens the density profile because the amount of dark matter
%in the central region will always be increased. 
%From the other side, one has to take account of the exchange of angular momentum between
%the baryons and the dark matter. Dynamical friction can result in a transfer
%of angular momentum from the baryons to the dark matter. Because the dark matter gains angular
%momentum, it moves further from the galactic center. This complicated interplay between different effects makes it necessary
%to repeat the mass modeling including a self-consistent treatment of the baryons
%and dark matter component.
Summarizing, the flattening of the inner slopes of haloes is produced by the role of angular momentum, dynamical friction 
and the interplay between dark matter and baryonic component.
We want to remark that the {\it Aquarius Project} showed that even in N-body simulations of $\Lambda$CDM haloes the mass 
profiles of haloes is not strictly universal (Navarro et al. 2008).   
%
%%%%or in the regions of the profile were dark matter is dominant, but in the inner parts of structures where baryons have an important role, one %%%%should not expect such a universality in dark matter profiles\footnote{As previously stressed, 
%%%%the situation is different for the total mass profile for which the NFW seems to give a good fit.
%%%%}, like one does not expect that ``universal" NFW or S\'ersic profiles adequately represents the very central profiles of elliptical galaxies %%%%with cores, the pointlike nuclei of some dwarf ellipticals (dE) galaxies, or the steep power-law density cusps observed in the inner few parsecs %%%%of nearby galaxies like M32 and the bulge of the Milky way (Merrit et al. 2005). 
%
}
Concluding, concerning the cusp/core problem, there is no real contradiction between observations and dissipationless simulations,  
the problem arises because we expect that dissipationless simulations predict the same density profile of a system whose physics is not just that of dark matter. This lead to the conclusion that the so called cusp/core problem is not a reason to reject the CDM model. 

%Again, there is no real contradiction between ............
%observations and simulations 

%\begin{figure}
%\psfig{file=ff4.ps,width=10cm,height=10cm}
%\caption[]{The mass accreted by a collapsed perturbation, in units of
%$10^{15}M_{\odot}$, taking into account non-radial motions, dynamical friction and a non-zero cosmological constant (dotted line)  
%compared to SIM mass (solid line).}
%\end{figure}

\section{Conclusions}

In this paper, we studied the cusp/core problem by means of an improved version of the SIM, taking into account, simultaneously and for the first time, the effects of ordered and random angular momentum, dynamical friction and adiabatic contraction. Initial conditions were introduced by means of the 
theory of Gaussian random fields. Angular momentum 
%acquisition by proto-structures 
was calculated through the standard theory of acquisition of angular momentum through tidal torques, while the random part of angular momentum was assigned to protostructures according to Avila-Reese et al. (1998) scheme as modified by Ascasibar et al. (2003). Dynamical friction was calculated dividing the gravitational field into an average and a random component generated by the clumps constituting hierarchical universes. The adiabatic contraction was taken into account by means of Gnedin et al. (2004) model and Klypin et al. (2002) model taking also account of exchange of angular momentum between baryons and dark matter. 
The improved SIM of the present paper, taking account the previous effects gives rise to haloes being characterized by log-log density slope that changes gradually 
within the virial radius and slopes of the inner part of haloes flattening with decreasing mass.
The density profiles of structure having masses smaller than $10^{11} M_{\odot}$ are well fitted by Burkert's profiles. We then calculated the time evolution of a dwarf galaxy having mass $10^{9} M_{\odot}$. {\bf The result showed an initial steepening of the profile due to adiabatic compression and a subsequent flattening till a slope $\alpha \simeq 0$ is reached, due to angular momentum and dynamical friction. If the effects of angular momentum and dynamical friction are not taken into account, the final profile is cuspy}. We then compared some of the rotational curves given by Gentile et al. (2004) with the rotational curves obtained by means of our model, obtaining a good agreement. 
In the case of clusters of galaxies the density profile evolution is similar to that observed on galactic scales with the difference that the final slope is steeper than in the dwarf galaxies case. However the total mass profile is still cuspy.
% and evolves slightly.
The behavior of the dark matter halo is in agreement with the analysis of Sand et al. (2002, 2004) who fitted the baryonic and dark matter profiles only in the very inner part of the cluster MS 2137-23 within $\simeq 50 h^{-1} $ kpc by means of a generalized NFW profile. 
The behavior of total mass is in agreement with X-ray observations by Chandra and XMM (Buote 2003, 2004; Lewis et al. 2003) weak lensing (Dahle et al. 2003) and strong lensing (Bartelmann 2002), which are consistent with a cusp having $\alpha=1$ or larger.
The previous results tell us that the apparent disagreement between dissipationless N-body simulations and observations of the central density profiles of galaxies and clusters can be resolved within the CDM paradigm (cosmogony). 
The cusp/core problem is simply due to the fact that we are trying to predict the dynamics of galaxies at small scales by using just the CDM model while at that scales other effects influence their dynamics.
%%dominate the dynamics.
%For example, it may be that our assumption of a single dominant component is simplistic.
 
\acknowledgements

We would like to thank Massimo Ricotti, Nicos Hiotelis, and Antonaldo Diaferio for their very helpful suggestions and comments.

%\begin{acknowledgements}

%%%%%%%%%%%%%%%%%%%%%%%%%%%%\acknowledgements

%\end{acknowledgements}

%\begin{thebibliography}{999}

\newpage

%%%-------------------------------APPENDIX------------------------------------

\appendix

\section{Model details}

As shown by spherical collapse model\footnote{A slightly overdense sphere, embedded in the Universe, is a useful non-linear model, as it behaves exactly as a closed sub-universe because of Birkhoff's theorem. The sphere is divided into spherical ``shells". A spherical ``shell" may be defined as the set of particles at a given radius that are all at the same phase in their orbits (see Le Delliou \& Henriksen 2003).} of Gunn \& Gott (1972), a bound mass shell having initial comoving
radius $x_i$ will expand to a maximum radius $x_m$ (named apapsis or turn-around radius 
%) of a shell 
%(or turn-around radius, 
$x_{ta}$):
\begin{equation}
%r_m=x/{\overline \delta(r)}
x_m=g(x_i)=x_i/{\overline \delta_i}
\label{eq:pee}
\end{equation} 
where the mean fractional density excess inside the shell, as measured at
current epoch $t_0$, assuming linear growth, 
can be calculated as:
\begin{equation}
{\overline \delta_i}=\frac{3}{x_i^3} \int_0^{x_{i}} \delta(y)y^2 dy
\label{eq:overd}
\end{equation}
At initial time $t_i$ and for a Universe with density parameter $\Omega_i$,
a more general form of
Eq. (\ref{eq:pee}) (Peebles 1980) is :
\begin{equation}
x_m=g(x_i)=x_i\frac{1+{\overline \delta_i}}{{\overline \delta_i}-(\Omega_i^{-1}-1)}
\label{xtr}
\end{equation}
The last equation must be regarded as the main essence of the SIM. It tells
us that the final time averaged radius of a given Lagrangian shell
does scale with its initial radius.
Expressing the scaling of the final radius, $x$, with the initial one
by relating $x$ to the turn around radius, $x_m$, it is possible to write:
\begin{equation}
x=f(x_i) x_m
\label{eq:rc} 
\end{equation}
where $f$ 
%is a costant that 
depends on $\alpha$:
\begin{equation}
f = f(\alpha) = 0.186+0.156 \alpha+0.013 \alpha^2+0.017 \alpha^3-0.0045 \alpha^4+0.0032 \alpha^5
\end{equation}
(Zaroubi, Naim \& Hoffman 1996).
%where
%the mean fractional density excess inside a given shell of radius $r$
%can be calculated as:
%\begin{equation}
%{\overline \delta}=\frac{3}{r^3} \int_0^r \delta(y)y^2 dy
%\end{equation}
If mass is conserved (i.e., $m(x_i)=m(x_m)$) and each shell is kept at its turn-around radius ($f=1$),
%$r_{vir}$ is $r_m/2$ (Hoffman 1988)
the shape of the density profile 
%at maximum of expansion is conserved, 
%after the virialization,
%and
%the relaxed density profile
is given by (Peebles 1980; HS; White \& Zaritsky 1992):
\begin{equation}
\rho_{ta}(x_m)=\rho_i (x_i) \left( \frac{x_i}{x_m} \right)^2 \frac{d x_i}{dx_m}
\label{eq:dturn}
\end{equation}
Using the Virial theorem one obtains a value for the collapse factor $f=0.5$ and 
%while 
the final density profile is obtained using again mass conservation:
\begin{equation}
\rho(x)x^2 dx=\rho_i x_i^2 d x_i
\label{eq:dturnn}
\end{equation}
If one approximates the initial density of the shell of radius $x_i$ as:
\begin{equation}
\rho_i(x_i)=\rho_{b,i} [1+ \delta_i(x_i)]
\end{equation}
%where $\rho_{b,i}$ is the background (critical) density 
and expand the right hand-side of Eq. (\ref{xtr}) in $\delta_i$ keeping only the linear term, one obtains the power-law density profile obtained by HS:
\begin{equation}
\rho(x) \propto x^{-3(n+3)/(n+4)}
\end{equation}

However, after reaching maximum radius, a shell collapses and will start oscillating and it will contribute to the inner 
shells and so even energy is not  an integral of motion anymore and the collapse factor, $f$, is no longer constant. The effect of the infalling outer
shells on the dynamics of a given shell can be described as follows.
As I previously told, one assumption of SIM is that the collapse is ``gentle". One can assume that the 
potential well near the center varies adiabatically (Gunn 1977, FG84).
This means that a shell near the center makes many oscillations before the potential changes significantly (Gunn 1977, FG84)
%and as a consequence 
or similarly the orbital period of 
the inner shell is much smaller than the collapse time of the outer shells (Zaroubi \& Hoffman 1993). 
%%%%In a slowly 
%%%%varying potential the action variables associated with the motion of a particle are invariant.
%%%%The angular action variable is the angular momentum, which thanks to spherical simmetry, is conserved independent of whether or not the %%%%potential changes adiabatically.
This implies that the radial action $\oint v(r) dr$ (being $v(r)$ the radial velocity) is an adiabatic invariant of the inner shell. 
As the outer shells collapse, the potential changes slowly and because of the above adiabatic invariant the inner shell shrinks. 
If a shell has an apapsis radius (i.e., apocenter) $x_m$ and initial radius $x_i$, then the mass inside $x_m$ is obtained summing the mass contained in shells 
with apapsis smaller than $x_m$ (permanent component, $m_p$) and 
%the second (additional mass, $m_{add}$) is 
the contribution of the outer shells passing
momentarily through the shell $x_m$ (additional mass $m_{add}$). 
Because of mass conservation, we have:
\begin{equation}
m_p(x_m)=m(x_i)=\frac{4}{3} \pi \rho_{b,i} x_i^3 (1+{\overline \delta_i})
\label{eq:mp}
\end{equation}
where $\rho_{b,i}$ is the constant density of the homogeneous Universe at the initial time.
%conditions.
The additional component $m_{add}(x_m)$ is:
\begin{equation}
m_{add}(x_m)=\int_{x_m}^{R} P_{r_m} (x) \frac{d m(x)}{dx} dx
\label{eq:madd}
\end{equation}
where $R$ is the radius of the system (the apapsis of the outer shell) 
and the distribution of mass $m(x)=m(x_m)$ is given by Eq. (\ref{eq:dturn}).
The total mass is so given by:
\begin{equation}
m_T(x_m)=m_p(x_m)+m_{add}(x_m)
\label{eq:mpp}
\end{equation}
while $P_{x_m}(x)$ is the probability to find the shell with apapsis $x$ inside radius $x_m$,
calculated as the ratio of the time the outer shell (with apapsis
$x$) spends inside radius $x_m$ to its period. 
This last quantity can be expressed as 
\begin{equation}
P_{x_m}(x)= \frac{
\int_{x_p}^{x_m} \frac{d \eta}{v_x(\eta)} }
{\int_{x_p}^{x} \frac{d \eta}{v_x(\eta)}
}
\end{equation}
where $x_p$ is the pericenter of the shell with apsis $x$ and $v_x (\eta)$ is the radial velocity of the shell with apapsis $x$ as it passes from radius $\eta$.
This radial velocity
%The radial velocity $v$ of a shell with apapsis .......
can be obtained by integrating the equation of motion of the shell:
\begin{equation}
\frac{dv_r}{dt}=\frac{h^2(r,\nu )+j^2(r, \nu)}{r^3}-G(r) -\mu \frac{dr}{dt}+ \frac{\Lambda}{3}r 
\label{eq:coll}
\end{equation}
%\begin{figure}
%\psfig{file=tc1.eps,width=11cm,height=11cm}
%\caption[]{The time of collapse of a shell of matter in units of the age of the
%universe $t_{o}$ for $\nu=3$ (dotted line) compared with Gunn \& Gott's
%model (solid line).}
%\end{figure}
where $h(r,\nu )$ \footnote{As defined in Appendix B, $\nu=\delta(0)/\sigma$, where $\sigma$ is the mass variance filtered on a scale $R_f$.}
is the ordered specific angular momentum generated by tidal torques, $j(r, \nu)$ the random angular momentum (see RG87 and the following of the present paper), $G(r)$ the acceleration, $\Lambda$ the cosmological constant and $\mu$ the coefficient of dynamical friction.
I shall discuss in the Appendices C and D, how to calculate angular momentum and $\mu$.
In the peculiar case of $\mu=0$, Eq. (\ref{eq:coll}) can be integrated to obtain the square of velocity:
\begin{equation}
v(r)^2=2 \left[\epsilon -G \int_0^r \frac{m_T(y)}{y^2} d y +\int_0^r \frac{h^2}{y^3} dy + \frac{\Lambda}{6} r^2 
\right]
\end{equation}
where $\epsilon$ is the specific binding energy of the shell that can be obtained from the previous equation at turn-around when, $dr/dt=0$. 

If $\mu\neq 0$, the previous equation must be substituted with: 
\begin{equation}
\frac{d v^2}{d t}+2 \mu v^2=2 \left[
\frac{h^2+j^2}{r^3} -G\frac{m_T}{r^2} + \frac{\Lambda}{3} r
\right] v
\label{eq:veloc}
\end{equation}
which can be solved numerically for $v$.

%{\bf AGGIUNGERE CALCOLO TEMPO E RAGGIO DI TURN AROUND}

Following Gunn (1977) and FG84, the collapse factor $f(x_i)$ of a shell with initial radius $x_i$ and apapsis $x_m$ is given by:
\begin{equation}
f(x_i)=\frac{m_p(r_m)}{m_p(r_m)+m_{add}(r_m)}
\label{eq:cfact}
\end{equation}
and the final density profile can be obtained using 
Eq. (\ref{eq:dturn}) together with Eq. (\ref{eq:rc}) as:
\begin{equation}
\rho(x)=\frac{\rho_{ta}(x_m)}{f^3} \left[1+\frac{d \ln f}{d \ln g} \right]^{-1}
\label{eq:dturnnn}
\end{equation}
%Using virial theorem, one obtains the result that $f=1/2$, but 
The collapse factor $f(x_i)$ as previously reported and as confirmed by N-body simulations (Voglis et al. 1995), is not a constant but 
it is related to the initial profile of the density perturbation and moreover increases with the initial radius. 
The radial collapse is recovered for $f \rightarrow 0$ and $x_i\rightarrow 0$ (Lokas 2000).
% giving rise to very condensed central regions with very steep density profiles (Lokas 2000). 
%
%%%{TOGLIERE o spostare, connessione con ultima paginaAscasibar \it The density profile is a function of three parameters: 
%%%the spectral index $n$, the density parameter $\Omega$, and the height of the density peak,
%%%$\nu$. In the limit $\nu>>1$, the overdensity $\delta(r)$ is proportional
%%%to the two-point correlation function and the density profile is
%%%a function of $n$ and $\Omega$ only, and then the expected profile is
%%%that by HS. In DP2000, we obtained the density profile in scale-free universes and 
%%%in CDM cosmologies. In the quoted paper, the collapse was purely radial, we did not take 
%%%account of angular momentum. In the folowing, I shall extend DP200 model in order to take account of 
%%%angular momentum, dynamical friction and baryonic dissipative collapse}.
%
The calculation of the collapse factor (Eq. \ref{eq:cfact}) requires that we find before the mass $m_{add}$ which namely means to evaluate the integral in Eq. (\ref{eq:madd}).
After changing varibles from the turn-around radius to the initial one (Lokas 2000, Hiotelis 2002), one can calculate it numerically.
By means of the quoted change of variables, Eq. (\ref{eq:madd}) can be written as:
\begin{equation}
m_{add}(r_m)=4\pi\rho_{b,i}\int_{x_i}^{x_b}P_{x_i}(x'_i)[1+\delta_i(x'_i)]x_i'^2\mathrm{d}x'_i,
\label{eqb10}
\end{equation}
where $P_{x_i}(x'_i)=I(x_i)/I(x'_i)$ with
\begin{equation}
I(r)=\int_{x'_p}^r\frac{1}{v_{g(x'_i)}(g(\eta))}\frac{\mathrm{d}g(\eta)}{\mathrm{d} \eta}\mathrm{d} \eta,\\
\label{eqb11}
\end{equation}
$r_m = g(x_i)$, $x'_p=g^{-1}(x_p)$, and where $x_p$ is the pericenter of the shell with
initial radius $x'_i$. The upper limit $x_b$ of the integral 
%in (\ref{eqb10}), 
is taken to be the initial radius of the sphere that has collapsed at the present
epoch. 
The same change of variables must be applied to Eq. (\ref{eq:veloc}).
So, the radial velocity $v$ of a shell with apapsis $x=g(x_i)$ as it reaches the
radius $r=g(r_i)$ is given by:
%by the conservation of the energy of the shell and is:
%\begin{equation}
%v^2_x(r)=2[\Psi(r)-\varepsilon_x]-\frac{j^2_x}{r^2},\\
%\end{equation}
\begin{equation}
\frac{d v^2_x(r)}{d t}+2 \mu v_x^2=2 \left[
\frac{h^2_x+j^2_x}{r^3} - \Psi(r)
%G\frac{m_T}{r^2} 
+ \frac{\Lambda}{3} r
\right] v_x(r)
\label{eq:vell}
\end{equation}
%where $\Psi$ equals minus the potential $\Phi$,$\varepsilon_x$
%equals minus the specific energy and $j_x$ is the specific angular
%momentum of the shell. 
The potential $\Psi$ after the change of
variables is given by the expression:
\begin{equation}
\Psi[g(r_i)]=\frac{Gm(x_b)}{g(x_b)}+G\int_{r_i}^{x_b}\frac{m(x_i)}{g^2(x_i)}
\frac{\mathrm{d}g(x_i)}{\mathrm{d}x_i}\mathrm{d}x_i,\\
\end{equation}
where the distribution of mass $m(x_i)$ is that at the initial
conditions. 
%The energy of the shell is calculated by:
%\begin{equation}
%\varepsilon_x=\Psi[g(x_i)]-\frac{j^2_x}{2g^2(x_i)}.
%\end{equation}
%The angular momentum is introduced by the following scheme:
%
%Each shell expands radially from its initial radius $x_i$ up to
%its maximum expansion radius $x$. At this stage a specific angular
%momentum $j_x$ is added, given by
%$j_x=\mathcal{L}\sqrt{M(x)x}=\mathcal{L}\sqrt{M(x_i)g(x_i)}$,
%where $\mathcal{L}$ is a constant. This way of introducing angular
%momentum is consistent with the angular momentum distribution in
%N-body simulations (e.g, Barnes \& Efstathiou \cite{barnes}) and
%does not introduce any additional physical scale. It has been used
%by Avila-Reese et al. (\cite{avila}) and recently by Nusser
%(\cite{nusser}).
%}

Summarizing, knowing the initial conditions, angular momentum and the coefficient of 
dynamical friction, for a given shell one integrates the equation of motions (Eq. \ref{eq:vell}), 
%till turn-araound epoch (??). 
then 
one calculates the probability $P_{x_i}$ (Eq. \ref{eqb11}), from this the contribution of the shell to $m_{add}$ (Eq. \ref{eqb10}), the collapse factor (Eq. \ref{eq:cfact}) and finally the density profile through Eq. (\ref{eq:dturnnn}) (see also Lokas 2000 (Sect. 4); Ascasibar et al. 2004 (Sect. 2.1)).
%one has to calculate the collapse factor $f$ and the........
%(Ascasibar pag. 3; Lokas pag. 427; Hiotelis pag. 4).

\section{Initial conditions}

In order to calculate the density profile it is necessary to calculate the initial overdensity $\overline \delta_i (x_i)$. 
%from a primordial fluctuation $\overline \delta_i (x_i)$. 
This can be calculated when the spectrum of perturbations is known. 
%The model described in Section 2 allows to calculate the density profile arising from a primordial fluctuation $\overline \delta_i (x_i)$. 
It is widely accepted that structure formation 
in the universe is generated through the growth and collapse 
of primeval density 
perturbations originated from quantum fluctuations (Guth \& Pi 1982; 
Hawking 1982; Starobinsky 1982; BBKS) in an inflationary 
phase of early Universe. 
The growth in time of small 
perturbations is due to gravitational instability. 
The statistics of 
density fluctuations originated in the inflationary era are Gaussian, and 
can be expressed entirely 
in terms of the power spectrum of the density fluctuations: 
\begin{equation}
P( k) = \langle |\delta_{{\bf k}}|^{2} \rangle 
\end{equation}
where 
\begin{equation}
\delta_{{\bf k}} =\int d^{3} k exp(-i {\bf k x}) \delta({\bf x})
\end{equation}
\begin{equation}
\delta({\bf x}) = \frac{ \rho ({\bf x}) - \rho_{b}}{ \rho_{b} }
\end{equation}
and $ \rho_{b} $ is the mean background density. 
In biased structure formation theory it is assumed that cosmic structures 
of linear scale $ R_f$ form around the peaks of the density field, 
$  \delta( {\bf x})$, smoothed on the same scale. 
HS suggested that, according to the hierarchical scenario of structure formation, 
haloes should collapse around maxima of the smoothed density field (see below).  
The statistics of peaks in a Gaussian random field has been studied in the classical paper by BBKS. 
A well known result is the expression for the radial density profile of a fluctuation centered on a primordial 
peak of arbitrary height $\nu$:
\begin{equation}
\langle \delta (r) \rangle =\frac{\nu \xi (r)}{\xi (0)^{1/2}}-\frac{\vartheta (\nu
\gamma ,\gamma )}{\gamma (1-\gamma ^2)}\left[ \gamma ^2\xi (r)+\frac{%
R_{\ast }^2}3\nabla ^2\xi(r) \right] \cdot \xi (0)^{-1/2} 
\label{eq:dens}
\end{equation}
(BBKS; RG87),
where $\nu= \delta(0)/\sigma $ (see the following for a definition of $\sigma$) is the height of a density peak, $\xi (r)$ is the two-point 
correlation function:
\begin{equation}
\xi(r)= \frac{1}{2 \pi^2 r} \int_0^{\infty} P(k) k \sin(k r) d k
\end{equation}
$\gamma $ and $R_{\ast}$ are two spectral parameters
given respectively by:
\begin{equation}
\gamma =\frac{\int k^4P(k)dk}{\left[ \int k^2P(k)dk\int k^6P(k)dk\right]
^{1/2}}
\label{eq:gammm}
\end{equation}
\begin{equation}
R_{*}=\left[ \frac{3\int k^4P(k)dk}{\int k^6P(k)dk}\right] ^{1/2}
\label{eq:rrr}
\end{equation}
while $ \vartheta (\gamma \nu ,\gamma )$ is: 
\begin{equation}
\theta (\nu \gamma ,\gamma )=\frac{3(1-\gamma ^2)+\left( 1.216-0.9\gamma
^4\right) \exp \left[ -\left( \frac \gamma 2\right) \left( \frac{\nu \gamma }%
2\right) ^2\right] }{\left[ 3\left( 1-\gamma ^2\right) +0.45+\left( \frac{%
\nu \gamma }2\right) ^2\right] ^{1/2}+\frac{\nu \gamma }2}
\label{eq:tet}
\end{equation}

Then $\overline \delta_i$ is calculated from Eq. (\ref{eq:dens}) similarly to Ascasibar et al. (2003) (their Section 2.2).
%by using:
%\begin{equation}
%{\overline \delta_i}=\frac{3}{x_i^3} \int_0^{x_{i}} \delta(y)y^2 dy
%\end{equation}
In order to calculate $\delta(r)$ we need a power spectrum, $P(k)$.
The CDM spectrum used in this paper is that of BBKS (equation~(G3)), with transfer function:
\begin{equation}
T(k) = \frac{[\ln \left( 1+2.34 q\right)]}{2.34 q}
\cdot [1+3.89q+
(16.1 q)^2+(5.46 q)^3+(6.71)^4]^{-1/4}
%
%T^2(k) &=& [\ln \left( 1+4.164k\right)]^2 \cdot (192.9+1340k+ \nonumber \\
%& + &  1.599\cdot 10^5k^2+1.78\cdot 10^5k^3+3.995\cdot
%10^6k^4)^{-1/2}
%
\label{eq:ma5}
\end{equation}
where 
%$ A$ is the normalizing constant and 
$q=\frac{k\theta^{1/2}}{\Omega_{\rm X} h^2 {\rm Mpc^{-1}}}$.
Here $\theta=\rho_{\rm er}/(1.68 \rho_{\rm \gamma})$
represents the ratio of the energy density in relativistic particles to
that in photons ($\theta=1$ corresponds to photons and three flavors of
relativistic neutrinos). The spectrum is connected to the transfer function through the equation:
\begin{equation}
P(k)=P_{CDM} e^{-1/2 k^2 R_f^2}
\end{equation}
where $R_f$ is the smoothing (filtering) scale and $P_{CDM}$ is given by:
\begin{equation}
P_{CDM}= A k T^2(k)
\end{equation}
where $A$ is the normalization constant. 
We normalized the spectrum by imposing that the mass variance of the density field
\begin{equation}
\sigma^2(M)=\frac{1}{2 \pi^2} \int_0^\infty dk k^2 P(k) W^2(kR)
\end{equation}
convolved with the top hat window 
\begin{equation}
W(kR)=\frac{3}{(kR)^3} (\sin kR-kR \cos kR)
\end{equation}
of radius 8 $h^{-1}$ $Mpc^{-1}$ is $\sigma _{8}=0.76$ (Romano-Diaz et al. 2008).
%%%%The variance $\sigma_8=0.76$ (Romano-Diaz et al. 2008)
%(0.9 in Gnedin ed altri) 
%%%%of the density field convolved with the top hat window of radius 8 $h^{-1}$ $Mpc^{-1}$ was used to normalize the power spectrum.
%
%%%%%%%%%%%%%%%%%%The power spectrum was normalized to reproduce the observed abundance of rich 
%%%%%%%%%%%%%%%%%%cluster of galaxies (e.g., Bahcal \& Fan 1998).
%%%%%%%%
%In the following, we restrict our study to an Einstein-De Sitter ($\Omega=1$)
%Universe with zero cosmological constant and scale-free density
%perturbation spectrum $P(k)$
%\begin{equation}
%P(k)=A k^n
%\end{equation}
%with a spectral index in the range $-1 \leq n \leq 0$,
%and also to a CDM Universe with 
%spectrum given by BBKS:
%\begin{eqnarray}
%P(k) = Ak^{-1}\left[ \ln \left( 1+4.164k\right) \right] ^2 \nonumber\\
%\left(192.9+1340k+1.599\times 10^5k^2+1.78\times 10^5k^3+3.995\times 
%10^6k^4\right) ^{-1/2}
%\end{eqnarray}
%We normalized the spectrum by  
%imposing that the mass variance at $8h^{-1}Mpc$ is $\sigma _{8}=0.63$. 
%%%%%%%%%%%%%%As described in Section 3.1, 
Throughout the paper we adopt a $\Lambda$CDM cosmology with WMAP3 parameters, $\Omega_m=1-\Omega_{\Lambda}=0.24$,  $\Omega_{\Lambda}=0.76$, $\Omega_b=0.043$ and $h=0.73$, where $h$ is the Hubble constant in units of 100 km $s^{-1}$ $Mpc^{-1}$. 

%, $\Omega_m=1-\Omega_{\Lambda}=0.24$,  $\Omega_{\Lambda}=0.76$, $\Omega_b=0.043$ and $h=0.73$, where $h$ is the Hubble constant in units of 100 km %$s^{-1}$ $Mpc^{-1}$. The variance $\sigma_8=0.76$ (0.9 in Gnedin) of the density field convolved with the top hat window of radius 8 $h^{-1}$ %$Mpc^{-1}$ (?) was used to normalize the power spectrum. 
%%%%
%For 
%the $\Lambda$CDM model ($\Omega_{\rm m}=0.3$, $\Omega_{\Lambda}=0.7$,
%$h=0.7$), I also use the BBKS spectrum normalized as
%$\sigma_8=1$. 

%CI VUOLE LO SPETTRO $\Lambda$CDM (vedi Ascasibar).

The mass enclosed in $R_f$ is calculated, as in RG87, as $M=4 \pi/3 \rho_b R_f^3$, so that for $R_f=0.12$ Mpc, $M \simeq 10^9 M_{\odot}$
\footnote{For precision sake, the mass scale $M$ is connected to the smoothing scale by:
%by $M \propto l^3$ (BBKS, specificare meglio; Hiotelis 2002;....):
%the mass enclosed by a Gaussian smoothing function is
$M_G=(2 \pi) ^(3/2) \b R_{f}^3$
for a Gaussian smoothing ($P(k,R_f)=e^{-R_f^2 k^2}P(k)$) and by
$M_{TH}=4 \pi/3 \b R_{TH}^3$
for top hat smoothing.
The mass enclosed by the smoothing function applied to the uniform background is the same for $R_f=0.64 R_{TH}$ (see BBKS). 
}. 
%and the virial mass, $M_v$ 
%is connected to the filtering scale by: 
%\begin{equation}
%M_v=4 \pi/3 \rho_c \Delta_v r_v^3 
%\end{equation}
Structure like Galaxies form from high peaks in the density field, high enough so that they stand out above the 
``noise" and dominate the infall dynamics of the surrounding matter. 
%What is the mean
%excess density distribution around a peak? 
Density profile around particles located at the local maxima and minima of the density field is given by 
$<\frac{\delta \rho}{\rho} (x)>=<\delta(x)> = \nu \xi(x)/\xi(0)^{1/2}$
%$\frac{\delta \rho}{\rho}(x)$
% =\delta(x) = \nu \xi(x)/\xi(0)^{1/2}$ 
(Peebles 1984), where $\xi$ is the two-point mass correlation function, and $x$ 
is the comoving separation. Eliminating minima, the density excess around centers of local density peaks
was derived by BBKS, and is given by Eq. (\ref{eq:dens}). 
%in Appendix A. 

The amplitude of any given peak is expressed in terms of its $\sigma$ deviation, where 
$\sigma=\xi(0)^{1/2}$. 
%, and $\xi(0)$ is the zero lag value of the two-point function. 
Thus the central density contrast of an $ \nu \sigma$ peak is $ \nu \xi(0)^{1/2}$ and the peak height is given by $\nu=\delta(0)/\sigma$.
Given that galaxies are rather common, they must have formed from 
peaks that are not very rare, say, 2-4 $\sigma$ peaks (RG87)). 
In Fig. 6, we plot density profiles for $\nu=2,3,$ and $4$. Light lines are density runs around
maxima or minima, and are proportional to $\xi(x)$. Heavy lines show density profiles calculated by BBKS.
{\bf Note that the plotted quantity is the initial excess density distribution around a peak linearly evolved to the present
day, namely $\delta(x)=\delta_i D(z_i)$, where $D(z)$ describes the growth of density perturbations (Peebles 1980).}
%$\delta_0(x)$, the initial excess density distribution around a peak linearly 
%evolved to the present day. The density profile of an initial, pre-collapse halo,
%at early times, $\delta_i(x)$, is related to $\delta_0(x)$
%by the linear growth factor: $\delta_i(x)=\delta_0(x)/(1+z_i)$. 
As previously reported in Section 2, the assumption that the initial density profile is proportional to the 
correlation function (light lines in Fig. 6), instead of using the BBKS initial density profile (heavy lines in Fig. 6) 
is one of the reasons causing the discrepancies between the SIM and some high resolution N-body simulations (DP2000).

\section{Calculation of the angular momentum}

\subsection{Ordered angular momentum}

%AGGIUNGERE anche j

The explanation of galaxies spins gain through tidal torques was pioneered
by Hoyle (1949). Peebles (1969) performed the first detailed calculation of the
acquisition of angular momentum in the early stages of protogalactic
evolution. More recent analytic computations (White 1984, Hoffman 1986,
R88; Eisenstein \& Loeb 1995; Catelan \& Theuns 1996a, b) and numerical simulations (Barnes \& Efstathiou 1987) have
re-investigated the role of tidal torques in originating galaxies angular
momentum. 

Following Eisenstein \& Loeb (1995), we separate the universe into two
disjoint parts: the collapsing region, characterized by having high density,
and the rest of the universe.
The boundary between these two regions is taken to be a
sphere centered on the origin.
As usual, in the following, we denote with $\rho({\bf x})$, being ${\bf x}$
the position vector, the density as function of space and
$\delta({\bf x})={\rho({\bf x})-\rho_{\rm b} \over \rho_{\rm b}}$.
%%This two regions are separated by a sphere
The gravitational force exerted on the spherical central region by the external
universe can be calculated by expanding the potential, $\Phi({\bf x})$, in spherical harmonics.
Assuming that the sphere has radius $R$, we have:
\begin{equation}
\Phi ({\bf x})=\sum_{l=0}^\infty {4\pi  \over 2l+1}%
\sum_{m=-l}^l a_{\rm lm}(x)Y_{\rm lm}(\theta ,\phi )x^l 
\end{equation}
where $Y_{\rm lm}$ are spherical harmonics and the tidal moments,
$a_{\rm lm}$, are given by:
\begin{equation}
a_{\rm lm}(x)=\rho_{\rm b}\int_R^\infty Y_{\rm lm}(\theta ,\phi )\rho ({\bf s})
s^{-l-1}d^3s 
\end{equation}

In this approach the proto-structure
is divided into a series
of mass shells and the torque on each mass shell is computed separately. The
density profile of each proto-structure is approximated by the superposition
of a spherical profile, $\delta (r)$, and a random CDM
distribution, ${\bf %
\varepsilon (r)}$, which provides the quadrupole moment of
the proto-structure.
To the first order, the initial density can be represented by:
\begin{equation}
\rho ({\bf r})=\rho _{\rm b}\left[ 1+\delta (r)\right] \left[ 1+\varepsilon ({\bf
r})\right]  
\label{eq:profil}
\end{equation}
where
%%$\rho_{\rm b}$ is the background density and
$ \varepsilon(\bf r)$
is given by:
\begin{equation}
\langle |\varepsilon _k|^2 \rangle = P(k) 
\end{equation}
being $ P(k)$ the power spectrum.
The torque on a thin spherical shell of internal radius $x$ is given by:
\begin{equation}
{\bf \tau}(x)=-{G M_{\rm sh} \over 4 \pi} \int \varepsilon({\bf x})
{\bf x} {\bf \times} {\bf \bigtriangledown} \Phi({\bf x}) d \Omega 
\label{eq:tauu}
\end{equation}
where $M_{sh}= 4 \pi \rho_{\rm b}\left[1+\delta(x)\right] x^2 \delta x$.
Before going on, I want to recall that we are interested in the
acquisition of angular momentum from the inner region, and
for this purpose we take account only
of the $l=2$ (quadrupole) term. In fact, the $l=0$ term produces no force, while the
dipole ($l=1$) cannot change the shape or induce any rotation of the
inner region. As shown by Eisenstein \& Loeb (1995), in the standard CDM
scenario the dipole is generated at large scales, so the object we are studying
and its neighborhood move as bulk flow with the consequence that the
angular distribution of matter will be very small, then the dipole terms can be
ignored. Because of the isotropy of the random field, $\varepsilon(\bf x)$,
Equation~(\ref{eq:tauu}) can be written as:
\begin{equation}
<|{\bf \tau}|^2>=\sqrt(30) {4 \pi G \over 5}
\left[<a_{2m}(x)^2><q_{2m}(x)^2>-
<a_{2m}(x) q^{\ast}_{2m}(x)>^2
\right]^{1/2}   
\label{eq:tauuu}
\end{equation}
where $<>$ indicates a mean value of the physical quantity considered.
%As stressed in the next section, following Eisenstein \& Loeb (1995),
%the integration of the equations of motion shall
%be ended at some time before the inner external tidal shell (i.e.,
%the innermost shell of the part of the universe outside the sphere containing
%the ellipsoid) collapses.
%
%%be performed till a time
%%at which the external tidal shells does not collapse before the integration
%%ends.
%Then the inner region behaves as a density peak. This last
%point is an important one in the development of the present paper.

%%In this paper, I'll consider only peaks with $\nu>2$.
%%%%%\end{Fv}
%%
%%%In any case, the result of the model shall be compared to that of
%%%Eisentein \& Loeb (1995) which calculates the acquisition
%%%of angular momentum without our previous approximation,
%%%and supposing that the collapsing region is an ellipsoid.
%%

In order to find the total angular momentum imparted to a mass shell by tidal
torques, it is necessary to know the time dependence of the torque.
This can be done connecting $q_{\rm 2m}$ and $a_{\rm 2m}$ to
parameters of the spherical collapse model (Eisenstein \& Loeb 1995
(equation~(32), R88 (equation~(32) and (34)). 
Following R88 we have:
\begin{equation}
q_{\rm 2m} (\theta )={1 \over 4} q_{\rm 2m,0}
\overline{\delta} _0^{-3} 
{\left( 1-\cos {\theta}\right) ^2 f_2 (\theta) \over
f_1(\theta )-\left({\delta _0 \over \overline{\delta} _0}\right)
f_2(\theta )}  
\label{eq:quadd}
\end{equation}
and
\begin{equation}
a_{\rm 2m} (\theta )=a_{\rm 2m,0} 
\left({4 \over 3}\right)^{4/3}
\overline{\delta}_0 (\theta-\sin{\theta})^{-4 \over 3} 
\label{eq:a2m}
\end{equation}
The collapse parameter $\theta$ is given
by:
\begin{equation}
t(\theta)={3 \over 4} t_0 \overline{\delta}_0^{-3/2}(\theta-\sin{\theta})
\end{equation}
Equation~(\ref{eq:quadd}) and (\ref{eq:a2m}),
by means of equation~(\ref{eq:tauuu}), give to us the tidal
torque:
\begin{equation}
\tau (\theta )=\tau _0 {1 \over 3}({4 \over 3})^{(1/3)}
\overline{\delta} _0^{-1}{
\left( 1-\cos {\theta }\right) ^2 \over (\theta -\sin {\theta })^{(4/3)}}
{f_2(\theta ) \over f_1(\theta )-
\left({\delta _0 \over \overline{\delta} _0}\right)
f_2(\theta )} 
\end{equation}
where $f_1(\theta)$ and $f_2(\theta)$ are given in R88 (Eq. 31), $\tau_0$
and $\delta_0={\rho-\rho_{\rm b} \over \rho_{\rm b}}$
are respectively the torque and the mean fractional density excess inside the shell,
as measured at current epoch $t_0$.
The angular momentum acquired during expansion can then be obtained integrating
the torque over time:
\begin{equation}
L=\int \tau(\theta) {d t \over d \theta} d\theta  
\label{eq:ang}
\end{equation}
%%
%%\end{Fv}
%%
%%%Eq.(\ref{eq:ang}) for $\theta<<1$ gives $L \propto t^{5/3}$.
%%%It is important to recall that in non-linear models the angular momentum
%%%acquired by protostructures is only a factor 1.3 than the corresponding
%%%linear estimate (Catelan \& Theuns 1996b). \\
As remarked in the Del Popolo et al. (2001) the angular momentum obtained from
Eq. (\ref{eq:ang}) is evaluated at the time of maximum expansion $t_{\rm m}$.
Then the calculation of the angular momentum can be solved by means
of Eq. (\ref{eq:ang}), once we have made a choice for the power spectrum.
With the power spectrum (filtered on a galactic scale) and the parameters given in Appendix A, 
%the next section 
for a $\nu=2$ peak of mass $\simeq 2 \times 10^{11} M_{\odot}$, the model gives a value of
$2.5 \times 10^{74}$ $\frac{g cm^2}{s}$.
%$2.5 \times 10^{74} {\rm g cm^2/s}$. 
%As previously quoted, we assume
%that from $t_{M}$ on, the ellipsoid has this constant angular momentum. 
%Following the procedures 1) and/or 2), we shall be able to get the
%time evolution of the density.
%%
%%This expression,
%%together
%%with an expression for
%%the magnitude of the shear $\Sigma^2=\Sigma_{\rm ij} \Sigma^{\rm ij}$, 
%%can be introduced into Eq. (\ref{eq:dens}) in order to
%%calculate the evolution of density.
%%The shear term was obtained using Hoffman (1986a) paper showing that:
%%$$
%%\Sigma_{\rm ij} \propto \bigtriangledown { \bf v_{\rm p}}
%%$$
%%where ${\bf v_{\rm p}}$ is the peculiar velocity. This behavior agrees
%%with the exact non-linear theory, where the shear grows with time only in
%%the non-linear regime (Peebles 1980). Also in the QL regime the shear
%%is almost constant (Hoffman 1986a).

It is interesting to compare with a different method like that of Catelan \& Theuns (1996), who calculated the
angular momentum at maximum expansion time (see their Eqs. (31)-(32)) and compared it with previous theoretical
and observational estimates. Assuming the same value of mass $\nu$ used to obtain our previously quoted result 
%and the same redshift (z\3) and 
same distribution of angular momentum ($l_f$) as adopted by Catelan \& Theuns (1996), we obtain a value for the
angular momentum of $2.0 \times 10^{74}$ $\frac{g cm^2}{s}$ in agreement with our result.
%. This last result
%is in good agreement with ours and is well in line with previous theoretical estimates (Peebles 1969; Heavens \&
%Peacock 1988) and numerical simulations (Fall 1983).

For what concerns the calculation of the random angular momentum, $j$, it was described in next subsection.
%Appendix.

\subsection{Random and ordered angular momentum}

As reported in the introduction, several authors have emphasized the effect of an isotropic velocity dispersion (thus of non-radial
motion) in the core of collisionless haloes (RG87; White \& Zaritski 1992; Avila-Reese et al. 1998; Hiotelis 2002; Nusser 2001; Le Delliou \& Enriksen 2003; Ascasibar et al. 2004; Williams et al. 2004). One common result of the previous studies is that larger amount of angular momentum leads to shallower final density profiles in the inner region of the halo. 
There are two sources of angular momentum of collisionless dark matter: (a) bulk streaming motions, and (b) random tangential motions.
The first one (ordered angular momentum (RG87)), as described in the previous subsection, arises due to tidal torques experienced by proto-halos, and is usually quantified as a dimensionless spin parameter $\lambda$ (Peebles 1969). The second one (random angular momentum (RG87)) is connected to random velocities (see RG87 and the following of the present paper).

%Our halos do not have angular momentum of Type (i), only of Type (ii), so the net 
%{\it vector} sum of angular momentum in our halos is zero. Because dark matter
%is assumed to be collisionless, it is the sum of the {\it magnitudes} of angular 
%momenta of particles that is relevant for the dynamics, not the vector sum. 
%In all the following discussions angular momentum will mean Type (ii) angular
%momentum.
%
%The total angular momentum of the system is used in this paper as a
%measure of the non-radiality of the orbits of collapsing particles, that
%is to define the eccentricity of  orbits. 
In the present paper, we took into account both types of angular momentum: random $j$, and ordered, $h$. Type (a), connected to tidal torques is obtained as described in Appendix C1,  
%Filtering the spectrum on cluster scales, $R_f=3h^{-1}Mpc$, 
obtaining the rms torque, $\tau (r)$, on a mass shell using Eq. (\ref{eq:tauu}) and then 
calculating the total specific angular momentum, $h(r,\nu )$, acquired during
expansion by integrating the torque over time (Ryden 1988a (hereafter R88), Eq. 35): 
\begin{equation}
h(r,\nu )=\frac 13\left( \frac 34\right) ^{2/3} 
%\nonumber \\
\frac{\tau _ot_0}{M_{sh}}
%{M_{sh}}
\overline{\delta }_o^{-5/2}\int_0^\pi \frac{\left( 1-\cos \theta \right) ^3}{%
\left( \vartheta -\sin \vartheta \right) ^{4/3}}\frac{f_2(\vartheta )}{%
f_1(\vartheta )-f_2(\vartheta )\frac{\delta _o}{\overline{\delta _o}}}%
d\vartheta   \label{eq:ang}
\end{equation}
%\begin{eqnarray}
%h(r,\nu )=\frac 13\left( \frac 34\right) ^{2/3} \nonumber \\
%\frac{\tau _ot_0}{M_{sh}}
%\overline{\delta }_o^{-5/2}\int_0^\pi \frac{\left( 1-\cos \theta \right) ^3}{%
%\left( \vartheta -\sin \vartheta \right) ^{4/3}}\frac{f_2(\vartheta )}{%
%f_1(\vartheta )-f_2(\vartheta )\frac{\delta _o}{\overline{\delta _o}}}%
%d\vartheta   \label{eq:ang}
%\end{eqnarray}
where $M_{sh}= 4 \pi \rho_{\rm b}\left[1+\delta(x)\right] x^2 \delta x$ is the mass in a thin spherical shell of internal radius $x$, $\tau_o$ is the tidal torque at time $t_0$, 
the functions $f_1(\vartheta )$, $f_2(\vartheta )$ are given by R88 (Eq. 31) while the mean over-density inside the shell, $\overline{%
\delta }(r)$, is given by Eq. (\ref{eq:overd}).
%Ryden (1988a): %(Eq. 9 and 13)
%\begin{equation}
%\overline{\delta }(r,\nu )=\frac 3{r^3}\int_0^\infty d\sigma \sigma ^2\delta
%(\sigma )
%\label{eq:denss}
%\end{equation}
%where $\delta(r)=\frac{\rho(r)-\rho_b}{\rho_b}$.

In Fig. 7, we show the variation of $h(r,\nu)$ with mass $M$ \footnote{As previously reported radius scales with mass as $r \propto M^{1/3}$.} 
%the distance $r$
for three values of the peak height $\nu=2$, 3, 4 for a power spectrum filtered on scale $R_f=0.12$ Mpc. The rms specific angular
momentum, $h(r,\nu)$, increases with distance $r$ while peaks
of greater $\nu$ acquire less angular momentum via tidal torques.
This is the angular momentum-density anti-correlation showed
by Hoffman (1986). This effect arises because the angular momentum
is proportional to the gain at turn around time, $t_m$,
which in turn is proportional to $\overline \delta$$(r, \nu)^{-3/2} \propto \nu ^{-3/2}$.

For what concerns the calculation of the random angular momentum, $j$, we have to recall that 
%I'm not quite sure about the validity of Ryden results regarding j estimate. 
this quantity is product of highly non-linear processes and it is difficult to calculate. 
An estimate has been performed  
%It has been calculated 
by RG87 and R88 assuming that 
the substructure of a peak can be described as the sum of the mean spherical profile, $\delta(x)$, and of a random Gaussian field with the same power spectrum as the parent density field (see Appendix C, Eq. \ref{eq:profil}). 
%RG87 and Williams et al. (2004) assumed that angular momenta arise from the random secondary perturbations and derived it from the same fluctuation %spectrum that generates the primary peak. 
RG87 showed that $j(r)=(2/3)^{1/2} r_{m} \Delta v(r,t_{m}/2)$ where $\Delta v$ is the r.m.s. velocity.
It is important to note that $j$, as $h$, are anti-correlated with the height of the peak since both $\Delta v$ and $r_{max}$ decrease with an increase in the height $\nu$ of the peak. To the lowest order in $r/R_f$ the random angular momentum $j$ is (Ryden 1988b):
\begin{equation}
\frac{j(r)}{(G \rho_b \sigma_0)^{1/2} R_f^2} \simeq \frac{1.5}{\nu^{3/2}} (\frac{r}{R_f})^2
\end{equation}
The ratio of ordered to random angular momentum, to the same order is:
\begin{equation}
\frac{h}{j} \simeq \frac{0.09}{\nu} (\frac{r}{R_f})^2 (n+3)
\end{equation}
The random angular momentum is important for the particle (shell) orbits since the larger it is, the larger is the orbital ellipticity, and then the shell penetrates less to the center, resulting in a flattening of the inner density profile.
In several of the previously quoted papers only the second type of angular momentum (the one generated by random motions) was taken into consideration. The usual approach (Nusser 2001; Hiotelis 2002; Ascasibar et al. 2003) consists in assigning a specific angular momentum at turn around:
\begin{equation}
j \propto \sqrt{GM r_m}  
\end{equation}
%where $r_m$ is the maximum radius of the shell. 
With this prescription, the orbital eccentricity $e$ is the same for all particles in the halo (Nusser 2001).
%This angular momentum, $j$, has been taken into account expressing it in terms of the eccentricity, $e$, of orbits (Avila-Reese et al. 1998; Nusser 2002; Ascasibar et al. %2003)....
Avila-Reese et al. (1998) expressed the specific angular momentum $j$ through the ratio $e_0=\left( \frac{r_{min}}{r_{max}} \right)_0$, 
%%%%%%($\frac{r_{max}-r_{min}}{r_{max}+r_{min}}$, Ascasibar).
and left this quantity as a free parameter. Processes related to mergers and tidal forces that could produce tangential perturbations in the collapsing matter were implicitly considered in their model trough the free parameter $e_0$. According to them, the detailed description of 
these processes is largely erased by the virialization process, remaining only through the value of $e_0$, which then they fixed to $e_0=0.3$. 
The value $e \simeq 0.2$ gives density profiles very close to the NFW profile (Avila-Reese et al. 1998, 1999).
%. They used values for $e$ obtained in numerical N-body simulations of halo collapse. For CDM halos, this value is around 0.1-0.3 (e.g., Ghigna et %al. 1999).
%
%%Williams
%%The average orbit eccentricity is approximately constant within the 
%%virialized halo, changing only from $\langle r_{peri}/r_{apo}\rangle \approx 0.3$
%%at $R_{vir}$ to $0.35$ at $0.01\,R_{vir}$, as orbits become more circular 
%%close to center. The constancy of $r_{peri}/r_{apo}$
%%was predicted by Nusser (2001) for halos with adiabatically varying potentials,
%%and power law radial profiles. It is not entirely surprising that our method, which 
%%conserves adiabatic invariants, but does not insist on power laws density profiles,
%%yields roughly constant $\langle r_{peri}/r_{apo}\rangle$ ratios. It is more 
%%surprising that very high resolution N-body simulations described in 
%%Ghigna \etal (1998) also produce constant $\langle r_{peri}/r_{apo}\rangle$
%%($\approx 0.2$) ratios of dark matter particles in virialized
%%halos. This constancy of $\langle r_{peri}/r_{apo}\rangle$ throughout the halo 
%%could suggest that the adiabatic approximation is largely valid for halos 
%%generated in N-body simulations (see also Jesseit \etal 2002).
%
This procedure is justified by N-body simulations of halo collapse:
%In order to take account of type b of angular momentum, one can follow Avila-Reese et al. (1998), and
%express $j$ through the ratio $e_0=\frac{r_{min}}{r_{max}}_0$ ($\frac{r_{max}-r_{min}}{r_{max}+r_{min}}$, Ascasibar).
%They used values for $e_0$ obtained in numerical N-body simulations of halo collapse. 
for CDM halos, N-body simulations produce constant $< \frac{r_{min}}{r_{max}}> \simeq 0.2$ ratios of dark matter particles in virialized haloes \footnote{The constancy of $< \frac{r_{min}}{r_{max}}>$ throughout the halo has been interpreted as a proof that the adiabatic approximation is largely valid for haloes generated in N-body simulations (Jesseit et al. 2002).}, 
and the range for this value is around 0.1-0.3 (e.g., Ghigna et al. 1999).
%, with  $e \simeq 0.2$ giving density profiles very close to the NFW profile (Avila-Reese et al. 1998, 1999).
Ascasibar et al. (2003) noticed that particle orbits are slightly more radial as we move out to the current turn around radius, $r_{ta}$. There is a dependence on the 
dynamical state: major mergers are well described by constant eccentricity up to the virial radius \footnote{For their dark matter haloes, $r_{v}/r_{ta}$ is typically of the order of 0.2-0.3.}, 
while relaxed systems are more consistent with a power-law 
profile. Minor mergers are in the middle. The average profile can be fitted by a power law, but the slope is shallower than for relaxed systems. A least-square fit to the relaxed population yields:
\begin{equation}
e(r_{max}) \simeq 0.8 (r_{max}/r_{ta})^{0.1}
\end{equation}
for $r_{max}< 0.1 r_{ta}$. 
In the present paper, we use Avila-Reese et al. (1998) method with the correction of Ascasibar et al. (2003).
It is important to notice that even using a constant value for the eccentricity angular momentum will change with radius and $\nu$.
In fact the specific angular-eccentricity relation is:
\begin{equation}
j = \sqrt{G M r_{max} (1-e)}
\end{equation}
which depends from mass (radius) and is anti-correlated with $\nu$ ($r_{m} \propto 1/\nu$). 
%(VERIFICARE)

In Fig. 8, we plot the specific random angular momentum for different values of $\nu$ showing a similar behavior to that of the ordered angular momentum. 
%A comparison of Fig. 2 and Fig. 3, reveals that 
Since the ordered angular momentum $h$ of a shell increases more rapidly with the initial radius of the shell than random angular momentum $j$ (R88),
%. As a consequence 
the inner regions of the halo are ``hotter" than the outer regions, in the sense of having a smaller ratio of ordered to random velocities. Moreover high peaks are hotter than low peaks. 
%(AGGIUNGERE FIGURA PER j, e $h/j$). 
This behaviour is plotted in Fig. 9, where we plot the ratio of ordered to random angular momentum. 
  
It is important to note that after turn-around, not only the random angular momentum $j$ will contribute to originate non-radial motions in the protostructure but also the ordered angular momentum.
In fact, as successive shells expand to their maximum radius they acquire angular momentum and then fall in on orbits determined by the angular momentum. Actual peaks are not spherical; thus the infall of matter will not be purely radial. Random substructure in the region surrounding the peak will divert infalling matter onto non-radial orbits.  
%Dissipative physics and the process of violent relaxation will eventually intervene and convert the kinetic energy of collapse into random motions %(virialization). 
The role of this random motions is of fundamental importance in the structure formation. 
%%%%%%%%DIRE CHE sia h che j contribuiscono ai moti non radiali: acquisto momento angolare in espansione (h) conversione in moti random.....
%This agrees with the previrialization conjecture (Peebles \& Groth 1976; Davis \& Peebles 1977; Peebles 1990), according to which initial %asphericities and tidal interactions between neighboring density fluctuations induce significant non-radial motions that oppose or stop the collapse.
%%%In RG87 picture, tangential velocity is provided by the field of random perturbations which is superposed on a large spherical perturbation %%%$\delta(x)$.

In the present paper, as in Nusser (2001), and Williams et al. (2004)\footnote{Note that in those papers, just the random angular momentum is taken into account.}, the contribution of $h$ and $j$ are evaluated analytically but are not imparted to the shell until it reaches turn-around.
In other words, as long as a given shell is expanding its dark matter particles' positions 
and velocities are not corrected for the effects of $h$ and $j$;
%secondary perturbations; 
it is assumed that during this relatively orderly evolutionary stage the effects
of random perturbations are small compared to what they will be later when
the halo starts to collapse.  
Finally, angular momentum is taken into account by adding the term $\frac{h^2+j^2}{r^3}$ in the equation of acceleration of the shell.

\section{Dynamical friction.}

The effect of dynamical friction on the central halo profile has been studied in several papers (e.g., EZ01; Tonini, Lapi \& Salucci 2006 (TLS)) using different approaches. EZ01 studied the changes induced by dynamical friction in the dark matter halo structure within the context of galaxy formation, where gas cools to form dense clumps. The orbital energy lost by the clumps to the dark matter background 
is sufficient to ``heat up" and flatten the dark matter density cusps. 
TLS performed two independent calculations over the NFW halo: 1) they perturbed its phase-space distribution function with angular momentum and 2) they followed the evolution of the baryonic angular momentum during the collapse inside the halo potential well. In both cases the final result 
was a core-like profile.
In the present paper, we took into account dynamical friction by introducing the dynamical friction force (see the following) in the equation of motions.

%\section{Dynamical friction and structure formation. } 

In a hierarchical structure formation model, the large scale cosmic
environment can be represented as a collisionless medium made of a hierarchy
of density fluctuations whose mass, $M$, is given by the mass function $%
N(M,z)$, where $z$ is the redshift. In these models matter is concentrated
in lumps, and the lumps into groups and so on.
%In what follows, we consider
%a shell of matter outside the main body of a perturbation which collapsed to
%form a cluster of galaxies.\\ The equation of motion of a shell of matter
%(as previously told composed of galaxies and substructure) around a maximum
%of the density field (neglecting tidal interactions and the substructure),
%can be expressed in the form: 
%\begin{equation}
%\frac{d^2r}{dt^2}=-\frac{GM}{r^2(t)}  \label{eq:pee}
%\end{equation}
%(Peebles 1980, Eq. ~19.9), where $M$ is the mass enclosed in the proper
%radius $r(t)$. Using Gunn \& Gott's notation the proper radius can be
%written as: 
%\begin{equation}
%r(r_i,t)=a(r_i,t)r_i
%\end{equation}
%where $r_i$ is the initial radius and $a(r_i,t)$ is the expansion parameter
%of the shell. At the initial time $t_i$ the initial condition is given by 
%\begin{equation}
%a(r_i,t_i)=1
%\end{equation}
%In presence of substructure it is necessary to modify the equation of
%motion, Eq. ~(\ref{eq:pee}), because the graininess of mass distribution in
%the system induces dynamical friction that at last introduces a frictional
%force term to Eq. ~(\ref{eq:pee}).\\
In such a material system, gravitational
field can be decomposed into an average field, ${\bf F}_0(r)$, generated
from the smoothed out distribution of mass, and a stochastic component, $%
{\bf F}_{stoch}(r)$, generated from the fluctuations in number of the field
particles. %first neighbours of a test one. 
The stochastic component of the gravitational field is specified assigning a
probability density, $W({\bf F})$, (Chandrasekhar \& von Neumann 1942). In
an infinite homogeneous unclustered system $W({\bf F})$ is given by
Holtsmark distribution (Chandrasekhar \& von Neumann 1942) while in
inhomogeneous and clustered systems $W({\bf F})$ is given by Kandrup (1980)
and Antonuccio-Delogu \& Barandela (1992) respectively. The stochastic
force, ${\bf F}_{stoch}$, in a self-gravitating system modifies the motion
of particles as it is done by a frictional force. In fact a particle moving
faster than its neighbors produces a deflection of their orbits in such a
way that average density is greater in the direction opposite to that of
traveling causing a slowing down in its motion. Following Chandrasekhar
\& von Neumann's (1942) method, the frictional force which is experienced by
a body of mass $M$ (galaxy), moving through a homogeneous and isotropic
distribution of lighter particles of mass $m$ (substructure), having a
velocity distribution $n(v)$ is given by: 
\begin{equation}
M\frac{d{\bf v}}{dt}=-4\pi G^2M^2n(v)\frac{{\bf v}}{v^3}\log \Lambda \rho 
\label{eq:cha}
\end{equation}
where $\log \Lambda $ is the Coulomb logarithm, $\rho $ the density of the
field particles (substructure). \\ 
A more general formula is that given by Kandrup(1980) in the hypothesis that there are no correlations among random
force and their derivatives: 
\begin{equation}
{\bf F}=-\mu {\bf v}=-\frac{\int W(F)F^2T(F)d^3F}{2<v^2>}{\bf v}
\end{equation}
where $F$ is dynamical friction force per unit mass,
$\mu $ is the coefficient of dynamical friction, $T(F)$ the average
duration of a random force impulse, $<v^2>$ the characteristic speed of a
field particle having a distance $r\simeq (\frac{GM}F)^{1/2}$ from a test
particle (galaxy). This formula is more general than Eq. (\ref{eq:cha})
because the frictional force can be calculated also for inhomogeneous
systems when $W(F)$ is given. If the field particles are distributed
homogeneously the dynamical friction force per unit mass is given by: 
\begin{equation}
F=-\mu v=-\frac{4.44G^2m_a^2n_a}{[<v^2>]^{3/2}}\log \left\{ 1.12\frac{<v^2>%
}{Gm_an_a^{1/3}}\right\} 
\end{equation}
(Kandrup 1980), 
where $m_a$ and $n_a$ are respectively the average mass and number density of the field particles.
%$N=\frac{4\pi }3R_{sys}^3n_a$, $n_{ac}=n_a\times a^3$ is the
%comoving number 
%density 
%of peaks of substructure 
%of field particles 
Using the virial theorem we also have: 
\begin{equation}
\frac{<v^2>}{Gm_an_a^{1/3}}\simeq \frac{M_{tot}}m_a\frac
1{n_a^{1/3}R_{sys}}\simeq N^{2/3}
\end{equation}
where $M_{tot}$ is the total mass of the system, $R_{sys}$ its radius 
and $N$ is the total number of field particles. 
The dynamical friction force per unit mass can be
written as follows: 
\begin{equation}
F=-\mu v=-\frac{4.44[Gm_an_{ac}]^{1/2}}N\log \left\{ 1.12N^{2/3}\right\}
\frac v{a^{3/2}}=-\epsilon_o \frac v{a^{3/2}}
\end{equation}
where $N=\frac{4\pi }3R_{sys}^3n_a$, 
%is the total number of objects generating the stochastic field, $R_{sys}$ the system radius,
%$m_a$ and $n_a$ are respectively the average mass and the 
%cumulative 
%number density of the subunits generating the fluctuating field,
%the average mass and
%density of the field particles, 
%peaks inside a protostructure of radius $R_{sys}$,
$a$ is the expansion parameter, connected to the proper radius of a shell by: 
\begin{equation}
r(r_i,t)=r_ia(r_i,t)
\end{equation}
$n_{ac}=n_a\times a^3$ is the comoving number 
%density 
density of field particles. 
%where $N=\frac{4\pi }3R_{sys}^3n_a$ and $n_{ac}=n_a\times a^3$ is the
%comoving number density of peaks of substructure of field particles. 
This last equation supposes that the field particles generating the stochastic
field are virialized. 
This is justified by the previrialization hypothesis
(Davis \& Peebles 1977). To calculate the dynamical evolution of the
galactic component of the cluster it is necessary to calculate the number
and average mass of the field particles generating the stochastic field. \\ %
The protocluster, before the ultimate collapse at $z\simeq 0.02$, is made of
substructure having masses ranging from $10^6-10^9M_{\odot }$ and from
galaxies. I suppose that the stochastic gravitational field is generated
from that portion of substructure having a central height $\nu $ larger than
a critical threshold $\nu _c$. This latter quantity can be calculated
(following Antonuccio-Delogu \& Colafrancesco 1994 (hereafter ADC)) using the condition that the peak radius, $r_{pk}(\nu \ge \nu
_c),$ is much less than the average peak separation $n_a(\nu \ge \nu
_c)^{-1/3}$, where $n_a$ is given by the formula of BBKS for the upcrossing
points: 
\begin{eqnarray}
n_{ac}(\nu \ge \nu _c)=\frac{\exp (\nu _c^2/2)}{(2\pi )^2}(\frac \gamma
{R_{*}})^3 [ \nu _c^2-1+ \nonumber \\
\frac{4\sqrt{3}}{5\gamma ^2(1-5\gamma ^2/9)^{1/2}}
\exp ( -5\gamma ^2\nu _c^2/18)] 
\end{eqnarray}
where $\gamma $, $R_{*}$ are parameters related to moments of the power
spectrum (BBKS Eq. ~4.6A). The condition $r_{pk}(\nu \ge \nu _c)<0.1n_a(\nu
\ge \nu _c)^{-1/3}$ ensures that the peaks of substructure are point like.
Using the radius for a peak: 
\begin{equation}
r_{pk}=\sqrt{2}R_{*}\left[ \frac 1{(1+\nu \sigma _0)(\gamma ^3+(0.9/\nu
))^{3/2}}\right] ^{1/3}
\end{equation}
(ADC), I obtain a value of $\nu _c=1.3$ and then we have $n_a(\nu \ge \nu
_c)=50.7Mpc^{-3}$ %and $ m_{a}= 8 \times 10^{8} M_{\odot}$ , 
($\gamma =0.4$, $R_{*}=50$ kpc) and $m_a$ is given by: 
\begin{equation}
m_a=\frac 1{n_a(\nu \ge \nu _c)}\int_{\nu _c}^\infty m_{pk}(\nu )N_{pk}(\nu
)d\nu =10^9M_{\odot }
\end{equation}
(in accordance with the result of ADC), where $m_{pk}$ is given in Peacock $\&
$ Heavens (1990) and $N_{pk}$ is the average number density of peak (BBKS
Eq. ~4.4).
Galaxies and Clusters of galaxies are correlated systems whose autocorrelation function, $\xi(r)$,
can be expressed, 
%%%%%%%%%in the range $10h^{-1} Mpc<r<60 h^{-1} Mpc$,
in a power law form (Peebles 1980; Bahcal \& Soneira 1983; Postman et al. 1986; Davis \& Peebles 1983; Gonzalez et al. 2002).
%
%%\begin{equation}
%%\xi_{cc}=(\frac{r_{o,c}}{r})^{\gamma}
%%\end{equation}
%%with $\gamma \simeq 1.8$ and a correlation length,
%%$r_{o,c} \simeq 25h^{-1} Mpc$ (Bahcal \& Soneira 1983; Postman et al. 1986).
%%The analysis of fair samples of galaxies gives for the galaxy autocorrelation
%%function the expression:
%%\begin{equation}
%%\xi_{gg}=(\frac{r_{o,g}}{r})^{\gamma}
%%\end{equation}
%%in the range $0.1h^{-1} Mpc<r<10 h^{-1} Mpc$ ($r_{o,g} \simeq 5 h^{-1} Mpc$,
%%$\gamma=1.77 \pm 0.03$ (Peebles 1980, Davis \& Peebles 1983)). 
%
The description of dynamical friction in these systems need to
use a distribution of the stochastic forces, $W(F)$, taking account of
correlations. In this last case the coefficient of dynamical friction,
$ \mu$, may be calculated using the equation:
\begin{equation}
\mu=\int  d^{3} {\bf F} W(F) F^{2} T(F)/(2<v^{2}>)
\end{equation}
and using Antonuccio-Delogu \& Atrio-Barandela (1992) distribution:
\begin{equation}
W(F)=\frac{1}{2 \pi^2 F} \int_{0}^{\infty} dk k sin(kF)A_{f}(k)
\end{equation}
where $A_{f}$, which is a linear integral function of the correlation function $ \xi(r)$,
is given in the quoted paper (Eq. 36). 

In Fig. 10, we plot the coefficient of dynamical friction of a system having $R_{sys} = 5 h^{-1}$ Mpc for different values of $\nu_c$ \footnote{We calculate the contribution to $\mu$ from small peaks of the density field, namely those peaks small enough to be considered ``pointlike", having their central height $\nu$ larger than a critical threshold $\nu_c$.} and $R_f$. The increase of $\mu$ with $\nu_c$ is mainly due to an increase of the average mass of the peaks generating the stochastic field and a simultaneous decrease of their number density.

In Fig. 11, we show the effects of non-radial motions (angular momentum) and dynamical friction, in the case of a $\nu=3$ peak. 
We plot the evolution of the expansion parameter, $a(t)$, 
with time expressed in terms of the collapse time of a pure SIM model (namely when tidal torques, dynamical friction and cosmological constant are not taken into account).
%the time of collapse, $T_c(r,\nu )$. 
As shown by Gunn \& Gott (1972), this last quantity 
is given by:
\begin{equation}
T_{c0}(r,\nu )=\frac{\pi}{H [\overline{\delta }(r,\nu )]^{3/2}}
\label{eq:beef}
\end{equation} 
where $H$ is Hubble constant.

In Fig. 11, the dashed line represents the effect of dynamical friction while the dotted line the cumulative effect of dynamical friction 
and angular momentum. 
%As shown 
The presence of non-radial motions and dynamical friction changes the dynamics of the shells 
producing an increase in the time of collapse.
%of a spherical shell. 
The collapse delay is larger for low value of $\nu $ and reduces for high values of $\nu$ for which 
the collapse time becomes equal to that of the pure spherical collapse.
%negligible for $\nu \geq 3$. 
This result is in
agreement with the angular momentum-density anticorrelation effect: density
peaks having low value of $\nu $ acquire a larger angular momentum than high 
$\nu $ peaks and consequently the collapse is more delayed with respect to
high $\nu $ peaks. Dynamical friction has a similar effect to that of angular momentum for shell evolution.

\section{Baryonic dissipative collapse}

%
%%%The shape of the central density profile is influenced by baryonic collapse: baryons drag dark matter in the so called adiabatic contraction (AC) 
%%%steepening the dark matter density slope.
%%%For a spherically symmetric protostructure that consists of a fraction $F << 1$ (this fraction is fixed using WMAP collaboration data (Spergel et %%%al. 2003)) of dissipational baryons and a fraction $1-F$ of dissipationless dark matter particles constituting the halo, Blumenthal et al. (1986) %%%described an approximate analytical  model to calculate the effects of AC. More recently Gnedin et al. (2004) 
%%%proposed a simple modification of the Blumenthal model, which describes numerical results more accurately. 
%%%For systems in which angular momentum is exchanged between baryons and dark matter to dark matter (e.g., through dynamical friction),
%%%Klypin et al. (2002) introduced a modification to Blumenthal's model. 
%%%In Appendix D, we describe the way AC is calculated in the present paper.
%
%RYDEN

The response of dark matter to baryonic infall has traditionally been
calculated using the model of adiabatic contraction.
Eggen et al. (1962) were the first to use adiabatic invariants of
particle orbits to estimate the effect of a changing potential in a
contracting proto-galaxy.  Zeldovich et al. (1980)  presented the first
analytical expressions for adiabatic contraction (AC) for purely
radial and circular orbits, as well as the numerical tests of such
model. The present standard form of the AC model was introduced and tested
numerically by Blumenthal et al. (1986) (see also RG87).  
Blumenthal et al. (1986) described an approximate analytical  model 
which has been checked by numerical simulations (Barnes 1987; Oh 1990) 
for calculating the radial redistribution of the dissipationless halo matter 
of a protostructure when its dissipational matter falls in toward the center. 
One starts by noting that for a particle moving in a periodic orbit 
$\oint  p dq$ (where $p$ is the canonical momentum conjugate to the coordinate $q$) 
is an adiabatic invariant. Provided that $M (r) $, the mass inside the orbital radius $r$, 
changes slowly compared with an orbital time (Steigman et al. 1978; Zeldovich et al. 1980; 
RG87) and that particles move in circular orbits about a spherically symmetric mass 
distribution, the mass invariant is $r M(r) $. For purely radial orbits $r_{m} M(r_{m})$
is also an adiabatic invariant, provided that $M(r )$ varies in a self-similar fashion.
Consider a spherically symmetric protostructure that consists of a fraction $F << 1$ (this fraction is fixed using WMAP collaboration data (Spergel et al. 2003)) of dissipational baryons and a fraction $1-F$ of dissipationless dark matter particles constituting the halo. One then assumes that the dissipational baryons and the halo particles are well mixed initially (i.e., the ratio of their densities is $F$ through the protostructure). 
Since there is more phase space for nearly circular orbits than for nearly radial orbits, one can assume that the dark matter particles move in circular orbits about the proto-structure center with almost random oriented angular momentum vectors. As the baryons dissipatively cool and fall 
into a final mass distribution $M_b( r)$, which is constrained by the initial angular momentum distribution, 
a dark matter particle initially at radius $r_i$ will move in to  radius $r< r_i$. 
The adiabatic invariant for such a particle orbits implies that:
\begin{equation}
r \left [ M_b( r) +M_{dm} ( r) \right] = r_i M_i (r_i)
\label{eq:ad1}
\end{equation}
%which can be solved iteratively for the final distribution of dissipationless halo particles 
%$M_{dm} ( r)$ given the initial total mass distribution $M_i (r_i)$ and the final baryon mass distribution $M_b( r)$.
where  $M_i (r_i)$ is the initial total mass distribution, $M_b( r)$ (as previously reported) is the final mass distribution of dissipational baryons and $M_{dm}$ is the final distribution of dissipationless halo particles. 
Assuming that the orbits of the halo particles do not cross, then 
\begin{equation}
M_ {dm} ( r)=(1-F) M_i (r_i)
\label{eq:ad2}
\end{equation}
Eqs. (\ref{eq:ad1}), (\ref{eq:ad2}) can be iteratively solved 
%used 
to calculate the final radial distribution of the halo particles once $M_i (r_i)$ and $M_b (r)$ are given.
If $F<<1$, then for a halo particle not too near to the center of the protostructure, the mass interior to its orbit will undergo a small fractional change in one orbital period, even if dissipation occurs rapidly provided the particle is fairly far from the protostructure center. Therefore, the adiabatic invariant given in Eq. (\ref{eq:ad1}) is expected to be a good approximation for all but the innermost particles.

I recall that the AC model has been studied further by R88, Ryden (1991), and
Flores et al. (1993). It is routinely used in mass modeling of
galaxies (Flores et al. 1993, Dalcanton et al. 1997, Mo et al. 1998,
Courteau \& Rix 1999, van den Bosch 2001, van den Bosch \& Swaters 2001, Klypin et al. 2002, Seljak 2002) and
clusters of galaxies (Treu \& Koopmans 2002).  
%The effect of the contraction of the dark matter distribution is important for
%studying star formation feedback on the centers of dark matter halos
%\citep{gnedin_zhao02} and for comparing the abundance of dark matter
%halos and galaxies as a function of circular velocity
%\citep[e.g.,][]{gonzalez_etal00,kochanek_white01}. It is particularly
%important in calculations of the dark matter annihilation signal from
%the Galactic center \citep[e.g.,][]{gnedin_primack04,prada_etal04}.

In order to use the adiabatic invariant, one has to know the initial mass distribution $M_i (r_i)$ and the final distribution of dissipational baryons $M_b (r)$. A usual assumption is that initially baryons had the same density profile as the dark matter (Mo  et al. 1998; Cardone \& Sereno 2005; Treu \& Koopmans 2002; Keeton 2001).
%and that initial dark matter density profile is described by a NFW profile.

The final baryons distribution is assumed to be a disk (for spiral galaxies) (Blumenthal et al. 1986; Flores et al. 1993; Mo et al. 1998; Klypin et al. 2002; Cardone \& Sereno 2005). In our calculations, we shall assume Klypin et al (2002) model for baryons distribution (see their subsection 2.1), when dealing with mass scales typical of spiral galaxies. In the case of elliptical galaxies and clusters a typical assumption (that we shall use in the paper) is that baryons collapse to a Hernquist configuration (Rix et al. 1997; Keeton 2001; Treu \& Koopmans 2002), having density profile:  
%On cluster and elliptical galaxies scales I will adopt for the baryons component a Hernquist (1990) model. This has a density profile
\begin{equation} \label{eq:hern}
  \rho(r) = {\rho_s \over {(r/r_s)(1+r/r_s)^3} }\ .
\end{equation}
This profile is described by a scale radius $r_s$, but the
projected profiles of elliptical galaxies are usually described by
an effective (or half-mass) radius $R_e$; they are related by $r_s
= 0.551 R_e$. The total mass of a Hernquist model is $M = 2\pi
\rho_s r_s^3$. 
%This adiabatic contraction formalism has often been applied to the
%problem of a disk galaxy in an NFW halo. 
%Rix et al.\ (1997) have computed adiabatic contraction for elliptical galaxies numerically,
As previously reported, in general, adiabatic contraction is computed numerically, 
but with a Hernquist model (Eq.~\ref{eq:hern}) 
the problem can be solved analytically (Keeton 2001). Each initial radius
$r_i$ maps to a unique final radius $r$ given by the solution of
the equation
\begin{equation} 
\label{eq:ri2r}
  f r^3 + (r+s_g)^2 \left[ (1-f) r - r_i \right] m_i(r_i) = 0,
\end{equation}
which is a cubic polynomial in $r$. Here 
%Note that I have take the
the galaxy scale radius $r_s$ from Eq. (\ref{eq:hern}) has been relabeled as
$s_g$. Also, $m_i(r_i) = M_i(r_i)/M_{v}$ is the initial mass
profile normalized by the virial mass.
%(the mass inside the virial radius $r_{200}$). 
In the limit $r >> s_g$, Eq. (\ref{eq:ri2r}) has the
simple asymptotic solution
\begin{equation}
r= \frac{r_i m_i(r_i)}{f+(1-f)m_i(r_i)}
%r = {r_i\,m_i(r_i) \over \fcool + (1-\fcool) m_i(r_i)}\ .
\end{equation}
The full general solution can also be found analytically,
although it cannot be written quite so compactly (see Keeton 2001). 
%Following Abramowitz \& Stegun (1981), solve a cubic equation of the form
%\begin{equation} \label{eq:cubic}
%  z^3 + a_2 z^2 + a_1 z + a_0 = 0
%\end{equation}
%by defining
%\begin{eqnarray}
%  p &=& {a_1 a_2 - 3 a_0 \over 6} - {a_2^3 \over 27}\,, \\
%  q &=& {a_1 \over 3} - {a_2^2 \over 9}\,, \\
%  s_1 &=& \left( p + \sqrt{q^3+p^2} \right)^{1/3} , \\
%  s_2 &=& \left( p - \sqrt{q^3+p^2} \right)^{1/3} .
%\end{eqnarray}
%There is always a real solution of Eq. (\ref{eq:cubic}) at
%\begin{equation}
%  z_1 = (s_1 + s_2) - {a_2 \over 3}\ .
%\end{equation}
%There are two other roots that may be real or complex, but because
%the $r_i \to r$ mapping under adiabatic contraction should be
%one-to-one, only the single real root is relevant. 
Once the cubic
equation has been solved to map $r_i$ to $r$, Eqs. (\ref{eq:ad1}, \ref{eq:ad2}) can
be used to write the total mass profile as
\begin{equation} \label{eq:Mtot}
  M_{tot}(r) \equiv M_b(r) + M_{dm}(r) = {r_i \over r}\,M_i(r_i)\,.
\end{equation}
This solution of adiabatic contraction by a Hernquist galaxy can be
used for any form of the initial halo, by simply inserting the
desired initial profile $M_i(r_i)$ into Eq. (\ref{eq:ri2r}) and Eq. (\ref{eq:Mtot}).

Recently Gnedin et al. (2004) 
%In this paper we consider the effect of dissipation on the dark matter
%distribution in high-resolution cosmological simulations. We also
presented the first test of the AC model in the self-consistent
simulations of hierarchical structure formation and propose a simple
modification which describes numerical results more accurately.
They proposed a modified adiabatic
contraction model based on conservation of the product of the current
radius and the mass enclosed within the orbit-averaged radius:
\begin{equation}
  M(\bar{r})r= {\rm const}.
  \label{eq:modified}
\end{equation}
where the orbit-averaged radius is
\begin{equation}
  \bar{r} = {2 \over T_r} \int_{r_p}^{r_a} r \, {dr \over v_r},
\end{equation}
where $T_r$ is the radial period, $r_a$ is the apocenter radius and $r_p$ the pericenter radius.  

The previous, classical AC model, assumes no angular momentum exchange between different components (e.g., 
baryons and dark matter). 
Models with exchange of angular momentum between the baryons and dark
matter are more complicated. The exchange probably happens at late
stages of the baryonic infall when the baryon density becomes large
and a non-axisymmetric component may develop due to the excitation of
spiral waves and/or bar-like modes.
%Giant molecular clouds may play some role in this process.
Dynamical friction can then result in a transfer of angular momentum
from the baryons to the dark matter.  
% Tolto
%%%Because the dark matter gains
%%%angular momentum, it moves further from the galactic center. Thus, the
%%%density of the dark matter in the central region decreases.  
%%%In the early stages of galaxy formation, when most of the baryons were still
%%%in gaseous form, we might expect that those non-axisymmetric features
%%%were more prevalent and more dynamically important then at the present
%%%time when most of the baryons are locked in stars.
%%%It is difficult to estimate the exact amount of angular momentum that
%%%would be lost by the baryons in such a situation, but we know that it
%%%cannot be very large. 
%%%For simplicity we assume that the formation of
%%%the disk happens in two stages. During the first stage, when the
%%%baryons experience most of the collapse, they preserve their angular
%%%momentum. During the second stage the disk shrinks further, losing
%%%some of its angular momentum to the dark matter.
%
The classical approach can then be used to compute the
state of the system at the end of the first stage of adiabatic
compression (in which angular momentum is conserved).  
For the later evolution, one can use a simple model of Klypin et al. (2002).
If one considers a spherical shell of dark matter with radius $r$, thickness
$dr$, density $\rho_{\rm dm}$, and specific angular momentum
\begin{equation}
j=r V_c = \sqrt{G\left[M_{\rm b}(r)+M_{\rm dm}(r) \right] r}.
\label{eq:jr}
\end{equation}
one can get an implicit equation for the final radius $r_f$:
\begin{eqnarray}
\label{eq:exchangeangularA}
j_f &=& j \left[1+\frac{A\Delta M}
                  {4\pi\rho_{\rm dm}r^3}\right],\phantom{mmm} \\
A &=& 1+\frac{r}{V_c}\frac{dV_c}{dr},\\
\Delta M &=&M_{\rm b,f}-M_{\rm b}.
\label{eq:exchangeangularB}
\end{eqnarray}
Klypin et al. (2002), where $M_{\rm b,f}$ is the final baryons mass and
$M=M_{\rm dm}+M_{\rm b}$ is the total mass inside a radius $r$.
Eq.~(\ref{eq:exchangeangularA}) is solved numerically. The solution
also gives the mass inside a final radius $r_f$.
Eq.~(\ref{eq:exchangeangularA}) has the same structure as
Eq.~(\ref{eq:ad1}). The only difference is the term on the
right-hand-side, which is the correction due to angular momentum
deposition.
% Tolto
%%%Using some simple assumptions (Klypin et al. 2002), one can show that:
%%%\begin{equation}
%%%   r_f \approx r_i\left(1+\frac{\Delta\rho}{3\rho_{\rm dm}}\right)
%%%\label{eq:effectang}
%%%\end{equation}
%%%So, during the initial stages of the collapse, when the
%%%density of the baryons was about ten times smaller than the density of
%%%the dark matter, the exchange of angular momentum had little impact on
%%%the dark matter: $r_f\approx r_i$. The effect peaks at around
%%%$\Delta\rho =3\rho_{\rm dm}$. At even larger values of the density
%%%ratio, the approximation fails because a small amount of dark matter
%%%cannot exert significant dynamical friction on large mass of
%%%baryons. Nevertheless, at the peak of its importance, the effect is
%%%potentially quite large, with $r_f\approx 2r_i$, resulting in a
%%%decrease of the dark matter density by a factor of ten.

\newpage

%FIGURES
%---------------------------------------------------------------------------------------------

\begin{figure}[tbp]
\centerline{\hbox{(a)
\psfig{figure=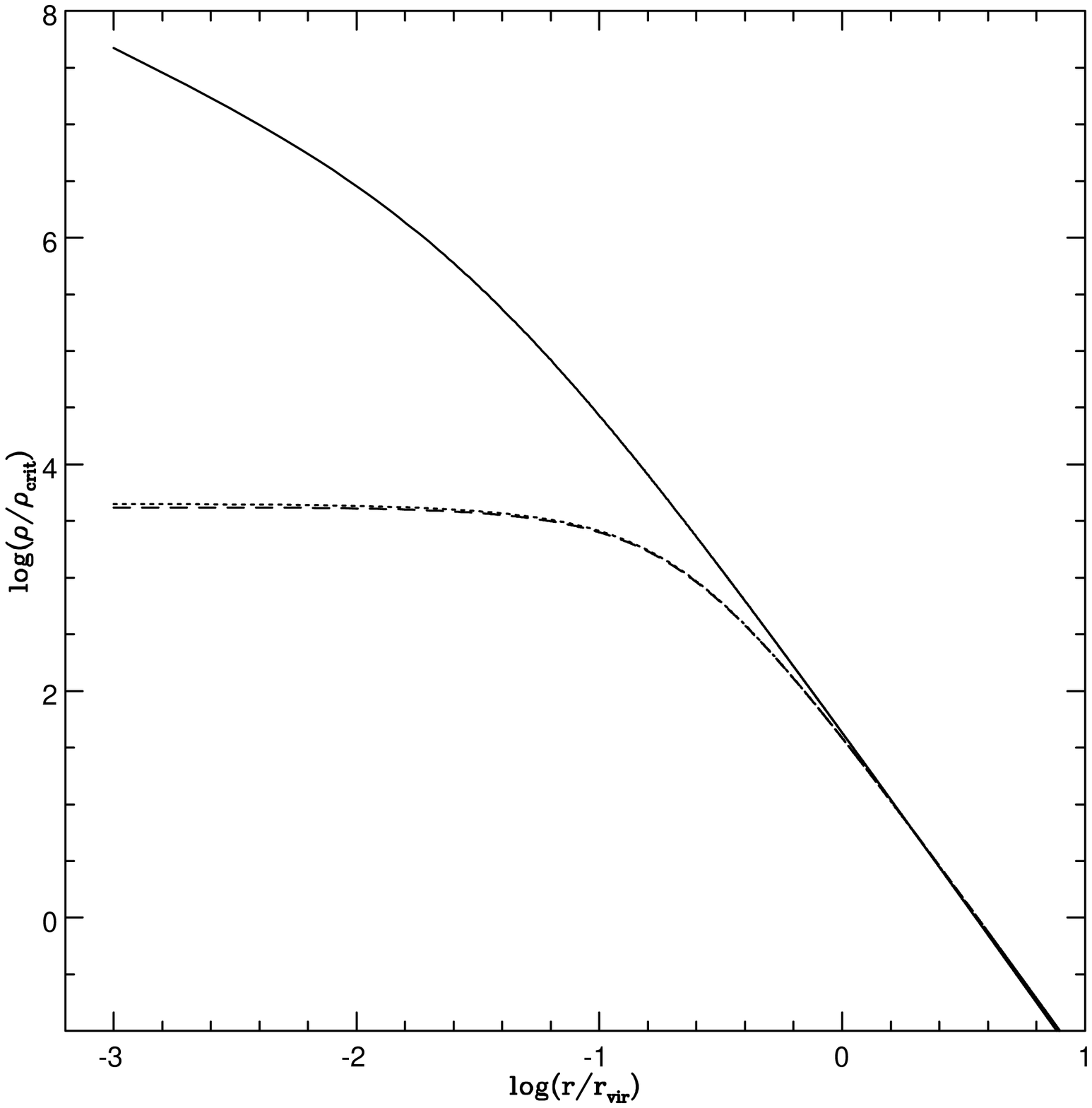,width=9.4cm} (b)
\hspace{0.1cm}
\psfig{figure=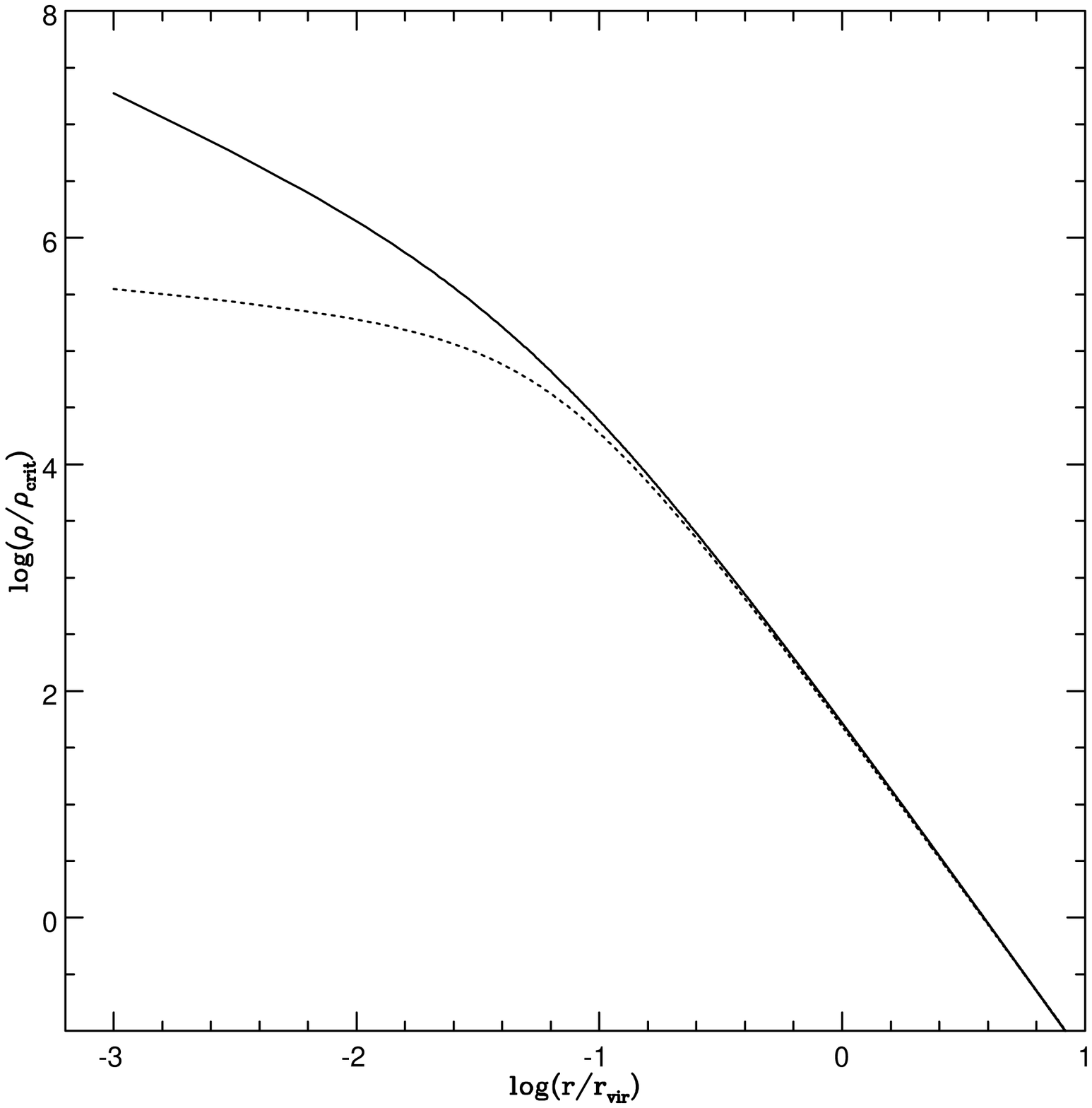,width=9.4cm}
}}
\centerline{\hbox{(c)
\psfig{figure=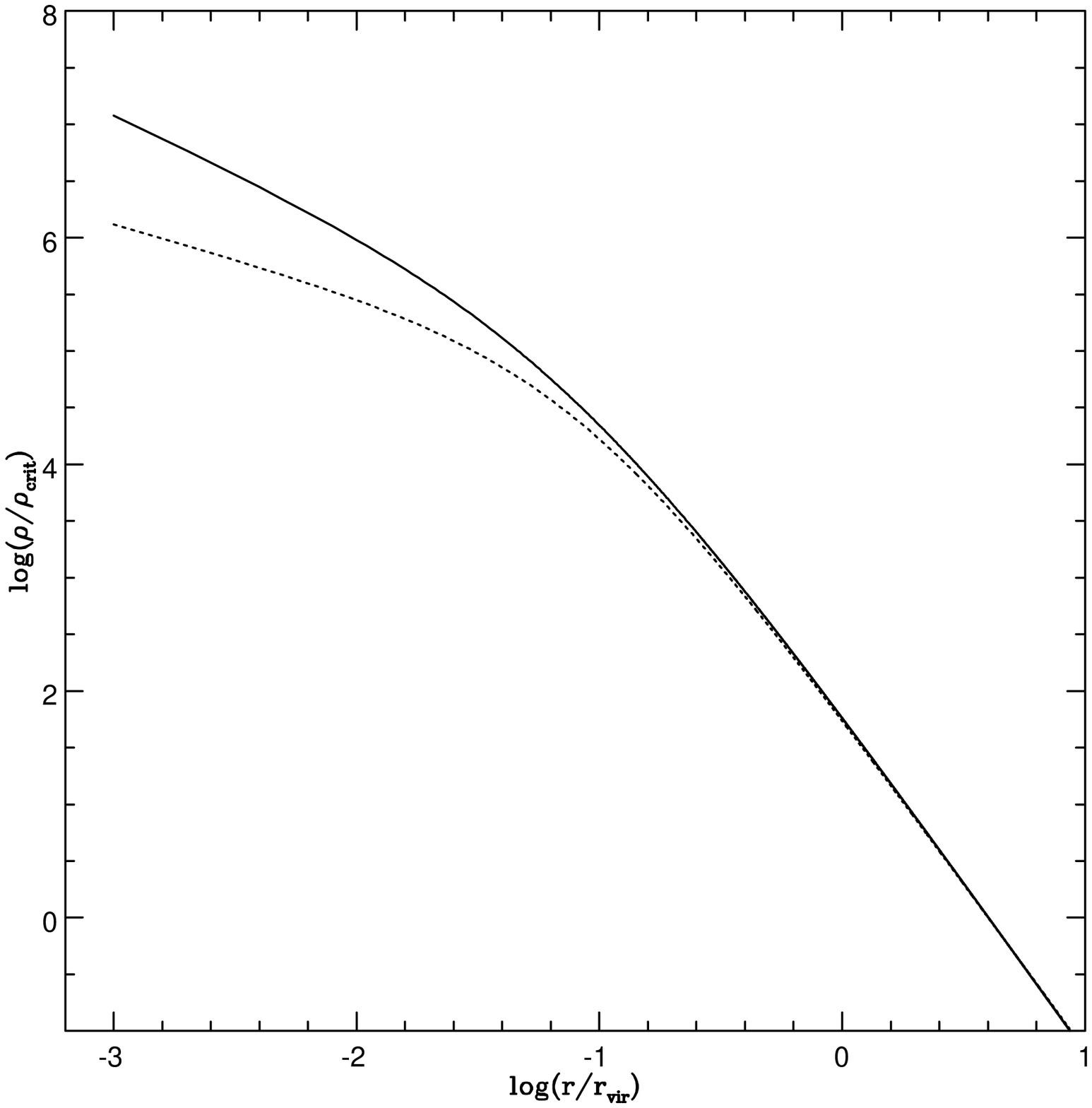,width=9.4cm} d)
\hspace{0.1cm}
\psfig{figure=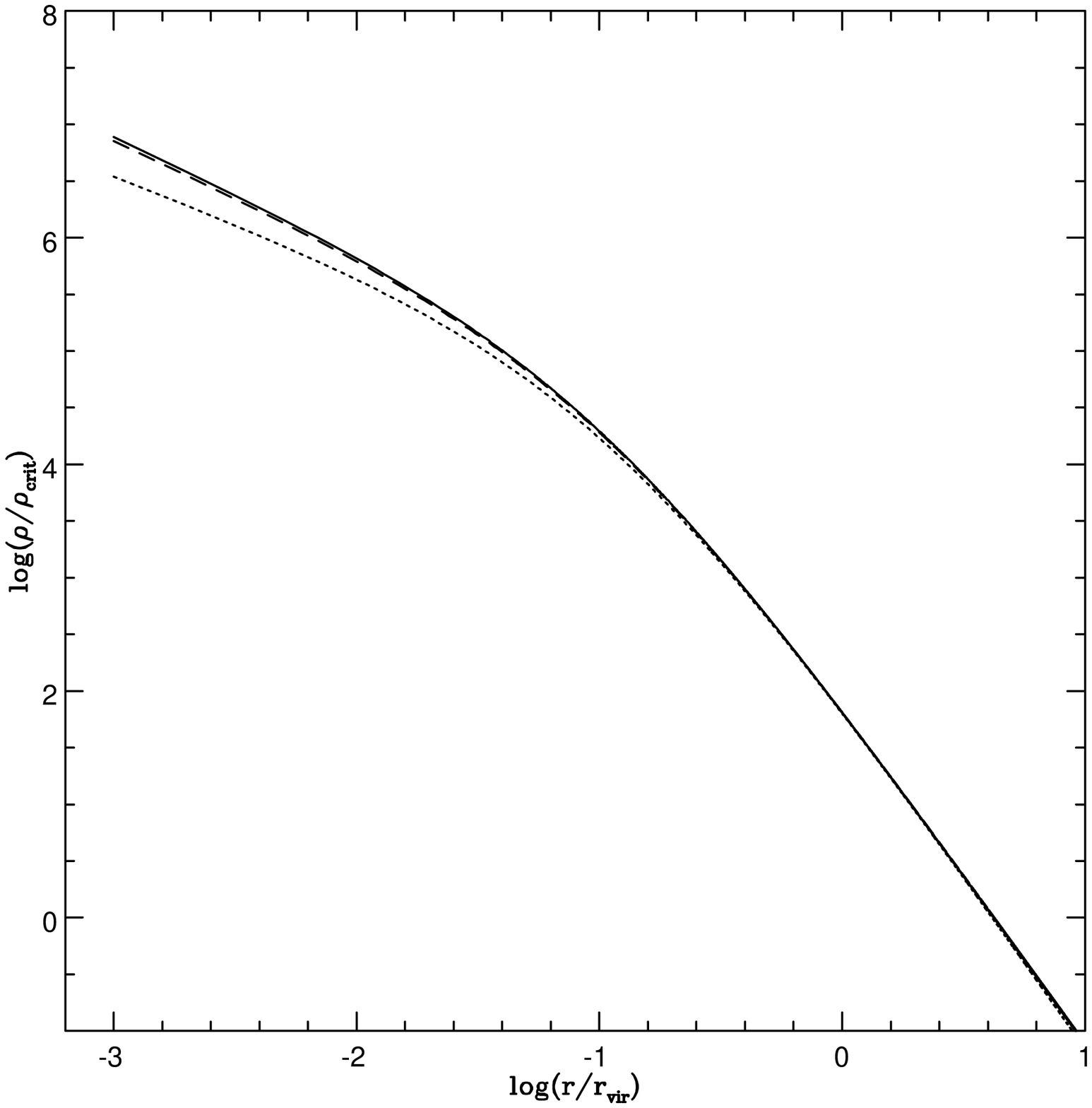,width=9.4cm}
}}
\caption{
Dark matter haloes generated with the model of Section 2. In panels (a)-(d) the solid line represents the NFW model while the dotted line the 
density profile obtained with the model of the present paper for masses $10^8 M_{\odot}$ (panel a), $10^{10} M_{\odot}$ (panel b), $10^{11} M_{\odot}$ (panel c) and $10^{12} M_{\odot}$ (panel d).
The NFW profile for the given mass was calculated with Eqs. (\ref{eq:cvirr}),(\ref{eq:navar}), (\ref{eq:navarr}) expressing the scaling radius $r_s$ in terms of the virial radius. The dashed line in panel (a) represents the Burkert fit to $10^8 M_{\odot}$ halo. 
The dashed line in panel (d) represents the density profile obtained reducing the magnitude of $h$, $j$ and $\mu$ as described in the text.
}
\end{figure}

% TOLTA
%%%%\begin{figure}[ht]
%%%%\psfig{file=dproff.ps,width=12.0cm}
%%%%\caption[]{Dark matter haloes generated with the model of Section 2.  
%%%%Haloes of different masses are shown:
%%%%$10^{11} M_{\odot}$ (short-dashed line), $5 \times 10^{10} M_{\odot}$ (long-dashed line),
%%%%$10^{10} M_{\odot}$ (dot-short-dashed line), $10^9 M_{\odot}$ (dot-long-dashed line). 
%%%%The short dashed-long dashed line (almost indistinguishable from the density profile of the $10^9 M_{\odot}$ halo) 
%%%%represents the Burkert fit to $10^9 M_{\odot}$ halo. 
%%%%The solid line represents NFW $c=10$ halo, and the dotted line the 
%%%%density profile obtained reducing the magnitude of $h$, $j$ and $\mu$ as described in the text.
%%%%}
%%%%\end{figure}

\begin{figure}[ht]
\psfig{file=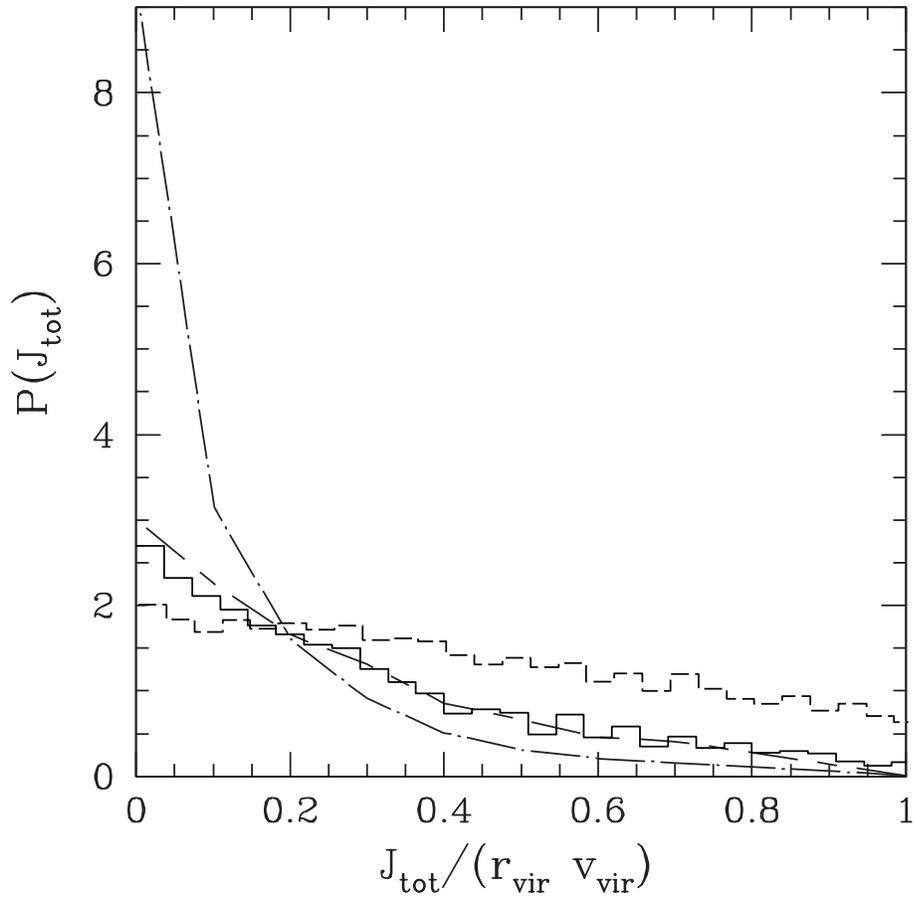,width=12.0cm}
\caption[]{Distribution of the total specific angular momentum, $J_{Tot}$. The dotted-dashed and dashed line represents the quoted 
distribution for the halo n. 170 and n. 081, respectively, of van den Bosch et al. (2002). The dashed histogram is the distribution obtained 
from our model for the $10^{12} M_{\odot}$ halo and the solid one the angular momentum distribution for the density profile reproducing the NFW halo, as descried in Section 4.
}
\end{figure}

\begin{figure}[ht]
\psfig{file=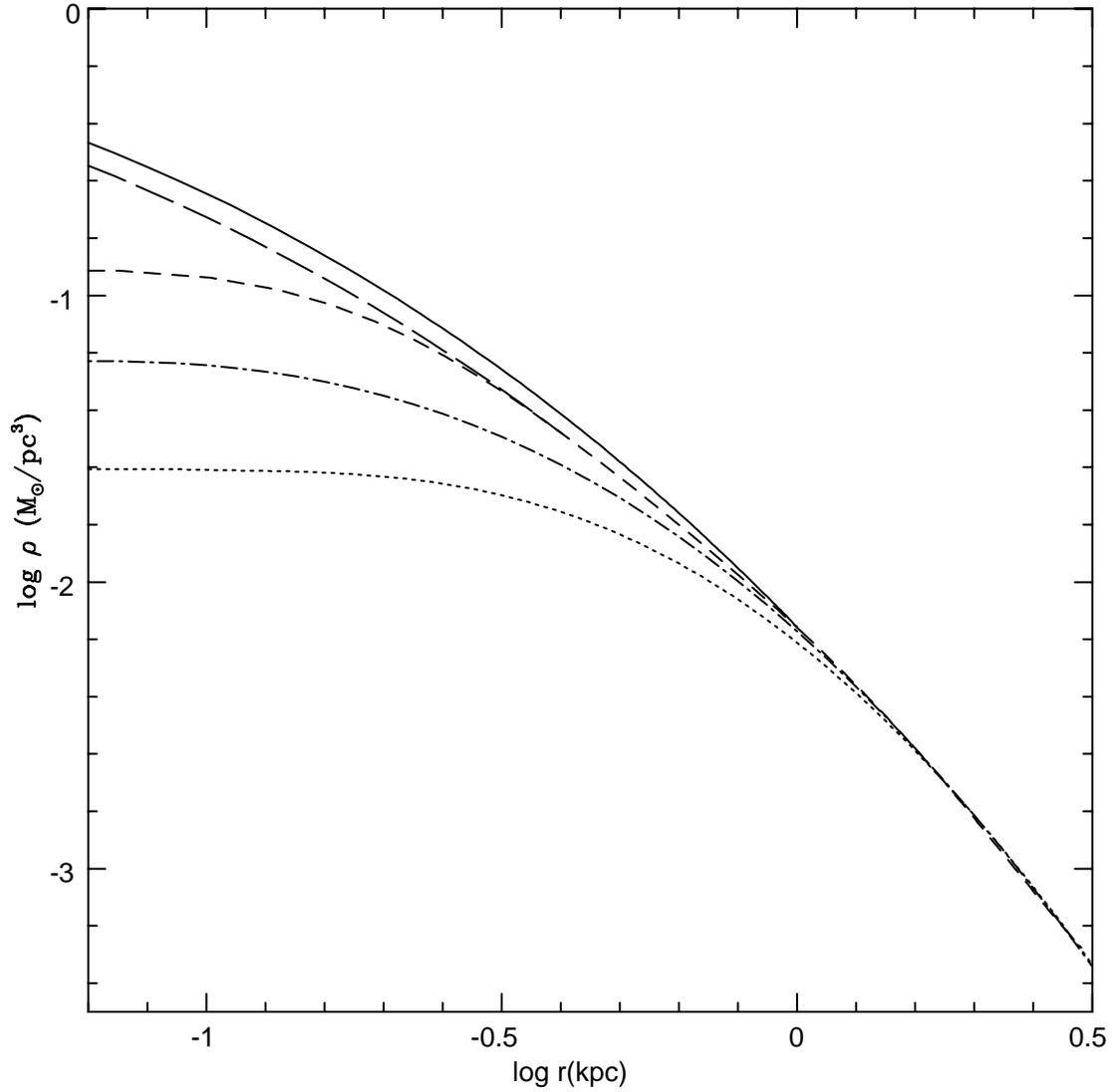,width=16.0cm}
%\psfig{file=deproff.eps,width=16.0cm}
%\picplace {2.0cm}
\caption[]{Density profile evolution of a $10^9 M_{\odot}$ halo.
The solid line represents the profile at $z=10$. The profile at $z=3$, $z=2$, $z=1$, $z=0$ is represented by the long-dashed line, short-dashed line, 
dot-dashed line, dotted line,
%long-dashed line, dotted line, and short-dashed line, 
respectively.
}
\end{figure}

\begin{figure}[ht]
\psfig{file=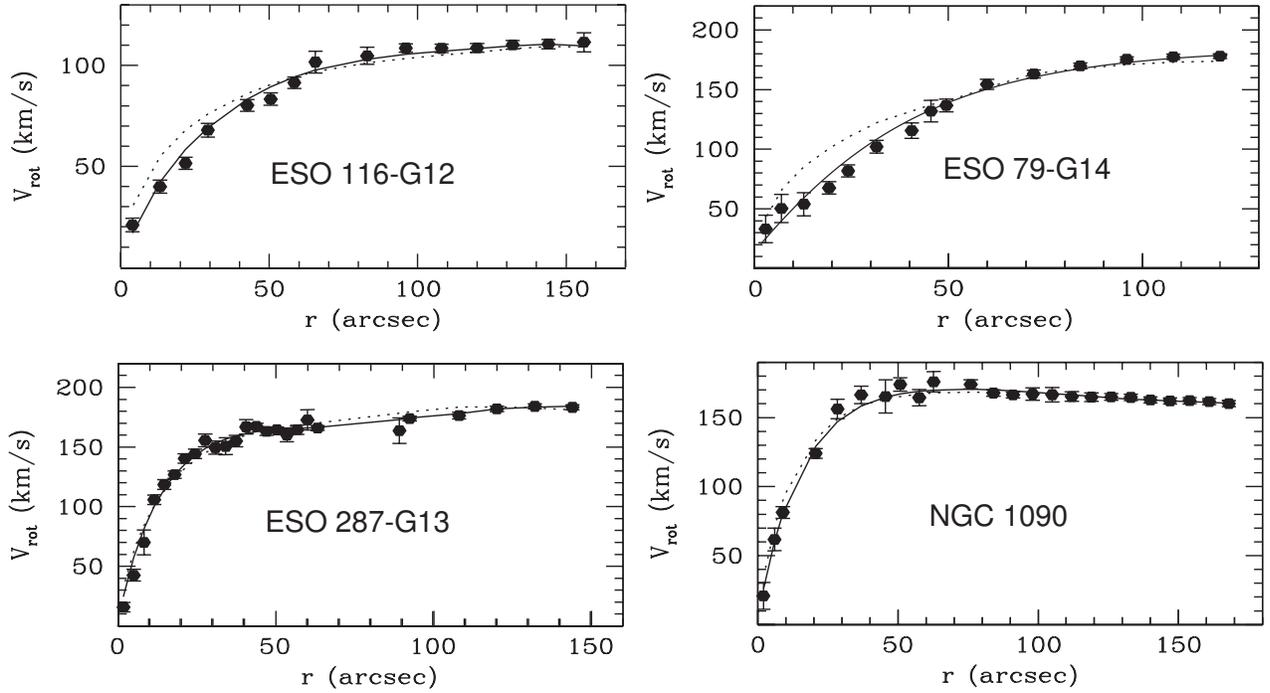,width=20.0cm}
%\picplace {2.0cm}
\caption[]{Comparison of the rotation curves obtained with the model in Section 2 (solid lines) with the rotation curves of four LSB galaxies  studied by Gentile et al. (2004). The dotted line represents the fit with NFW model (see Section 4 for details). 
}
\end{figure}

%\begin{figure}[ht]
%\psfig{file=ezant.ps,width=12.0cm}
%\caption[]{Density profile evolution of a $10^{14} M_{\odot}$ halo.
%The solid line represents the NFW initial profile at $z=3$. 
%The profile at $z=2$, $z=1.5$, $z=1$ and $z=0$ is represented by the dot-short-dashed line, dotted line, short-dashed line, 
%and long-dashed line respectively.
%}
%\end{figure}

\begin{figure}[ht]
\centerline{\hbox{(a)
\psfig{figure=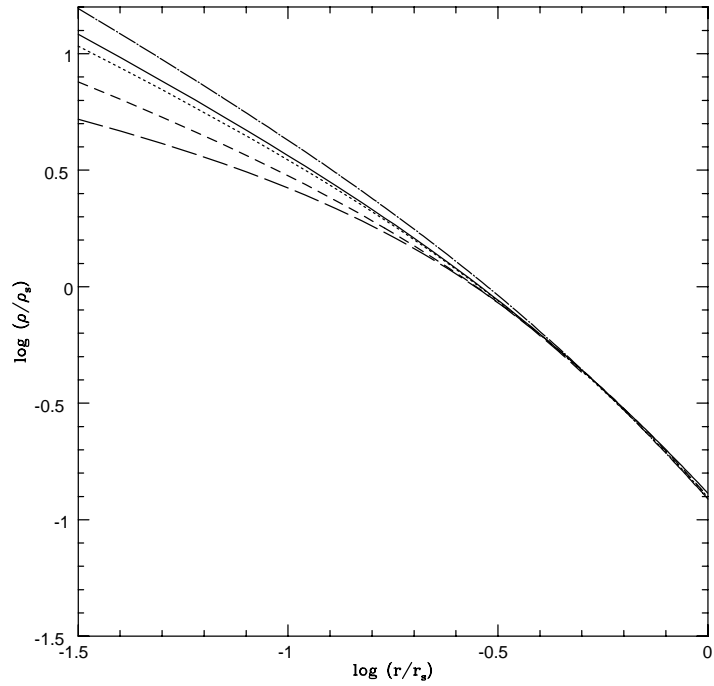,width=10cm} 
}}
%\centerline{\hbox{(b)
%\psfig{figure=ezantt.ps,width=10cm}
%}}
\caption[]{The density profile evolution of a $10^{14} M_{\odot}$ halo.
The (uppermost) dot-dashed line represents the total density profile of a $10^{14} M_{\odot}$ halo at $z=0$.
The profile at $z=3$, $z=1.5$, $z=1$ and $z=0$ is represented by the solid line, dotted-line, short-dashed-line, long-dashed-line,
%dot-short-dashed line, dotted line, short-dashed line, 
%and long-dashed line 
respectively.
%Panel (b) plots the evolution of the total density profile of a $10^{14} M_{\odot}$ halo.
%The solid line represents the NFW initial profile at $z=3$. The profile at $z=0$ is represented by the dotted line.
}
\end{figure}

\begin{figure}[ht]
\psfig{file=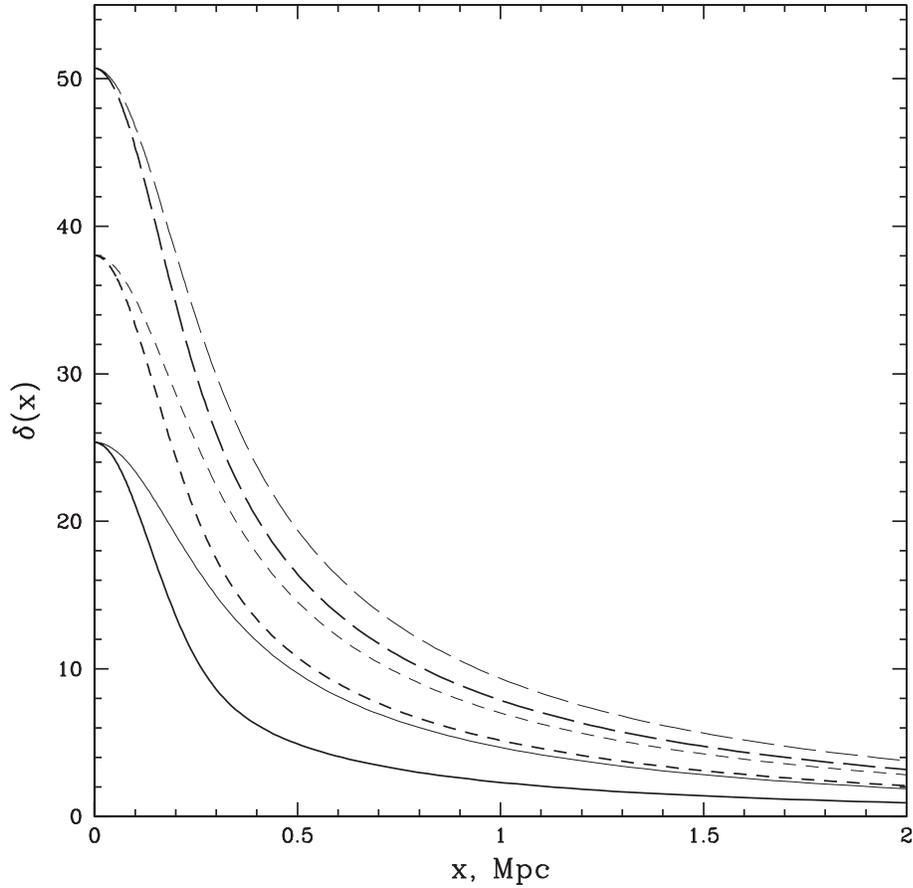,width=12.0cm}
\caption{Initial density profiles of dark matter halos linearly evolved to the present
day (see text).
%Note that the plotted quantity is the initial excess density distribution around a peak linearly evolved to the present
%day, namely $\delta(x)=\delta_i D(z_i)$, where the $D(z)$ describes the growth of density perturbations (Peebles 1980).
Heavy lines are density profiles around maxima only (BBKS). Light lines are proportional to 
the two-point correlation function, and represent density run averaged around 
local density maxima or minima in the initial Gaussian random field. 
Solid, short-dashed, and long-dashed lines represent 2, 3, and 4 $\sigma$ peaks.}
\end{figure}

\begin{figure}[ht]
\psfig{file=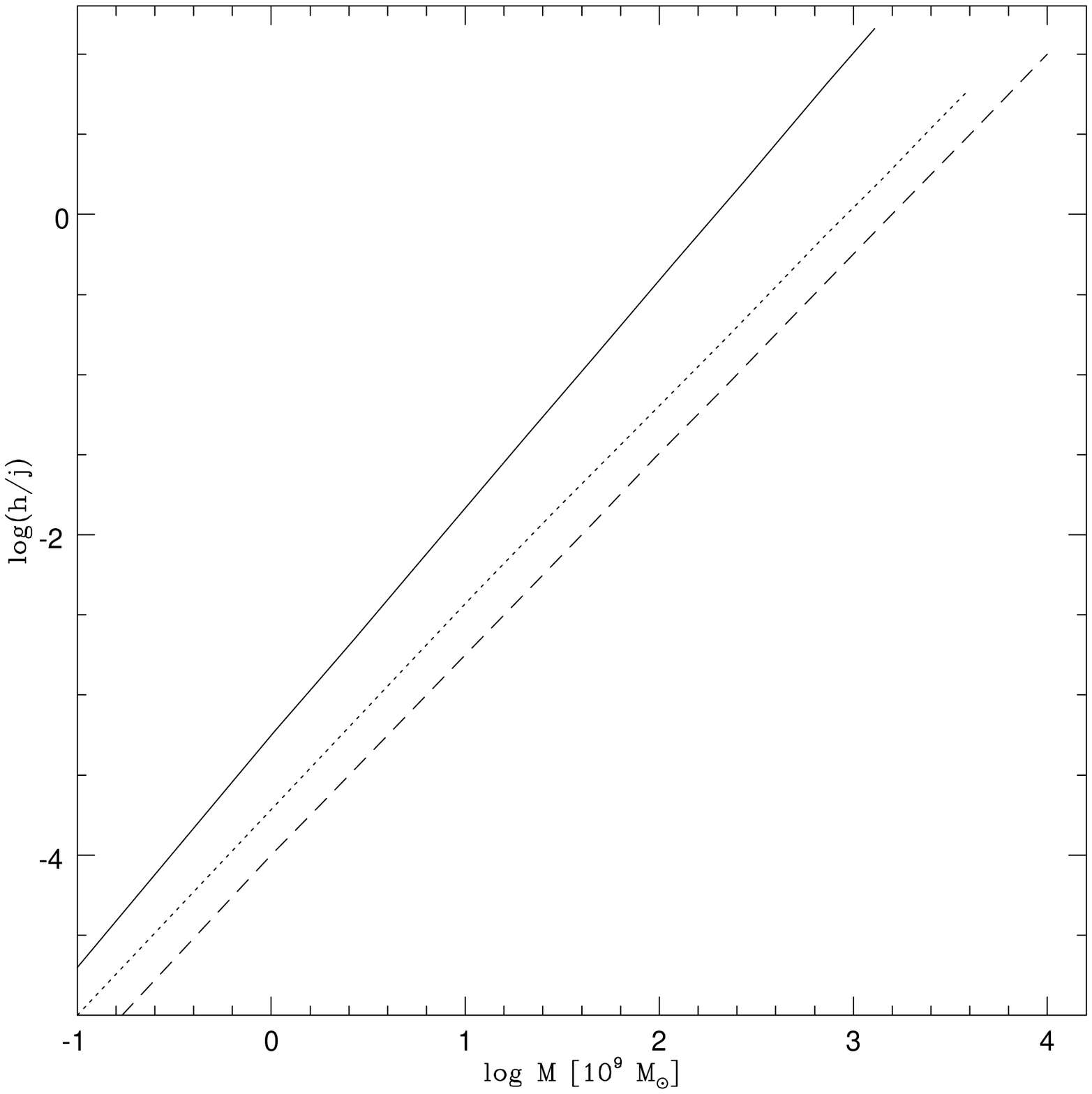,width=12.0cm}
\caption[]{The specific angular momentum for three values of the parameter $\nu$ ($\nu=2$ solid line, $\nu=3$
dotted line, $\nu=4$ dashed line) and for $R_{f}=0.12 h^{-1}Mpc$. The radius $r$ is connected to the mass $M$
as described in the text. 
}
\end{figure}

%\begin{figure}[ht]
%\psfig{file=figtid1.ps,width=20.0cm}
%\caption[]{The specific angular momentum, in units of $M_{\odot}$, Mpc and 
%the Hubble time, $t_{o}$, for
%three values of the parameter $\nu$ ($\nu=2$ dotted line, $\nu=3$
%solid line, $\nu=4$ dashed line)  and for $R_{f}=3h^{-1}Mpc$.}
%\end{figure}

\begin{figure}[ht]
\psfig{file=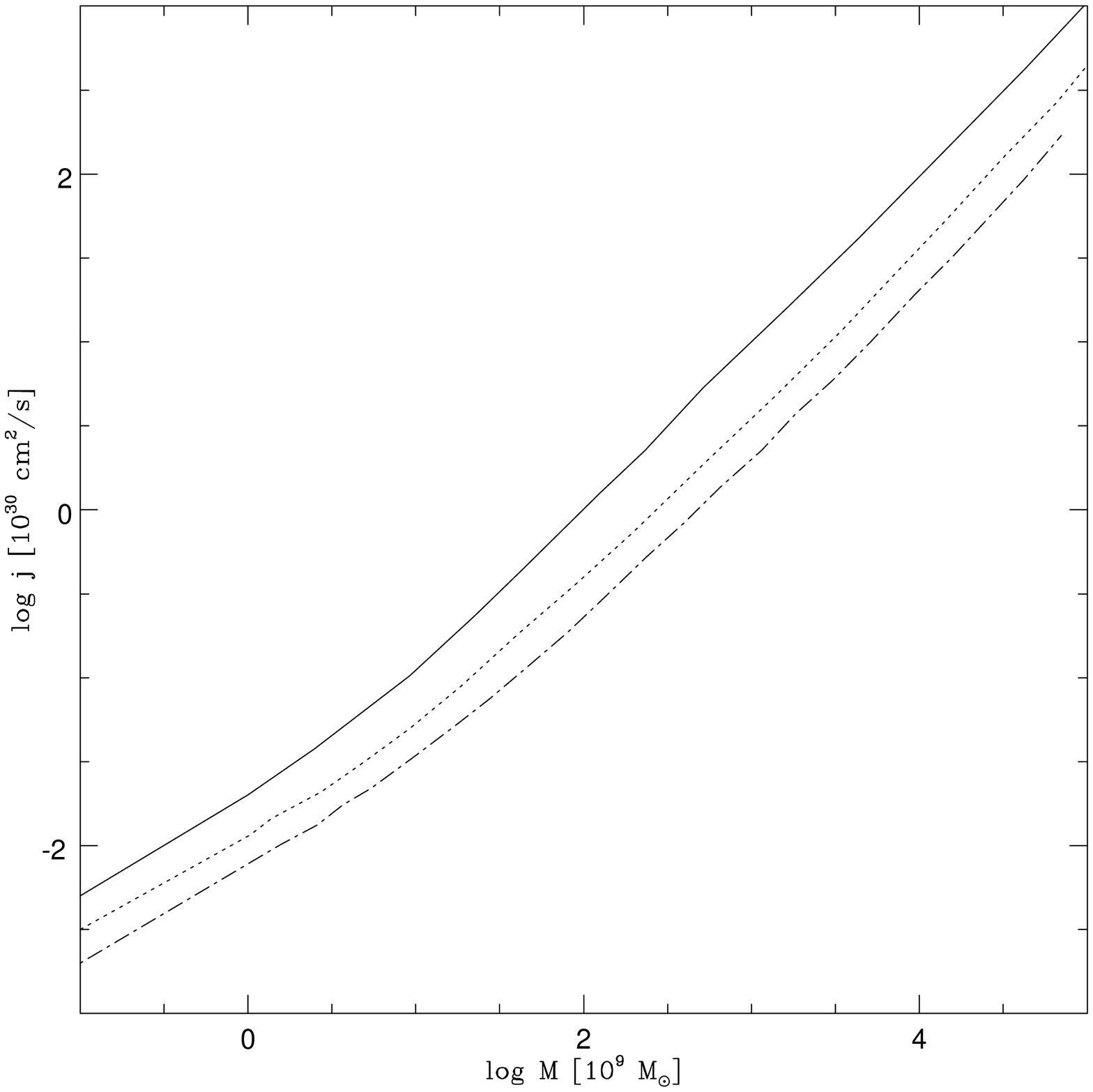,width=12.0cm}
\caption[]{The random specific angular momentum, $j$
for three values of the parameter $\nu$ ($\nu=2$ solid line, $\nu=3$
dotted line, $\nu=4$ dot-dashed line)
%dashed line) 
and for $R_{f}=0.12 h^{-1}Mpc$. The radius $r$ is connected to the mass $M$
as described in the text. 
}
\end{figure}

\begin{figure}[ht]
\psfig{file=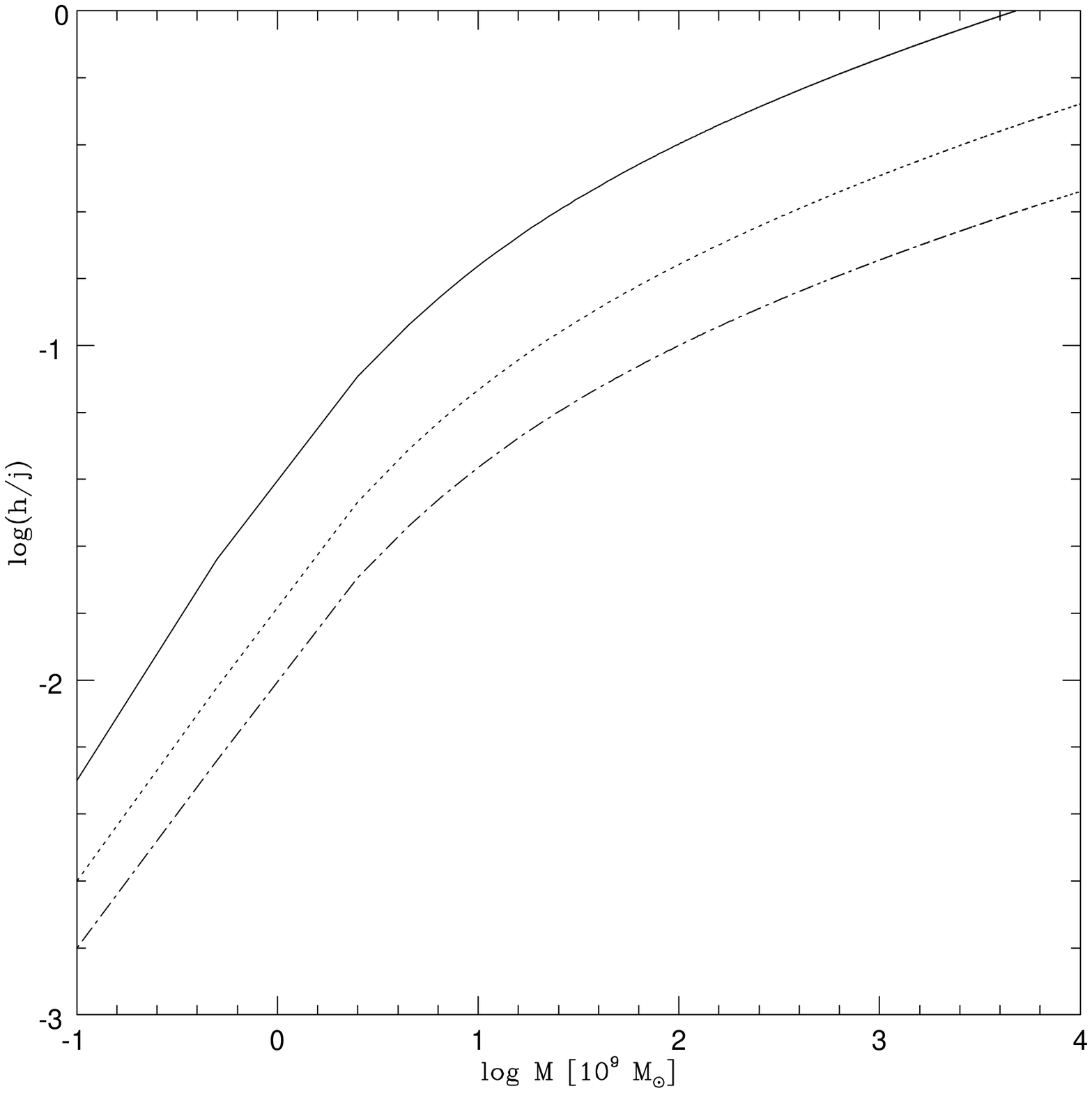,width=12.0cm}
\caption[]{The ratio of the ordered, $h$, to random secific angular momentum, $j$
for three values of the parameter $\nu$ ($\nu=2$ solid line, $\nu=3$
dotted line, $\nu=4$ dot-dashed line)
%dashed line) 
and for $R_{f}=0.12 h^{-1}Mpc$. The radius $r$ is connected to the mass $M$
as described in the text. 
}
\end{figure}

\begin{figure}[ht]
\psfig{file=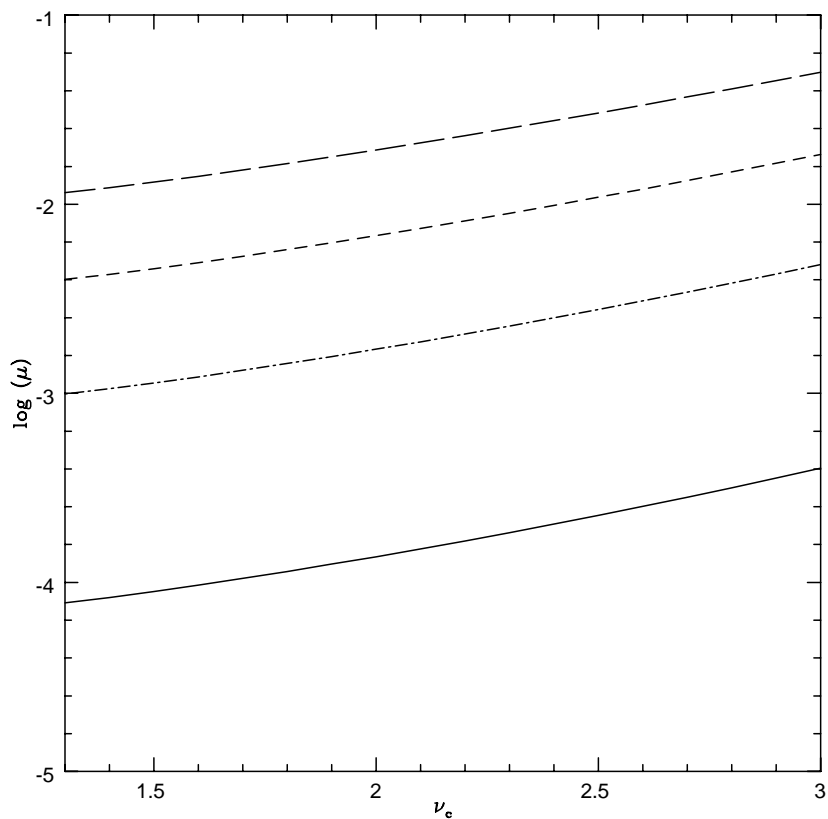,width=12cm,height=12cm}
\caption[]{Variation of the dynamical friction coefficient $\mu$ with $\nu_c$ for a system
having $R_{sys} = 5 h^{-1}$ Mpc and for $R_f=0.12$ Mpc (solid line), 
$R_f=0.3$ Mpc (dot-dashed line), $R_f=0.5$ Mpc (short-dashed line), $R_f=0.7$ Mpc (long-dashed line). In Eq. (\ref{eq:dynfric})
units have been expressed in terms of $M_{\odot}$, Mpc, $T_{c0}$.}
\end{figure}

\begin{figure}[ht]
\psfig{file=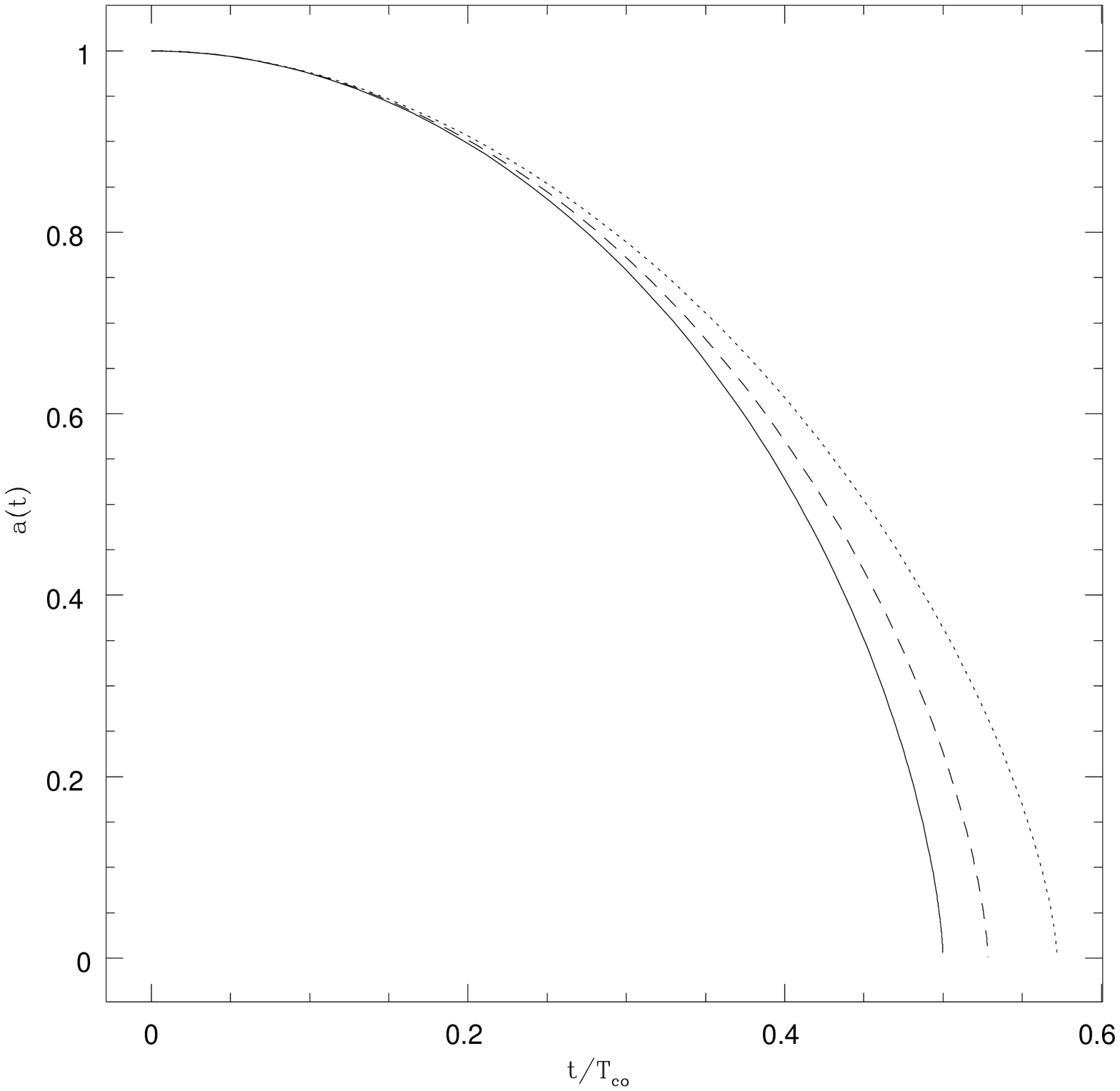,width=12cm,height=12cm}
\caption[]{The time evolution of the expansion parameter. The solid 
line is a(t) for the pure radial collapse (Gunn \& Gott 1972); the dashed line is a(t) taking account only
dynamical friction; the dotted line is a(t) taking account of the cumulative
effect of non-radial motions and dynamical friction in the case of a $\nu=3$ peak.}
\end{figure}

\begin{thebibliography}{}







\bibitem{}  Abramowitz M., Stegun I. A., 1981, Handbook of Mathematical Functions with Formulas, Graphs, and Mathematical Tables (Washington: National Bureau of Standards)
\bibitem{} Antonuccio-Delogu, V., Atrio-Barandela, F., 1992, ApJ 392,
403
\bibitem{}Antonuccio-Delogu V., Colafrancesco S., 1994, ApJ 427, 72 (ADC)
\bibitem{} Arabadjis J. S., Bautz M. W., Garmire G. P., 2002, ApJ,
572, 66
\bibitem{} Ascasibar Y., Yepes G., Gottlober S., 2004, MNRAS 352, 1109A
\bibitem{} Ascasibar Y.,  Hoffman Y., Gottl\"ober S., 2007, MNRAS 376, 393
\bibitem{} Audit, E., Teyssier, R. and Alimi, J. M., 1997, \aap, 325, 439
\bibitem{} Avila-Reese V., Firmani C., Hernandez X., 1998, ApJ, 505,
37
\bibitem{} Avila-Reese V., Firmani C., Klypin A., Kravtsov A., 1999, MNRAS 310, 527
\bibitem{} Avila-Reese V., Colin P., Valenzuela O., D'Onghia E. \& Firmani, C. 2001, ApJ 559, 516
\bibitem{} Bahcal, N. A., Soneira, R. M., 1983, ApJ 270, 20
%\bibitem{} Bahcal \& Fan 1998....
\bibitem{} Barnes, J., Efstathiou, G., 1987, ApJ 319, 575
\bibitem{} Barnes J., 1987, in Nearly Normal Galaxies, ed. Faber (NY: Springer), 154
\bibitem{} Barrow, J.D. and Silk, J., 1981, ApJ 250, 432
\bibitem{} Bartelman M., Meneghetti M., 2004, A\&A 418, 413
%\bibitem{} Blumenthal G.~R., Faber S.~M., Flores R., \& Primack J.~R. 1986, ApJ 301, 27
\bibitem{} Bernardeau, F., 1994, ApJ 427, 51
\bibitem{} Bertschinger E., 1985, ApJS 58, 39
\bibitem{} Binney J., Knebe A., 2002, MNRAS, 333, 378
\bibitem{} Blumenthal G. R., Faber S. M., Flores R., Primack J. R.,
1986, ApJ, 301, 27
\bibitem{} Bond, J.R. and Myers, S.T., 1993a, preprint CITA/93/27   
\bibitem{} Bond, J.R. and Myers, S.T., 1993b, preprint CITA/93/28   
\bibitem{} Bonometto S.A., Lucchin F., 1978, A\&A, 67, L7
\bibitem{} Borriello A., Salucci P., 2001, MNRAS, 323, 285
\bibitem{} Bothun, G., Impey, C. \& McGaugh, S. 1997, PASP, 109, 745  %OK
\bibitem{} Bryan G. L., Norman M. L., ApJ 495, 80
\bibitem{} Bullock, J.S., Dekel, A., Kolatt, T. S., Kravtsov, A.V., Klypin, A. A., 
Porciani, C. \& Primack, J. R. 2001, ApJ, 555, 240 
\bibitem{} Burkert A., 1995, ApJ, 447, L25
\bibitem{} Cardone V. F., Sereno M., 2005, A\&A 438, 545
\bibitem{}Catelan P., Theuns T., 1996, MNRAS2 82, 436
%\bibitem{}Catelan P., Theuns T., 1996b, MNRAS 282, 455
\bibitem{} Cen R., 2001, ApJ, 546, L77
\bibitem{} Cen R. Y., Dong F., Bode P., Ostriker J. P., 2004, astro-ph/0403352
\bibitem{} Chandrasekhar, S., von Neumann, J., 1942, ApJ, 95, 489
\bibitem{} Cole S., Lacey C., 1996, MNRAS 281, 716
\bibitem{} Colin P., Avila-Reese V., Valenzuela O., 2000, ApJ 542,
622
\bibitem{} Courteau S. \& Rix H. 1999, ApJ 513, 561
\bibitem{} Dahle H., Hannestad S., Sommer-Larsen J., 2003, ApJ, 588,
L73
\bibitem{} Dalal N., Keeton C. R., 2003, astro-ph/0312072
\bibitem{} Dalcanton J.~J., Spergel D.~N., \& Summers, F.~J. 1997, ApJ 482, 659
\bibitem{} Davis M., Peebles P.J.E., 1977, ApJS, 34, 425
\bibitem{} Davis, M., Peebles, P. J. E., 1983, Ap J, 267, 465
\bibitem{} Dav\'e R., Spergel D. N., Steinhardt P. J., Wandelt B. D.,
2001, ApJ, 547, 574
\bibitem{} de Blok, W.~J.~G., McGaugh, S.~S., Bosma, A., \& Rubin,
V.~C. 2001a, ApJ, 552, L23
\bibitem{} de Blok, W.~J.~G., McGaugh, S.~S., \& Rubin, V.~C. 2001b,
AJ, 122, 2396
\bibitem{} de Blok, W.~J.~G., Bosma, A., \& McGaugh, S.\ 2003, MNRAS
340, 657
\bibitem{} de Blok W. J. G., McGaugh S. S., Bosma A., Rubin V. C.,
2001, ApJ 552, L23
\bibitem{} de Blok W. J. G., Bosma A., 2002, A\&A 385, 816
\bibitem{} de Blok W. J. G.,  2003, in dark matter in Galaxies, ASP Conference series, Vol. 220, 2003, S. Ryder, D. J. Pisano, M. Walker, and K. C. Freeman, eds. 
\bibitem{} Del Popolo A., Gambera M., 1996, 308, 373
\bibitem{} Del Popolo A., E.N. Ercan, Z. Q Xia, 2001, AJ 122, 487
\bibitem{}  Del Popolo A., Gambera M., Recami E., Spedicato E., 2000, A\&A 353, 427 (DP2000)
\bibitem{} Del Popolo A., 2001, MNRAS 325, 1190
%\bibitem{} Del Popolo A., Ercan, E. N., Xia, Z. Q.,  2001, AJ 122, 487
\bibitem{} Del Popolo A., 2002, A\&A 387, 759
\bibitem{} Diemand J., Moore B., Stadel J., Kazantzidis S., 2004a, MNRAS 348, 977 
%astro-ph/0304549
\bibitem{} Diemand J., Moore, B., Stadel J., 2004b, MNRAS 353, 624
% ......astro-ph/0402267
\bibitem{} Dubinski J., Carlberg R., 1991, ApJ 378, 496
\bibitem{} Efstathiou G., Frenk C.S., White S.D.M., Davis M., 1988, MNRAS 235, 715
\bibitem{} Eggen O.~J., Lynden-Bell D., \& Sandage, A.~R. 1962, ApJ, 136, 748
\bibitem{} Eisenstein D.J., Loeb A., 1995, ApJ 439, 250
\bibitem{} El-Zant A., Shlosman I., Hoffman Y., 2001, ApJ, 560, 636 (EZ01)
\bibitem{} El-Zant A., Hoffman Y., Primack J., Combes F., Shlosman I., 2004, ApJ, ApJ 607, 75
%submitted (astro-ph/0309412)
\bibitem{} Ettori S., Fabian A. C., Allen S. W., Johnstone R. M.,
2002, MNRAS, 331, 635 
\bibitem{} Fall S. M. 1983, in IAU Symp. 100, Internal Kinematics and Dynamics of Galaxies, ed. E. Athanassoula (Dordrecht: Reidel), 391

\bibitem{} Filmore J.A., Goldreich P., 1984, ApJ 281, 1
\bibitem{} Flores R. A., Primack J. R., 1994, ApJ, 427, L1
\bibitem{} Flores R., Primack J.~R., Blumenthal, G.~R., \& Faber, S.~M. 1993, ApJ 412, 443

\bibitem{} Fry J. N., 1984, ApJ 279, 499
\bibitem{} Fukushige T. \& Makino, J, 2001, ApJ 557, 533
\bibitem{} Fukushige T., Kawai A., Makino J. 2004, ApJ 606, 625 
\bibitem{} Ghigna, S., Moore, B., Governato, F., Lake, G., Quinn, T., \& Stadel, J. 1998, MNRAS, 300, 146 
\bibitem{} Ghigna S., Moore B., Governato F., Lake G., Quinn T., Stadel J., 2000, ApJ, 544, 616 
\bibitem{} Gao L., White S.D.M., 2007, MNRAS 377, 5
\bibitem{} Gavazzi R., Fort B., Mellier Y., Pello R., Dantel-Fort M.,
2003, A\&A, 403, 11
\bibitem{} Gavazzi R., 2003, in Impact of Gravitational Lensing on Cosmology Proceedings IAU Symposium N. 225, Mellier Y., \& Meylan G. eds.
\bibitem{} Gelato S., Sommer-Larsen J., 1999, MNRAS 303, 321
\bibitem{} Gentile G., Salucci P., Klein U., Vergani D., Kalberla P., 2004, MNRAS 351, 903
\bibitem{} Gentile G., Tonini C. \& Salucci P., 2007, MNRAS 378, 41
%.....astro-ph/0701550
\bibitem{} Gonzalez A. H., Zaritsky D., Wechsler R., 2002 ApJ 571, 129
\bibitem{} Goodman J., 2000, New Astronomy 5, 103
\bibitem{} Gott J.R., 1975, ApJ 201, 296 
\bibitem{} Graham, A. W., Merritt D., Moore B., Diemand J. Terzic B., 2006, AJ 132, 2711 
\bibitem{} Gunn J.E., Gott J.R., 1972, ApJ 176, 1
\bibitem{} Gunn J.E., 1977, ApJ 218, 592
\bibitem{} Gurevich A. V., Zybin K. P., 1988a, Zhurnal Eksperimental noi i Teoreticheskoi Fiziki, 94, 3
\bibitem{} Gurevich A. V., Zybin K. P., 1988b, Zhurnal Eksperimental noi i Teoreticheskoi Fiziki, 94, 5
\bibitem{} Guth A.H., Pi S.Y., 1982, Phys.Rev.Lett. 49, 1110
\bibitem{} Hayashi E., Navarro J. F., Power C., Jenkins A., Frenk C. S., White S. D. M., Springel V., Stadel J., Quinn T. R., 2004 MNRAS, 355, 794
\bibitem{} Hawking S.W., 1982, Phys.Lett B 115, 295
\bibitem{} Heavens A., \& Peacock J. 1988, MNRAS 232, 339
\bibitem{}  Hernquist L., 1990, ApJ 356, 359
\bibitem{} Henriksen R. N., Widrow L. M., 1999, MNRAS 302, 321
\bibitem{} Hiotelis, N. 2002, A\&A 383, 84
\bibitem{} Hoeft M., Mucket J. P., Gottlober S., 2004, ApJ 602, 162
\bibitem{} Hoffman, Y.: (1986), ApJ 301, 65
\bibitem{} Hoffman Y., 1988, ApJ 328, 489
\bibitem{} Hoffman Y., Shaham J., 1985, ApJ 297, 16 (HS)
%\bibitem{} Huss A., Jain B., Steinmez M., 1998, preprint SISSA astro-ph/9803117
\bibitem{} Hoyle, F.: (1949), in IAU and International Union of Theorethicaland Applied Mechanics Symposium, p. 195
\bibitem{} Hu W., Barkana R., Gruzinov A., 2000, Physical Review
Letters, 85, 1158
\bibitem{} Huss, A., Jain, B. Steinmetz, M. 1999a, ApJ, 517, 64 %astro-ph/9803117  %OK
\bibitem{} Huss, A., Jain, B. Steinmetz, M. 1999b, MNRAS, 308, 1011
\bibitem{} Jesseit R., Naab T. \& Burkert A. 2002, ApJL 571, L89 
\bibitem{} Jimenez R., Verde L., Oh S. P., 2003, MNRAS 339, 243 
\bibitem{} Jing, Y. P. \& Suto, Y. 2000, ApJ, 529, L69  %OK
\bibitem{} Kandrup, H.E., 1980, Phys. Rep. 63, n 1, 1
\bibitem{} Kaplinghat M., Knox L., Turner M. S., 2000, Physical Review Letters, 85, 3335
\bibitem{} Keeton C. R., 2001, ApJ 561, 46 
\bibitem{} Kleyna J. T., Wilkinson M. I., Gilmore Gerard, Evans N. W., 2003 ApJ 588, L21
\bibitem{} Klypin, A., Kravtsov, A. V., Bullock, James S. \& Primack, J. R. 2001, ApJ, 554, 903 %OK
\bibitem{} Klypin A., Zhao H., \& Somerville R.~S. 2002, ApJ 573, 597
\bibitem{} Koopmans L. V. E., Treu T., 2003, ApJ 583, 606
\bibitem{} Kravtsov A. V., Klypin A. A., Bullock J. S., Primack J. R.,
1998, ApJ, 502, 48
\bibitem{} Kriessler J.R., Beers T.C., Odewahn S.C., 1995, Bull. AAS 186, 0702
\bibitem{} Kull A., 1999, ApJ 516, L5
\bibitem{} Lacey C.G., Cole S.M., 1993, MNRAS 262, 627
\bibitem{} Le Delliou M., Henriksen R. N., 2003, A\&A 408, 27
\bibitem{} Lemson, G., 1995, Ph.D. thesis, R\"{\i}ksuniversiteit Groningen  
\bibitem{} Lewis A. D., Buote D. A., Stocke J. T., 2003, ApJ 586,
135
\bibitem{} Lifshitz, E. M.; Pitaevskii, L. P., 1981,  Title: Physical kinetics , Publication: Course of theoretical physics, Oxford: Pergamon Press, 1981
\bibitem{} Lokas, E.L., Juskiewicz, R., Bouchet, F.R. and Hivon, E., 1996, ApJ 467, 1
\bibitem{} Lokas, E. L. 2000, MNRAS, 311, 432
\bibitem{} Lokas, E. L. \& Hoffman, Y. 2000, ApJ 542, L139
\bibitem{} Magorrian J., 2003, in The Mass of Galaxies at Low and High Redshift: Proceedings of the ESO Workshop Held in Venice, Italy, 24-26 October 2001, ESO ASTROPHYSICS SYMPOSIA. ISBN 3-540-00205-7. Edited by R. Bender and A. Renzini. Springer-Verlag, 2003, p. 18
\bibitem{} Mamon G.A., Lokas E. L., MNRAS 2005, 362, 95
%...astro-ph/0405466
\bibitem{} Manrique A., Raig A., Salvador-Sol\'e E., Sanchis T., Solanes
J. M., 2003, ApJ, 593, 26
\bibitem{} Marchesini D., D'Onghia E., Chincarini G., Firmani C.,
Conconi P., Molinari E., Zacchei A., 2002, ApJ, 575, 801
\bibitem{} Mashchenko S., Couchman H. M. P., Wadsley J., Nature 442, 539
\bibitem{} Mashchenko S. \& Sills,  2005, ApJ 619, 258.
\bibitem{} McGaugh S. S., de Blok W. J. G., 1998, ApJ 499, 41
\bibitem{} Merrit D., Navarro J. F., Ludlow A., Jenkins A., 2005, ApJ 624, L85
\bibitem{} Mo H.~J., Mao S., \& White S.~D.~M. 1998, MNRAS 295, 319
\bibitem{} Moore B., 1994, Nature, 370, 629
\bibitem{} Moore, B., Governato, F., Quinn, T., Stadel, J. Lake, G. 1998, ApJ, 499, L5 
\bibitem{} Moore, B., Quinn, T., Governato, F.,  Stadel, J. Lake, G. 1999, MNRAS 310, 1147
\bibitem{} Moore, B., 2001, in 20-th Texas Symposium on relativistic astrophysics, Austin, Texas, 
%10-15 Decem ber 2000, Melville, NY: American Institute of Physics, 2001, xix, 938 p. 
AIP conference proceedings, Vol. 586, p. 73. Edited by J. Craig Wheeler and Hugo Martel. 
\bibitem{} Navarro J.F., Frenk C.S., White S.D.M., 1995, MNRAS 275, 720
\bibitem{} Navarro J.F., Frenk C.S., White S.D.M., 1996, ApJ 462, 563
\bibitem{} Navarro J.F., Frenk C.S., White S.D.M., 1997, ApJ 490, 493 (NFW) 
\bibitem{} Navarro, J. F. \& Steinmetz, M. 2000, ApJ, 538, 477
\bibitem{} Navarro J. F., Hayashi E., Power C., Jenkins A. R., Frenk C. S., White S. D. M., Springel V., Stadel J., Quinn T. R, 2004,MNRAS 349, 1039
\bibitem{} Navarro J. F., Ludlow A., Springel V., Wang J., Vogelsberger M., White S. D. M., Jenkins A., Frenk C. S., Helmi A., 2008, astro-ph/0810.1522 
%\bibitem{} Nusser A., Dekel A., 1992, ApJ 362, 14
%\bibitem{} Nusser A., Sheth R.K., 1998, preprint SISSA astro-ph/9803281
\bibitem{} Nusser A., Sheth R. K., 1999, MNRAS, 303, 685
\bibitem{} Nusser A., 2001, MNRAS 325, 1397
\bibitem{} Oh K. S., 1990, Ph.D. thesis, Univ. California-Santa Cruz
\bibitem{} Ostriker J. P., Steinhardt P., 2003, Science 300,.1909 
\bibitem{} Peacock, J.A., Heavens,A.F., 1990, MNRAS 243, 133
\bibitem{} Peebles, P. J. E., 1969, ApJ 155, 393
\bibitem{} Peebles P.J.E., 1974, ApJ 189, L51
\bibitem{} Peebles P.J.E., Groth E.J., 1976, A\&A 53, 131
\bibitem{} Peebles P.J.E., 1980, The large scale structure of the Universe, Priceton University Press
\bibitem{} Peebles P.J.E., 1984, ApJ 277, 470
\bibitem{} Peebles, P.J.E., 1990, ApJ 365, 27
\bibitem{} Peebles P. J. E., 2000, ApJ 534, L127
\bibitem{} Postman, M., Geller, M. J., Huchra, J. P., 1986, ApJ 91, 1267
\bibitem{} Power, C., Navarro, J. F., Jenkins, A., Frenk, C. S., White, S. D. M., Springel V., Stadel J., Quinn T., 2003, MNRAS 338, 14
\bibitem{} Quinn P.J., Zurek W.H., 1988, ApJ 331, 1
\bibitem{} Quinn P.J., Salmon J.K., Zurek W.H., 1986, Nature, 322, 329
\bibitem{} Ricotti M., 2003, MNRAS 344, 1237
\bibitem{} Ricotti M., Wilkinson M. I., 2004, MNRAS 353, 867 
\bibitem{} Ricotti M., Pontzen A., Viel M., 2007, ApJ 663, 53
\bibitem{}  Rix, Hans-Walter; de Zeeuw, P. Tim; Cretton, Nicolas; van der Marel, Roeland P.; Carollo, C. Marcella,  1997, ApJ 488, 702 
\bibitem{} Romano-Diaz E., Shlosman I., Hoffman Y., Heller C., 2008, astro-ph 0808.0195
\bibitem{} Romeo A. B., Agertz O., Moore B., Stadel J., 2008, astro-ph/0804.0294
\bibitem{} Ryden B.S., Gunn J.E., 1987, ApJ 318, 15 (RG87)
\bibitem{} Ryden, B.S., 1988a, ApJ 329, 589 (R88)
\bibitem{} Ryden, B.S., 1988b, ApJ 333, 78
\bibitem{} Ryden B. S., 1991, ApJ 370, 15
\bibitem{} Salucci P., Burkert A., 2000, ApJ 537, L9 
\bibitem{} Salvador-Sol\'e E., Solanes J. M., Manrique A., 1998, ApJ 499, 542
\bibitem{} Sand D. J., Treu T., Ellis R. S., 2002, ApJ 574, L129
\bibitem{} Sand D. J., Treu T., Smith G. P., Ellis R. S., 2004, ApJ 604, 88
%submitted (astro-ph/0309465)
\bibitem{} Seljak U. 2002, MNRAS 334, 797
\bibitem{} Sikivie P., Tkachev I.I, Wang Y., 1997, Phys. Rev. D. 56, 1863
\bibitem{} Simon J. D., Bolatto A.D., Leroy A., Blitz L., 2003a, ApJ 596, 957
\bibitem{} Simon J. D., Bolatto A. D., Leroy A. Blitz L., 2003b, in Satellites and Tidal Streams, ASP conference Series 2003
%, astro-ph/0310193
\bibitem{} Stadel J., Potter D., Moore B., Diemand J., Madau P., Zemp M., Kuhlen M., Quilis V, 2008, astro-ph/0808.2981
\bibitem{} Syer D., White S. D. M., 1998, MNRAS 293, 337
\bibitem{} Sommer-Larsen J., Dolgov A., 2001, ApJ 551, 608
\bibitem{} Spekkens K. Giovanelli R., Haynes M. P., 2005 AJ 129, 2119 
\bibitem{} Spergel D .N., et al. 2003, ApJS 148, 175
\bibitem{} Spergel D. N., Steinhardt P. J., 2000, Physical Review Letters 84, 3760
\bibitem{} Starobinsky A.A., 1982, Phys.Lett. B 117, 175
\bibitem{} Steigman G., Sarazin C. L., Quintana H., Faulkner J., 1978, AJ 83, 1050 
\bibitem{} Subramanian K., Cen R., Ostriker J. P., 2000, ApJ 538,
528
\bibitem{} Swaters, R.~A., Madore, B.~F., van den Bosch, F.~C., \& Balcells, M. 2003, ApJ 583, 732 
%(SMVB)
\bibitem{} Szalay, A.S. and  Silk J., 1983, ApJ, 264, L31
\bibitem{} Taylor J. E., Navarro J. F., 2001, ApJ 563, 483
\bibitem{} Taylor J. E., Silk J., Babul A., 2004, IAUS no. 220, held 21 - 25 July, 2003 in Sydney, Australia. Eds: S. D. Ryder, D. J. Pisano, M. A. Walker, and K. C. Freeman. San Francisco: Astronomical Society of the Pacific., p.91
\bibitem{} Teyssier R., Cieze J.P, Alimi J.M, 1997, ApJ 480, 36
Tormen G., Bouchet F.R., White S.D.M., 1997, MNRAS 286, 865
\bibitem{} Tonini C., Lapi A., Salucci P., 2006, ApJ 649, 591
\bibitem{} Toth G., Ostriker J. P., 1992, ApJ 389, 5
\bibitem{} Treu T. \& Koopmans L.~V.~E., 2002, ApJ 575, 87
\bibitem{} Treu T. \& Koopmans L.~V.~E., 2004, ApJ 611, 739
%%%%%%%\bibitem[{{Treu} \& {Koopmans}(2004)}]{treu_koopmans04} ---. 2004, \apj, in press (astro-ph/0401373)
\bibitem{} van den Bosch, F.~C., Lewis G.~F., Lake G., \& Stadel J. 1999, ApJ 515, 50
%\bibitem{} van den Bosch, F.~C., \& Swaters, R.~A., 2000.............
\bibitem{} van den Bosch, F.~C., \& Swaters, R.~A.\ 2001, \mnras 325,
1017
\bibitem{} van den Bosch, F.~C., Robertson, B.~E., Dalcanton, J.~J., \& de Blok, W.~J.~G.  2000, AJ 119, 1579
\bibitem{} van den Bosch, F. C., Abel, T., Croft, R. A. C., Hernquist, L.  \& White S. D. M. 2002, ApJ 576, 21
\bibitem{} Villumsen, J.V. and Davis, M., 1986, ApJ 308. 499
\bibitem{} Voglis N., Hiotelis N., Harsoula M., 1995, Ap\&SS 226, 213
\bibitem{} Yoshida N., Springel V., White S. D. M., Tormen G., 2000, ApJ 544, L87
\bibitem{} Weinberg M. D., Katz N., 2002, ApJ 580, 627
\bibitem{} West M.J., Dekel A., Oemler A., 1987, ApJ 316, 1
\bibitem{} White, S. D. M., 1984, ApJ 286, 38
\bibitem{} White S.D.M., Zaritsky D., 1992, ApJ 394, 1
%\bibitem{} Yano T., Masahiro N., Gouda N., 1996, ApJ 466, 1
\bibitem{} Williams L. L. R.,  Babul A.,  Dalcanton, J. J., 2004, ApJ 604, 18
\bibitem{} Zaroubi S., Hoffman Y., 1993, ApJ 416, 410
\bibitem{} Zaroubi S., Naim A., Hoffman Y., 1996, ApJ 457, 50
\bibitem{} Zeldovich Y.~B., Klypin A.~A., Khlopov M.~Y., \& Chechetkin V.~M., 1980, Soviet J. Nucl. Phys. 31, 664
\end{thebibliography}
\end{document}